\documentclass[journal]{IEEEtran}
\usepackage{latexsym}
\usepackage{graphicx}
\usepackage{amsfonts,amssymb,amsmath}
\usepackage{bm}
\usepackage{hyperref}
\usepackage{xcolor}
\usepackage{verbatim}
\usepackage{booktabs}
\usepackage{tikz}
\usetikzlibrary{matrix,chains,positioning,decorations.pathreplacing,arrows,arrows.meta,external,shapes, shapes.geometric,shapes.misc,calc,external}
\usepackage{tikz-3dplot} 
\usepackage{pgfplots}
\usepackage{multirow}

\usepackage[justification=RaggedRight]{caption}
\usepackage[font=footnotesize]{subfig}

\usepackage{cite}

\usepackage{float}

\graphicspath{{photos/}}

\tikzstyle{atipstyle}=[>={Stealth[inset=0pt,length=6pt,angle'=40,round]}]
\tikzset{linestyle/.style={thick, rounded corners}}
\tikzset{softmax/.style={ellipse, text centered, inner sep=4pt, draw=black, thick}}
\tikzset{model/.style={rectangle, rounded corners, text centered, inner sep=5pt, text width=2cm, minimum width=3cm, draw=black, thick}}

\pgfplotsset{compat=1.8}
\usepgfplotslibrary{statistics}

\newcommand{\desboxplot}[9]{%
 \addplot[ boxplot prepared={%
  lower whisker={#1}, lower quartile={#2}, median={#3},
  upper quartile={#4}, upper whisker={#5}, average={#6},
  box extend=0.5,  
  whisker extend=0.5, 
  every box/.style={thin,draw=black,fill=gray!50},
  every whisker/.style={black,thick},
  every median/.style={black,thick},
  every average/.style={draw=black, /tikz/mark=* },
  /pgf/number format/precision=2 } ]
  coordinates {}
  node[right,black] at
  (boxplot box cs: \boxplotvalue{upper whisker},0.5)
  {\scriptsize {#7}-{#8}-{#9}}
  ;
} 

\newcommand{\desboxplotgrouped}[8]{%
 \addplot[ boxplot prepared={%
  lower whisker={#1}, lower quartile={#2}, median={#3},
  upper quartile={#4}, upper whisker={#5}, average={#6},
  box extend=0.5,  
  whisker extend=0.5, 
  draw position={#8},
  every box/.style={thin,draw=black,fill=gray!50},
  every whisker/.style={black,thick},
  every median/.style={black,thick},
  every average/.style={draw=black, /tikz/mark=* },
  /pgf/number format/precision=2 },
  ]
  coordinates {}
  node[right,black] at
  (boxplot box cs: \boxplotvalue{upper whisker},0.5)
  {\scriptsize {#7}}
  ;
} 

\newcommand{\desboxplottotal}[9]{%
 \addplot[ boxplot prepared={%
  lower whisker={#1}, lower quartile={#2}, median={#3},
  upper quartile={#4}, upper whisker={#5}, average={#6},
  box extend=0.7,  
  whisker extend=0.35, 
  every box/.style={thin,draw=black,fill=gray!50},
  every whisker/.style={black,thick},
  every median/.style={black,thick},
  every average/.style={draw=red, /tikz/mark=* },
  /pgf/number format/precision=2 } ]
  coordinates {}
  node[right,black] (stats) at
  (boxplot box cs: \boxplotvalue{upper whisker},0.5)
  {\scriptsize {#7}-{#8}-{#9}}
  node[above] (min) at
    (boxplot box cs: \boxplotvalue{lower whisker}+4,0.5)
    {\small min}
  node[above] (q1) at
    (boxplot box cs: \boxplotvalue{lower quartile}-0.5,1.0)
    {\small 1st quartile}
  node[left] (median) at
    (boxplot box cs: \boxplotvalue{median}-2,-0.1)
    {\small median}
  node[above] (q3) at
    (boxplot box cs: \boxplotvalue{upper quartile}+15,-0.35)
    {\small 3rd quartile}
  node[above] (max) at
   (boxplot box cs: \boxplotvalue{upper whisker}-5,0.5)
   {\small max}
  node[above] (avg) at
   (boxplot box cs: \boxplotvalue{average}+13,0.7)
   {\small average}
  node[above] (nsamples) at
    (boxplot box cs: \boxplotvalue{upper whisker}-12,0.0)
    {\small \#samples}
  node[above] (ndatasets) at
    (boxplot box cs: \boxplotvalue{upper whisker}+7,+0.8)
    {\small \#datasets}
  node[above] (nstudies) at
    (boxplot box cs: \boxplotvalue{upper whisker}+7,-0.2)
    {\small \#studies};
  \draw[->] (nsamples)--(axis cs:77,0.95);
  \draw[->] (ndatasets)--(stats);
  \draw[->] (nstudies)--(axis cs:88,0.9);
  \draw[->] (q3)--(axis cs:13.0,0.8);
  \draw[->] (q1.south)--(axis cs:3.2,1.3);
  \draw[->] (avg.south)--(axis cs:10.5,1.0);
  \draw[->] (median.north)--(axis cs:6.4,0.8);
  ;
} 

\newcommand{\nstudies}{47}
\newcommand{\ndsets}{38}
\newcommand{\npoints}{356}
\newcommand{\repourl}{\url{https://github.com/pswietojanski/ojsp_adaptation_review_2020}}

\begin{document}

\title{Adaptation Algorithms for Neural Network-Based Speech Recognition: An Overview}

\author{%
Peter Bell, \IEEEmembership{Member, IEEE},
Joachim Fainberg, \IEEEmembership{Member, IEEE},
Ondrej Klejch,  \IEEEmembership{Member, IEEE},\\
Jinyu Li,  \IEEEmembership{Member, IEEE},
Steve Renals, \IEEEmembership{Fellow, IEEE}, and 
Pawel Swietojanski, \IEEEmembership{Member, IEEE}
\thanks{Authors have made equal contributions and are listed alphabetically.}
\thanks{Peter Bell, Ondrej Klejch, and Steve Renals are with the Centre for Speech Technology Research, University of Edinburgh, Edinburgh EH8 9AB, UK (email: peter.bell@ed.ac.uk, o.klejch@ed.ac.uk, s.renals@ed.ac.uk).}
\thanks{Joachim Fainberg did this work at the Centre for Speech Technology Research, University of Edinburgh, Edinburgh EH8 9AB, UK. He is now with JPMorgan Chase, New York, NY 10017, USA (email: j.fainberg@ed.ac.uk).}
\thanks{Jinyu Li is with the Microsoft Corporation, Redmond, WA 98052, USA (e-mail: jinyli@microsoft.com)}
\thanks{Pawel Swietojanski did this work at the School of Computer Science and Engineering, University of New South Wales, Sydney, NSW 2052, Australia. He is now with Apple, Cambridge, UK (email: p.swietojanski@unsw.edu.au).}
\thanks{This work was partially supported by the  EPSRC Project EP/R012180/1 (SpeechWave), a PhD studentship funded by Bloomberg, and the EU H2020 project ELG (grant agreement 825627).}
\thanks{This research is based upon work supported in part by the Office of the Director of National Intelligence (ODNI), Intelligence Advanced Research Projects Activity (IARPA), via Air Force Research Laboratory (AFRL) contract \#FA8650-17-C-9117. The views and conclusions contained herein are those of the authors and should not be interpreted as necessarily representing the official policies, either expressed or implied, of ODNI, IARPA, AFRL or the U.S. Government. The U.S. Government is authorized to reproduce and distribute reprints for governmental purposes notwithstanding any copyright annotation therein.}
}



\newcommand{\PAWEL}[1]{}
\newcommand{\PETER}[1]{}
\newcommand{\ONDREJ}[1]{}
\newcommand{\JOACHIM}[1]{}
\newcommand{\JINYU}[1]{}
\newcommand{\STEVE}[1]{}
\newcommand{\TODO}[2]{}

\newcommand{\eq}{\textit{Eq.}}
\newcommand{\eg}{\textit{e.g.~}}
\newcommand{\ie}{\textit{i.e.~}}
\newcommand{\etal}{\textit{et al.}}

\newcommand\ignore[1]{}


\newcommand{\fig}{Fig.}
\newcommand{\figref}[1]{Fig.~\ref{#1}}
\newcommand{\secref}[1]{Sec.~\ref{#1}}

\newcommand{\cf}{\textit{cf.~}}
\newcommand{\etc}{\textit{etc.~}}
\newcommand*\diff{\mathop{}\!\mathrm{d}}
\newcommand{\ds}{\diff \s}
\newcommand{\dx}{\diff x}
\newcommand{\dy}{\diff y}

\newcommand{\transpose}{^\mathrm{T}}
\newcommand{\inv}{^{-1}}

\def \spanof{\mathop{\rm span}\nolimits}
\def \clip{\mathrel{\rm clip}}
\def \det{\mathop{\rm det}\nolimits}
\def \dim{\mathop{\rm dim}\nolimits}
\def \op{\mathrel{\rm op}}
\def \outcode{\mathop{\rm outcode}\nolimits}
\def \pixel{\mathop{\rm pixel}\nolimits}
\def \rank{\mathop{\rm rank}\nolimits}
\def \round{\mathrel{\rm round}}
\def \sgn{\mathop{\rm sgn}\nolimits}
\def \sinc{\mathrel{\rm sinc}}
\def \spur{\mathop{\rm spur}\nolimits}
\def \stencil{\mathop{\rm stencil}\nolimits}
\def \supp{\mathop{\rm supp}\nolimits}
\def \weci{\mathop{\rm wec}\nolimits}
\def \wec{\mathop{\rm WEC}\nolimits}
\def \zbuff{\mathop{\rm zBuffer}\nolimits}
\def \ease{\mathop{\rm ease}\nolimits}
\def \path{\bp C}
\def \profile{\bp S}


\newcommand{\bfa}{{\mathbf{a}}}
\newcommand{\bfb}{{\mathbf{b}}}
\newcommand{\bfc}{{\mathbf{c}}}
\newcommand{\bfd}{{\mathbf{d}}}
\newcommand{\bfe}{{\mathbf{e}}}
\newcommand{\bff}{{\mathbf{f}}}
\newcommand{\bfg}{{\mathbf{g}}}
\newcommand{\bfh}{{\mathbf{h}}}
\newcommand{\bfi}{{\mathbf{i}}}
\newcommand{\bfj}{{\mathbf{j}}}
\newcommand{\bfk}{{\mathbf{k}}}
\newcommand{\bfl}{{\mathbf{l}}}
\newcommand{\bfm}{{\mathbf{m}}}
\newcommand{\bfn}{{\mathbf{n}}}
\newcommand{\bfo}{{\mathbf{o}}}
\newcommand{\bfp}{{\mathbf{p}}}
\newcommand{\bfq}{{\mathbf{q}}}
\newcommand{\bfr}{{\mathbf{r}}}
\newcommand{\bfs}{{\mathbf{s}}}
\newcommand{\bft}{{\mathbf{t}}}
\newcommand{\bfu}{{\mathbf{u}}}
\newcommand{\bfv}{{\mathbf{v}}}
\newcommand{\bfw}{{\mathbf{w}}}
\newcommand{\bfx}{{\mathbf{x}}}
\newcommand{\bfy}{{\mathbf{y}}}
\newcommand{\bfz}{{\mathbf{z}}}
\newcommand{\bfell}{{\bolsymbol{ell}}}


\newcommand{\bfA}{{\mathbf{A}}}
\newcommand{\bfB}{{\mathbf{B}}}
\newcommand{\bfC}{{\mathbf{C}}}
\newcommand{\bfD}{{\mathbf{D}}}
\newcommand{\bfE}{{\mathbf{E}}}
\newcommand{\bfF}{{\mathbf{F}}}
\newcommand{\bfG}{{\mathbf{G}}}
\newcommand{\bfH}{{\mathbf{H}}}
\newcommand{\bfI}{{\mathbf{I}}}
\newcommand{\bfJ}{{\mathbf{J}}}
\newcommand{\bfK}{{\mathbf{K}}}
\newcommand{\bfL}{{\mathbf{L}}}
\newcommand{\bfM}{{\mathbf{M}}}
\newcommand{\bfN}{{\mathbf{N}}}
\newcommand{\bfO}{{\mathbf{O}}}
\newcommand{\bfP}{{\mathbf{P}}}
\newcommand{\bfQ}{{\mathbf{Q}}}
\newcommand{\bfR}{{\mathbf{R}}}
\newcommand{\bfS}{{\mathbf{S}}}
\newcommand{\bfT}{{\mathbf{T}}}
\newcommand{\bfU}{{\mathbf{U}}}
\newcommand{\bfV}{{\mathbf{V}}}
\newcommand{\bfW}{{\mathbf{W}}}
\newcommand{\bfX}{{\mathbf{X}}}
\newcommand{\bfY}{{\mathbf{Y}}}
\newcommand{\bfZ}{{\mathbf{Z}}}


\providecommand{\BA}{{\boldsymbol{A}}}
\providecommand{\BB}{{\boldsymbol{B}}}
\providecommand{\BC}{{\boldsymbol{C}}}
\providecommand{\BD}{{\boldsymbol{D}}}
\providecommand{\BE}{{\boldsymbol{E}}}
\providecommand{\BF}{{\boldsymbol{F}}}
\providecommand{\BG}{{\boldsymbol{G}}}
\providecommand{\BH}{{\boldsymbol{H}}}
\providecommand{\BI}{{\boldsymbol{I}}}
\providecommand{\BJ}{{\boldsymbol{J}}}
\providecommand{\BK}{{\boldsymbol{K}}}
\providecommand{\BL}{{\boldsymbol{L}}}
\providecommand{\BM}{{\boldsymbol{M}}}
\providecommand{\BN}{{\boldsymbol{N}}}
\providecommand{\BO}{{\boldsymbol{O}}}
\providecommand{\BP}{{\boldsymbol{P}}}
\providecommand{\BQ}{{\boldsymbol{Q}}}
\providecommand{\BR}{{\boldsymbol{R}}}
\providecommand{\BS}{{\boldsymbol{S}}}
\providecommand{\BT}{{\boldsymbol{T}}}
\providecommand{\BU}{{\boldsymbol{U}}}
\providecommand{\BV}{{\boldsymbol{V}}}
\providecommand{\BW}{{\boldsymbol{W}}}
\providecommand{\BX}{{\boldsymbol{X}}}
\providecommand{\BY}{{\boldsymbol{Y}}}
\providecommand{\BZ}{{\boldsymbol{Z}}}


\newcommand{\BGamma}{\mathbf{\Gamma}}
\newcommand{\bdelta}{\mathbf{\Delta}}
\newcommand{\btheta}{\mathbf{\Theta}}
\newcommand{\BLambda}{\mathbf{\Lambda}}
\newcommand{\BXi}{\mathbf{\Xi}}
\newcommand{\BPi}{\mathbf{\Pi}}
\newcommand{\BPsi}{\mathbf{\Psi}}
\newcommand{\BSigma}{\mathbf{\Sigma}}
\newcommand{\BUpsilon}{\mathbf{\Upsilon}}
\newcommand{\bphi}{\mathbf{\Phi}}
\newcommand{\BOmega}{\mathbf{\Omega}}


\newcommand{\nablabf}{\boldsymbol{\nabla}}
\newcommand{\alphabf}{\boldsymbol{\alpha}}
\newcommand{\betabf}{\boldsymbol{\beta}}
\newcommand{\gammabf}{\boldsymbol{\gamma}}
\newcommand{\deltabf}{\boldsymbol{\delta}}
\newcommand{\epsilonbf}{\boldsymbol{\epsilon}}
\newcommand{\varepsilonbf}{\boldsymbol{\varepsilon}}
\newcommand{\zetabf}{\boldsymbol{\zeta}}
\newcommand{\etabf}{\boldsymbol{\eta}}
\newcommand{\thetabf}{\boldsymbol{\theta}}
\newcommand{\varthetabf}{\boldsymbol{\vartheta}}
\newcommand{\iotabf}{\boldsymbol{\iota}}
\newcommand{\kappabf}{\boldsymbol{\kappa}}
\newcommand{\lambdabf}{\boldsymbol{\lambda}}
\newcommand{\mubf}{\boldsymbol{\mu}}
\newcommand{\nubf}{\boldsymbol{\nu}}
\newcommand{\xibf}{\boldsymbol{\xi}}
\newcommand{\pibf}{\boldsymbol{\pi}}
\newcommand{\varpibf}{\boldsymbol{\varpi}}
\newcommand{\rhobf}{\boldsymbol{\rho}}
\newcommand{\varrhobf}{\boldsymbol{\varrho}}
\newcommand{\sigmabf}{\boldsymbol{\sigma}}
\newcommand{\varsigmabf}{\boldsymbol{\varsigma}}
\newcommand{\taubf}{\boldsymbol{\tau}}
\newcommand{\upsilonbf}{\boldsymbol{\upsilon}}
\newcommand{\phibf}{\boldsymbol{\phi}}
\newcommand{\varphibf}{\boldsymbol{\varphi}}
\newcommand{\chibf}{\boldsymbol{\chi}}
\newcommand{\psibf}{\boldsymbol{\psi}}
\newcommand{\omegabf}{\boldsymbol{\omega}}
\newcommand{\Gammabf}{\boldsymbol{\Gamma}}
\newcommand{\Deltabf}{\boldsymbol{\Delta}}
\newcommand{\Thetabf}{\boldsymbol{\Theta}}
\newcommand{\Lambdabf}{\boldsymbol{\Lambda}}
\newcommand{\Xibf}{\boldsymbol{\Xi}}
\newcommand{\Pibf}{\boldsymbol{\Pi}}
\newcommand{\Psibf}{\boldsymbol{\Psi}}
\newcommand{\Sigmabf}{\boldsymbol{\Sigma}}
\newcommand{\Upsilonbf}{\boldsymbol{\Upsilon}}
\newcommand{\Phibf}{\boldsymbol{\Phi}}
\newcommand{\Omegabf}{\boldsymbol{\Omega}}


\newcommand{\sfa}{{\mathsf{a}}}
\newcommand{\sfb}{{\mathsf{b}}}
\newcommand{\sfc}{{\mathsf{c}}}
\newcommand{\sfd}{{\mathsf{d}}}
\newcommand{\sfe}{{\mathsf{e}}}
\newcommand{\sff}{{\mathsf{f}}}
\newcommand{\sfg}{{\mathsf{g}}}
\newcommand{\sfh}{{\mathsf{h}}}
\newcommand{\sfi}{{\mathsf{i}}}
\newcommand{\sfj}{{\mathsf{j}}}
\newcommand{\sfk}{{\mathsf{k}}}
\newcommand{\sfl}{{\mathsf{l}}}
\newcommand{\sfm}{{\mathsf{m}}}
\newcommand{\sfn}{{\mathsf{n}}}
\newcommand{\sfo}{{\mathsf{o}}}
\newcommand{\sfp}{{\mathsf{p}}}
\newcommand{\sfq}{{\mathsf{q}}}
\newcommand{\sfr}{{\mathsf{r}}}
\newcommand{\sfs}{{\mathsf{s}}}
\newcommand{\sft}{{\mathsf{t}}}
\newcommand{\sfu}{{\mathsf{u}}}
\newcommand{\sfv}{{\mathsf{v}}}
\newcommand{\sfw}{{\mathsf{w}}}
\newcommand{\sfx}{{\mathsf{x}}}
\newcommand{\sfy}{{\mathsf{y}}}
\newcommand{\sfz}{{\mathsf{z}}}

\newcommand{\sfA}{{\mathsf{A}}}
\newcommand{\sfB}{{\mathsf{B}}}
\newcommand{\sfC}{{\mathsf{C}}}
\newcommand{\sfD}{{\mathsf{D}}}
\newcommand{\sfE}{{\mathsf{E}}}
\newcommand{\sfF}{{\mathsf{F}}}
\newcommand{\sfG}{{\mathsf{G}}}
\newcommand{\sfH}{{\mathsf{H}}}
\newcommand{\sfI}{{\mathsf{I}}}
\newcommand{\sfJ}{{\mathsf{J}}}
\newcommand{\sfK}{{\mathsf{K}}}
\newcommand{\sfL}{{\mathsf{L}}}
\newcommand{\sfM}{{\mathsf{M}}}
\newcommand{\sfN}{{\mathsf{N}}}
\newcommand{\sfO}{{\mathsf{O}}}
\newcommand{\sfP}{{\mathsf{P}}}
\newcommand{\sfQ}{{\mathsf{Q}}}
\newcommand{\sfR}{{\mathsf{R}}}
\newcommand{\sfS}{{\mathsf{S}}}
\newcommand{\sfT}{{\mathsf{T}}}
\newcommand{\sfU}{{\mathsf{U}}}
\newcommand{\sfV}{{\mathsf{V}}}
\newcommand{\sfW}{{\mathsf{W}}}
\newcommand{\sfX}{{\mathsf{X}}}
\newcommand{\sfY}{{\mathsf{Y}}}
\newcommand{\sfZ}{{\mathsf{Z}}}


\newcommand{\cala}{{\mathcal{a}}}
\newcommand{\calb}{{\mathcal{b}}}
\newcommand{\calc}{{\mathcal{c}}}
\newcommand{\cald}{{\mathcal{d}}}
\newcommand{\cale}{{\mathcal{e}}}
\newcommand{\calf}{{\mathcal{f}}}
\newcommand{\calg}{{\mathcal{g}}}
\newcommand{\calh}{{\mathcal{h}}}
\newcommand{\cali}{{\mathcal{i}}}
\newcommand{\calj}{{\mathcal{j}}}
\newcommand{\calk}{{\mathcal{k}}}
\newcommand{\call}{{\mathcal{l}}}
\newcommand{\calm}{{\mathcal{m}}}
\newcommand{\caln}{{\mathcal{n}}}
\newcommand{\calo}{{\mathcal{o}}}
\newcommand{\calp}{{\mathcal{p}}}
\newcommand{\calq}{{\mathcal{q}}}
\newcommand{\calr}{{\mathcal{r}}}
\newcommand{\cals}{{\mathcal{s}}}
\newcommand{\calt}{{\mathcal{t}}}
\newcommand{\calu}{{\mathcal{u}}}
\newcommand{\calv}{{\mathcal{v}}}
\newcommand{\calw}{{\mathcal{w}}}
\newcommand{\calx}{{\mathcal{x}}}
\newcommand{\caly}{{\mathcal{y}}}
\newcommand{\calz}{{\mathcal{z}}}

\newcommand{\calA}{{\mathcal{A}}}
\newcommand{\calB}{{\mathcal{B}}}
\newcommand{\calC}{{\mathcal{C}}}
\newcommand{\calD}{{\mathcal{D}}}
\newcommand{\calE}{{\mathcal{E}}}
\newcommand{\calF}{{\mathcal{F}}}
\newcommand{\calG}{{\mathcal{G}}}
\newcommand{\calH}{{\mathcal{H}}}
\newcommand{\calI}{{\mathcal{I}}}
\newcommand{\calJ}{{\mathcal{J}}}
\newcommand{\calK}{{\mathcal{K}}}
\newcommand{\calL}{{\mathcal{L}}}
\newcommand{\calM}{{\mathcal{M}}}
\newcommand{\calN}{{\mathcal{N}}}
\newcommand{\calO}{{\mathcal{O}}}
\newcommand{\calP}{{\mathcal{P}}}
\newcommand{\calQ}{{\mathcal{Q}}}
\newcommand{\calR}{{\mathcal{R}}}
\newcommand{\calS}{{\mathcal{S}}}
\newcommand{\calT}{{\mathcal{T}}}
\newcommand{\calU}{{\mathcal{U}}}
\newcommand{\calV}{{\mathcal{V}}}
\newcommand{\calW}{{\mathcal{W}}}
\newcommand{\calX}{{\mathcal{X}}}
\newcommand{\calY}{{\mathcal{Y}}}
\newcommand{\calZ}{{\mathcal{Z}}}

\newcommand{\bbA}{{\mathbb{A}}}
\newcommand{\bbB}{{\mathbb{B}}}
\newcommand{\bbC}{{\mathbb{C}}}
\newcommand{\bbD}{{\mathbb{D}}}
\newcommand{\bbE}{{\mathbb{E}}}
\newcommand{\bbF}{{\mathbb{F}}}
\newcommand{\bbG}{{\mathbb{G}}}
\newcommand{\bbH}{{\mathbb{H}}}
\newcommand{\bbI}{{\mathbb{I}}}
\newcommand{\bbJ}{{\mathbb{J}}}
\newcommand{\bbK}{{\mathbb{K}}}
\newcommand{\bbL}{{\mathbb{L}}}
\newcommand{\bbM}{{\mathbb{M}}}
\newcommand{\bbN}{{\mathbb{N}}}
\newcommand{\bbO}{{\mathbb{O}}}
\newcommand{\bbP}{{\mathbb{P}}}
\newcommand{\bbQ}{{\mathbb{Q}}}
\newcommand{\bbR}{{\mathbb{R}}}
\newcommand{\bbS}{{\mathbb{S}}}
\newcommand{\bbT}{{\mathbb{T}}}
\newcommand{\bbU}{{\mathbb{U}}}
\newcommand{\bbV}{{\mathbb{V}}}
\newcommand{\bbW}{{\mathbb{W}}}
\newcommand{\bbX}{{\mathbb{X}}}
\newcommand{\bbY}{{\mathbb{Y}}}
\newcommand{\bbZ}{{\mathbb{Z}}}

\newcommand{\Rd}{\bbR^d}



\maketitle

\vspace{-1\baselineskip}\noindent
\begin{abstract}
We present a structured overview of adaptation algorithms for neural network-based speech recognition, considering both hybrid hidden Markov model / neural network systems and end-to-end neural network systems, with a focus on speaker adaptation, domain adaptation, and accent adaptation.  The overview characterizes adaptation algorithms as based on embeddings, model parameter adaptation, or data augmentation.  We present a meta-analysis of the performance of speech recognition adaptation algorithms, based on relative error rate reductions as reported in the literature.
\end{abstract}
\noindent
\vspace{1\baselineskip}

\section{Introduction}
\label{sec:intro}

\IEEEPARstart{T}{he} performance of automatic speech recognition (ASR) systems has improved dramatically in recent years thanks to the availability of larger training datasets, the development of neural network based models, and the computational power to train such models on these datasets \cite{hinton2012deep,seide2011conversational,saon2017english,chiu2018state}.  However, the performance of ASR systems can still degrade rapidly  when their conditions of use (test conditions) differ from the training data.  There are several causes for this, including speaker differences, variability in the acoustic environment, and the domain of use.   

Adaptation algorithms attempt to alleviate the mismatch between the test data and an ASR system's training data.  Adapting an ASR system is a challenging problem since it requires the modification of large and complex models, typically using only a small amount of target data and without explicit supervision.  Speaker adaptation -- adapting the system to a target speaker -- is the most common form of adaptation, but there are other important adaptation targets such as the domain of use, and the spoken accent.   Much of the work in the area has focused on speaker adaptation: it is the case that many approaches developed for speaker adaptation do not explicitly model speaker characteristics, and can be applied to other adaptation targets.  Thus our core treatment of adaptation algorithms is in the context of speaker adaptation, with a later discussion of particular approaches for domain adaptation and accent adaptation. Specifically, domain adaptation in this paper refers to the task of adapting the models to the target domain that has either acoustic or content mismatch from the source domain in which the models were trained. 

This overview focuses on the adaptation of neural network (NN) based speech recognition systems, although we briefly discuss earlier approaches to speaker adaptation of hidden Markov model (HMM) based systems.  NN-based systems~\cite{morgan1995neural,hinton2012deep,chan2016listen} have revolutionized the field of speech recognition, and there has been intense activity in the development of adaptation algorithms for such systems.  Adaptation of NN-based speech recognition is an exciting research area for at least two reasons: from a practical point of view, it is important to be able to adapt state-of-the-art systems;  and from a theoretical point of view the fact that NNs require fewer constraints on the input than a Gaussian-based system, along with the gradient-based discriminative training which is at the heart of most NN-based speech recognition systems, opens a range of possible adaptation algorithms.

\begin{figure*}[tb]

\centering

\newcommand{\offset}{0.50cm}

\centering
\subfloat[\label{fig:architectures_a}]{
\begin{tikzpicture}[atipstyle, node distance=1.8cm]
\node (encoder) [model] {Encoder};
\node (softmax) [softmax, above of= encoder] {Softmax};

\draw [linestyle, ->] ($(encoder.south)+(0,-\offset)$) node[below] {$x_t$} -- (encoder)  ;
\draw [linestyle, ->] (encoder) -- (softmax) node[midway, fill=white] {$h_t^{enc}$};
\draw [linestyle, ->] (softmax.north) -- ($(softmax.north)+(0,\offset)$) node[above] {$P(y_t \mid x_t)$};

\clip ($(current bounding box.south west)+(0,-1.5)$) rectangle ($(current bounding box.north east)+(0,0)$);

\end{tikzpicture}
}
\centering
\subfloat[\label{fig:architectures_b}]{
\begin{tikzpicture}[atipstyle, node distance=1.8cm]




\node (softmax_rnnt) [softmax] {Softmax};
\node (joint_rnnt) [model, below of=softmax_rnnt] {Joint};
\node (prediction_rnnt) [model, below of=joint_rnnt] {Prediction};
\node (encoder_rnnt) [model, below of=joint_rnnt, xshift=-3.5cm] {Encoder};

\draw[linestyle]     ($(joint_rnnt.north west)+(-0.25,0.25)$) rectangle ($(prediction_rnnt.south east)+(0.25,-0.25)$);
\draw [line width=2pt, white] ($(joint_rnnt.north)+(-0.25,0.25)$) -- ($(joint_rnnt.north)+(0.25,0.25)$);
\draw [line width=2pt, white] ($(prediction_rnnt.south)+(-0.25,-0.25)$) -- ($(prediction_rnnt.south)+(0.25,-0.25)$);
\draw [line width=2pt, white] ($(prediction_rnnt.west)+(-0.25,1.25)$) -- ($(prediction_rnnt.west)+(-0.25,0.75)$);

\node (bn_label) [right of=joint_rnnt, rotate=270, yshift=0.3cm, xshift=0.9cm] {Decoder};

\draw [linestyle, ->] ($(encoder_rnnt.south)+(0,-\offset)$) node[below] {$x_t$} -- (encoder_rnnt);
\draw [linestyle, ->] ($(prediction_rnnt.south)+(-\offset,-\offset-0.25cm)$) node[left] {$y_{u-1}$} -- ($(prediction_rnnt.south)+(0,-\offset-0.25cm)$) -- (prediction_rnnt)  ;
\draw [linestyle, ->] (encoder_rnnt) -- +(0, 1)  -| (joint_rnnt) node[pos=0.1, fill=white] {$h_t^{enc}$};
\draw [linestyle, ->] (prediction_rnnt) -- (joint_rnnt) node[pos=0.25, fill=white] {$h_u^{pre}$} ;

\draw [linestyle, ->] (joint_rnnt) -- (softmax_rnnt) node[midway, fill=white] {$h_{t,u}^{joint}$};
\draw [linestyle, ->] (softmax_rnnt.north) -- ($(softmax_rnnt.north)+(0,\offset)$) node[above] {$P(y_u \mid X_{1:t}, Y_{1:u-1})$};

\clip ($(current bounding box.south west)+(-.5,-0.8)$) rectangle ($(current bounding box.north east)+(.5,0)$);

\end{tikzpicture}
}
\centering
\subfloat[\label{fig:architectures_c}]{
\begin{tikzpicture}[atipstyle, node distance=1.8cm]
\node (encoder_aed) [model, right of=encoder_rnnt, xshift=3cm] {Encoder};
\node (attention_aed) [model, above of=encoder_aed] {Attention};
\node (decoder_aed) [model, above of=attention_aed] {Decoder};
\node (softmax_aed) [softmax, above of=decoder_aed] {Softmax};

\draw [linestyle, ->] ($(encoder_aed.south)+(0,-\offset)$)  node[below] {$x_t$} -- (encoder_aed) ;
\draw [linestyle, ->] (encoder_aed) -- (attention_aed) node[pos=0.25, fill=white] {$h_{1:T}^{enc}$};
\draw [linestyle, ->] ($(attention_aed.south)+(-\offset,-\offset)$) node[left] {$h_{u-1}^{dec}$} -- ($(attention_aed.south)+(0,-\offset)$) -- (attention_aed)  ;

\draw [linestyle, ->] (attention_aed) -- node[pos=0.25, fill=white] {$h^{att}_u$} (decoder_aed)  ;
\draw [linestyle, ->] ($(decoder_aed.south)+(-\offset,-\offset)$) node[left] {$y_{u-1}$} -- ($( decoder_aed.south)+(0,-\offset)$) -- (decoder_aed)  ;

\draw [linestyle, ->] (decoder_aed) -- (softmax_aed) node[midway, fill=white] {$h_u^{dec}$};
\draw [linestyle, ->] (softmax_aed.north) -- ($(softmax_aed.north)+(0,\offset)$) node[above] {$P(y_u \mid X_{1:T}, Y_{1:u-1})$};	

\end{tikzpicture}
}

\caption{NN architectures used for hybrid NN/HMM and end-to-end (CTC, RNN-T, AED) speech recognition systems: (a) Scheme of NN architecture used for NN/HMM hybrid systems and for connectionist temporal classification (CTC); (b) architecture for the RNN Transducer (RNN-T); (c) architecture for attention based encoder-decoder (AED) end-to-end systems.  Input acoustic feature vectors are denoted by $x_t$; hidden layers are denoted by $h_t,h_u$ and output labels by $y_t,y_u$ depending on whether they are indexed by time $t$ (in hybrid and CTC systems) or only by output label $u$ (in parts of RNN-T and AED systems).  In practice, the encoders use a wide temporal context as input, even the whole acoustic sequence in the case of most CTC and AED models.}
\label{fig:architectures}
\end{figure*}
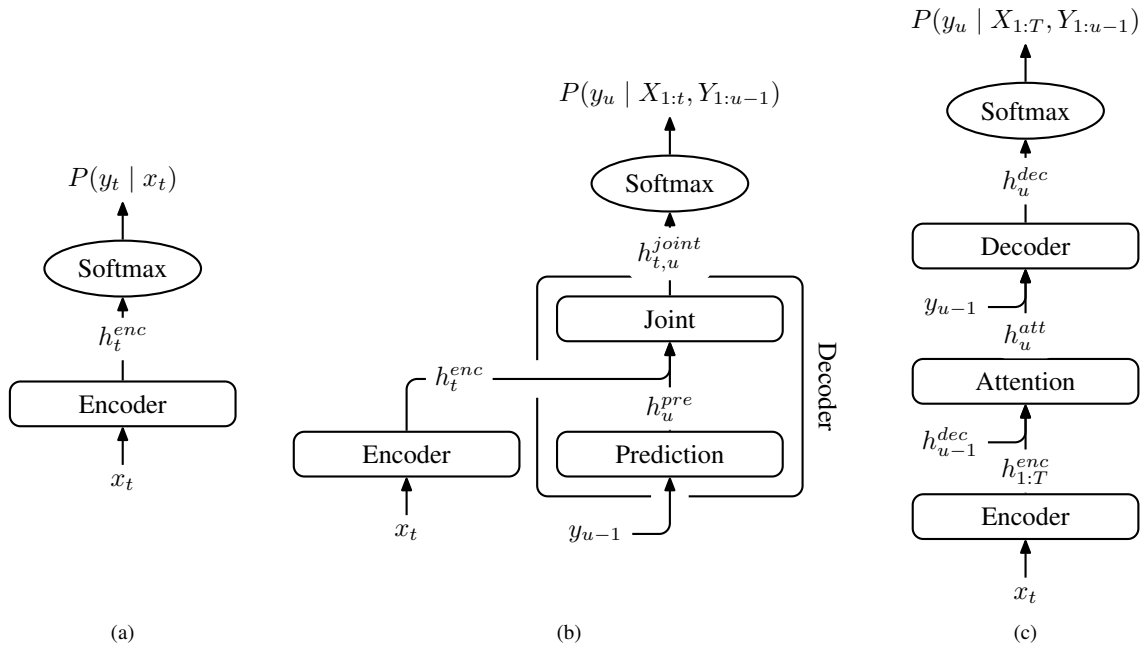

\subsection{NN/HMM Hybrid systems} 

Neural networks were first applied to speech recognition as so-called NN/HMM \emph{hybrid systems}, in which the neural network is used to estimate (scaled) likelihoods that act as the HMM state observation probabilities \cite{morgan1995neural} (\figref{fig:architectures_a}). During the 1990s both feed-forward networks \cite{morgan1995neural} and recurrent neural networks (RNNs) \cite{robinson1996use} were used in such hybrid systems and close to state-of-the-art results were obtained \cite{robinson-specom02}. These systems were largely context-independent, although context-dependent NN-based acoustic models were also explored \cite{kershaw1996context}.  

The modeling power of neural network systems at that time was computationally limited, and they were not able to achieve the precise levels of modeling obtained using context-dependent GMM-based HMM systems which became the dominant approach.  However, increases in computational power enabled deeper neural network models to be learned along with context-dependent modeling using the same number of context-dependent HMM tied states (senones) as GMM-based systems \cite{hinton2012deep, seide2011conversational}.  This led to the development of systems surpassing the accuracy of GMM-based systems.  This increase in computational power also enabled more powerful neural network models to be employed, in particular time-delay neural networks (TDNNs) \cite{waibel1989phoneme,peddinti2015time}, convolutional neural networks (CNNs) \cite{lecun1998gradient,abdel2014convolutional},  long short-term memory (LSTM) RNNs \cite{hochreiter1997long,graves2013speech}, and bidirectional LSTMs \cite{schuster1997bidirectional,graves2013hybrid}.

\subsection{End-to-end systems}

Since 2015, there has been a significant trend in the field moving from hybrid HMM/NN systems to end-to-end (E2E) NN modeling \cite{Attention-speech-chorowski2015,miao2015eesen, lu2015study, chan2016listen, rao2017exploring,  battenberg2017exploring, chiu2018state,  he2019streaming, Li2019RNNT} for ASR. E2E systems are characterized by the use of a single model transforming the input acoustic feature stream to  a target stream of output tokens, which might be constructed of characters, subwords, or even words.  E2E models are optimized using a single objective function, rather than comprising multiple components (acoustic model, language model, lexicon) that are optimized individually.   Currently, the most widely used E2E models are  connectionist temporal classification (CTC) \cite{graves2006connectionist,hannun2017sequence},  the RNN  Transducer (RNN-T) model \cite{Graves-RNNSeqTransduction,rao2017exploring}, and the  attention-based encoder-decoder (AED) model \cite{Attention-speech-chorowski2015, chan2016listen}.  

CTC and the RNN-T both map an input speech feature sequence to an output label sequence, where the label sequence (typically characters) is considerably shorter than the input sequence.  Both of these architectures use an additional blank output token to deal with the sequence length differences, with
an objective function which sums over all possible alignments using the forward backward algorithm~\cite{rabiner1989tutorial}.  CTC is an earlier, and simpler, method which assumes frame independence and functions similarly to the acoustic model in hybrid systems without modeling the linguistic dependency across words;  its architecture is similar to that of the neural network in the hybrid system (\figref{fig:architectures_a}).

An RNN-T (\figref{fig:architectures_b}) combines an additional prediction network with the acoustic encoder.  The prediction network is an RNN modeling linguistic dependencies whose input is the previously  output symbol.  It is possible to initialize some of its layers from an external language model trained on additional text data.  The acoustic encoder and the prediction network are combined using a feed-forward joint network followed by a softmax to predict the next output token given the speech input and the linguistic context.

Together, the RNN-T's prediction and joint networks may be regarded as a decoder, and we can view the RNN-T as a form of encoder-decoder system.  The AED architecture (\figref{fig:architectures_c}) enriches the encoder-decoder model with an additional attention network which interfaces the acoustic encoder with the decoder.  The attention network operates on the entire sequence of encoder representations for an utterance, offering the decoder considerably more flexibility.  A detailed comparison of popular E2E models in both streaming and non-streaming modes with large scale training data was conducted by Li et al. \cite{Li2020Comparison}. It is worth noting that with the recent success in machine translation, there is a trend of using the transformer model \cite{vaswani2017attention} to replace LSTM for both the AED \cite{dong2018speech,  Zhaoi2019speechtransformer, Karita2019Transformer} and RNN-T models \cite{yeh2019transformer, zhang2020transformer, chen2020developing}.


\subsection{Adaptation and transfer learning in related fields}

Adaptation and transfer learning have become important and intensively researched topics in other areas related to machine learning, most notably computer vision and natural language processing (NLP).  In both these cases the motivation is to train powerful base models using large amounts of training data, then to adapt these to specific tasks or domains, for which considerably less training data is available.  

In computer vision, the base model is typically a large convolutional network trained to perform image classification or object recognition using the ImageNet database \cite{deng2009imagenet,russakovsky2015imagenet}.  The ImageNet model is then adapted to a lower resource task, such as computer-aided detection in medical imaging \cite{shin2016deep}.  Kornblith et al. \cite{kornblith2019imagenet} have investigated empirically how well ImageNet models transfer to different tasks and datasets.  

Transfer learning in NLP differs from computer vision, and from the speech recognition approaches discussed in this paper, in that the base model is trained in an unsupervised fashion to perform language modeling or a related task, typically using web-crawled text data.  Base models used for NLP include the bidirectional LSTM \cite{peters2018deep} and Transformers which make use of self-attention \cite{devlin2019bert,raffel2020exploring}.  These models are then trained on specific NLP tasks, with supervised training data, which is specified in a common format (\eg text-to-text transfer \cite{raffel2020exploring}), often trained in a multi-task setting. Earlier adaptation approaches in NLP focused on feature adaptation (e.g. \cite{cer2018universal}), but more recently better results have been obtained using model-based adaptation, for instance ``adapter layers'' \cite{houlsby2019parameter,raffel2020exploring}, in which trainable transform layers are inserted into the pretrained base model.

More broadly there has been extensive work on domain adaptation and transfer learning in machine learning, reviewed by  Kouw and Loog \cite{kouw2019review}.  This includes work on few-shot learning \cite{ravi2016optimization,snell2017prototypical,li2020towards} and normalizing flows \cite{rezende2015variational,papamakarios2019normalizing}.  Normalizing flows  which provide a probabilistic framework for feature transformations, were first developed for speech recognition as Gaussianization \cite{chen2001gaussianization}, and more recently have been applied to speech synthesis \cite{prenger2019waveglow} and voice transformation \cite{serra2019blow}.

\subsection{Structure of this review}

We begin by considering the issues of identifying suitable data and target labels to adapt to in \secref{sec:adaptation_targets}.
After discussing speaker adaptation of non NN-based HMM systems in Section~\ref{sec:early}, we present a general framework for adaptation of NN-based speech recognition systems (both hybrid and E2E) in \secref{sec:adaptNN}, where we organize adaptation algorithms into three general categories:  embedding-based approaches (discussed in \secref{sec:embeddings}), model-based approaches (discussed in Secs.~\ref{sec:structured_xforms}--\ref{sec:objective_functions}), and data augmentation approaches (discussed in \secref{sec:data_augmentation}).

As mentioned above, most of our treatment of adaptation algorithms is in the context of speaker adaptation.  In Secs.~\ref{sec:accent_adaptation} and \ref{sec:domain_adaptation} we discuss specific approaches to accent adaptation and domain adaptation respectively.

Our primary focus is on the adaptation of acoustic models and end-to-end models.  In \secref{sec:lmadapt} we provide a summary of work in language model (LM) adaptation, mentioning both n-gram and neural network language models, and the use of LM adaptation in E2E systems.

Finally we provide a meta analysis of  experimental studies using the main adaptation algorithms that we have discussed (\secref{sec:meta_analysis}).  The meta-analysis is based on experiments reported in \nstudies~papers, carried out using \ndsets~datasets, and is primarily based on the relative error rate reduction arising from adaptation approaches.  In this section we  analyze the performance of the main adaptation algorithms across a variety of adaptation target types (for instance speaker, domain, and accent), in supervised and unsupervised settings,  in six different languages, and  using six different NN model types in both hybrid and E2E settings. Raw data, aggregated results and the corresponding scripts are available at \repourl.


\section{Identifying adaptation targets}
\label{sec:adaptation_targets}

Adaptation aims to reduce the mismatch between training and test conditions.  For an adaptation algorithm to be effective, the distribution of the adaptation data should be close to that  encountered in test conditions.  For this reason it is important to ensure that the target labels adapted to form coherent classes.  For the task of acoustic adaptation this requirement is typically satisfied by forming the adaptation data from one or more speech segments from known testing conditions (\ie the same speaker, accent, domain, or acoustic environment).
While for some tasks labels ascribed to speech segments may exist, allowing segments to be grouped into larger adaptation clusters, it is unrealistic to assume the availability of such metadata in general. However, depending on the application and the operating regime of the ASR system, it may be possible to derive reasonable proxies. 

\emph{Utterance}-level adaptation derives adaptation statistics using a single speech segment \cite{tan2016learning}. This waives the requirement to carry information about speaker identity between utterances, which may simplify deployment of recognition systems -- in terms of both engineering and privacy -- as one does not need to estimate and store offline speaker-specific information. On the other hand, owing to the small amounts of data available for adaptation the gains are usually lower than one could obtain with speaker-level clusters. While many approaches use utterances to directly extract corresponding embeddings to use as an auxiliary input for the acoustic model~\cite{saon2013speaker,senior2014improving, karanasou2014adaptation, vesely2016sequence}, one can also build a fixed inventory of speakers, domains, or topic codes~\cite{doulaty2015lda} or embeddings~\cite{pan2020memory, sari2020unsupervised} when learning the acoustic model or acoustic encoder, and then use the test utterance to select a combination of these at test stage. The latter approach alleviates the necessity of estimating an accurate representation from small amounts of data.
It may be possible to relax the utterance-level constraint by iteratively re-estimating adaptation statistics using a number of preceding segment(s)~\cite{senior2014improving}. Extra care usually needs to be taken to handle silence and speech uttered by different speakers, as failing to do so may deteriorate the overall ASR performance~\cite{zhang2000line, huang2015investigation,sari2020unsupervised}. 

\emph{Speaker}-level adaptation aggregates statistics across two or more segments uttered by the same talker, requiring a way to group adaptation utterances produced by different talkers.  In some cases -- for example lecture recordings and telephony -- speaker information may be available.  In other cases potentially inaccurate metadata is available, for instance in the transcription of television or online broadcasts.  In many cases (for instance, anonymous voice search) speaker metadata is not available.  The generic approach to this problem relies on a speaker diarization system~\cite{anguera2012diarisation}, that can identify speakers and accordingly assign their identities to the corresponding segments in the recordings. This is often used in the offline transcription of meetings or broadcast media.  Alternative clustering approaches can be used to define the adaptation classes  \cite{gales2000cluster,tan2016cluster}. 

\emph{Domain}-level adaptation broadens the speaker-level cluster by including speech produced by multiple talkers characterized by some common characteristic such as accent, age, medical condition, topic, \etc. This typically results in more adaptation material and an easier annotation process (cluster labels need to be assigned at batch rather than segment level). As such, domain adaptation can usually leverage adaptation transforms with greater capacity, and thus offer better adaptation gains.

Depending on whether adaptation transforms are estimated on held out data, or adaptation is iteratively derived from test segments, we will refer to these as \emph{enrolment} or \emph{online} modes, respectively.  In enrolment mode, the adaptation data would be ideally labeled with a gold-standard transcription, to enable supervised learning algorithms to be used for adaptation.  However, supervised data is rarely available:  small amounts may be available for some domain adaptation tasks (for example, adapting a system trained on typical speech to disordered speech \cite{christensen2012comparative}).  In the usual case, where supervised adaptation data is not available, supervised training algorithms can still be used with ``pseudo-labels'' that are automatically obtained from a seed model, a process which is a type of \textit{semi-supervised training} \cite{liao2013large}. Alternatively,  unsupervised training can be applied to  learn embeddings for the different adaptation classes, such as i-vectors \cite{saon2013speaker} or bottleneck features extracted from an auto-encoder neural network \cite{hsu2017unsupervised}.  
A \emph{two-pass} system is a special case for which the statistics are estimated from test data using the first pass decoding with a speaker-independent model in order to obtain adaptation labels, followed by a second pass with the speaker-adapted model.

For semi-supervised approaches, it is possible to further filter out regions with low-confidence to avoid the reinforcement of potential errors~\cite{mathias2005discriminative,liu2007investigating,walker2017semi}. There is some evidence in the literature that, for some limited-in-capacity transforms estimated in a semi-supervised manner, the first pass transcript quality has a small impact on the adapted accuracy as long as these are obtained with the corresponding speaker-independent model~\cite{miao2015speaker, swietojanski2016learning}. In \emph{lattice supervision} multiple possible transcriptions are used in a semi-supervised setting by generating a lattice, rather than the one-best transcription \cite{padmanabhan2000lattice,fraga2011lattice,manohar2018semi,klejch2019lattice}.  

\section{Adaptation algorithms for HMM-based ASR} 
\label{sec:early}

Speaker adaptation of speech recognition systems has been investigated since the 1960s \cite{suzuki1967acoustic,gerstman1968classification}.  In the mid-1990s, the influential maximum likelihood linear regression (MLLR) \cite{leggetter1995maximum} and maximum a posteriori (MAP) \cite{gauvain1994maximum} approaches to speaker adaptation for HMM/GMM 
systems were introduced.  These methods, described below, stimulated the field leading an intense activity in algorithms for the adaptation of HMM/GMM systems, reviewed by Woodland~\cite{woodland2001speaker} and Shinoda \cite{shinoda2011speaker}, as well as in section 5 of Gales and Young's broader review of HMM-based speech recognition~\cite{gales2008application}.  As we later discuss, some of the algorithms developed for HMM-based systems, in particular feature transformation approaches have been successfully applied to NN-based systems.  In this section we review  MAP,  MLLR, and related approaches to the adaptation of HMM/GMM systems, along with earlier approaches to speaker adaptation.  

\subsection{Speaker normalisation}

Many of these early approaches were designed to normalize speaker-specific characteristics, such as vocal tract length, building on linguistic findings relating to speaker normalization in speech perception \cite{johnson2005speaker}, often casting the problem as one of spectral normalization.  This work included formant-based frequency warping approaches  \cite{suzuki1967acoustic,gerstman1968classification,paliwal1985dynamic} and the estimation of linear projections to normalize the spectral representation to a speaker-independent form \cite{grenier1980speaker,choukri1986adaptation}.

Vocal tract length normalization (VTLN) was introduced by Wakita \cite{wakita1977normalization} (and again by Andreou \cite{andreou1994experiments}) as a form of frequency warping with the aim to compensate for vocal tract length differences across speakers. VTLN was extensively investigated for speech recognition in the 1990s and 2000s  \cite{eide1996parametric,lee1996speaker,kim2004using,garau2005applying}, and is discussed further in \secref{sec:embeddings}.

\subsection{Model based approaches}

In model based adaptation, the speech recognition model is used to drive the adaptation.  In work prefiguring subspace models, Furui \cite{furui1980training} showed how speaker specific models could be estimated from small amounts of target data in a dynamic time warping setting, learning linear transforms between preexisting speaker-dependent phonetic templates, and templates for a target speaker.  Similar techniques were developed in the 1980s by adapting the vector quantization (VQ) used in discrete HMM systems. Shikano, Nakamura, and Abe \cite{shikano1991speaker} showed that mappings between speaker dependent codebooks could be learned to model a target speaker (a technique widely used for voice conversion \cite{abe1990voice}); Feng et al. \cite{feng1988improved} developed a VQ-based approach in which speaker-specific mappings were learned between codewords in a speaker-independent codebook, in order to maximize the likelihood of the discrete HMM system. Rigoll  \cite{rigoll1989speaker} introduced a related approach in which the speaker-specific transform took the form of a Markov model.  A continuous version of this approach, referred to as probabilistic spectrum fitting, which aimed to adjust the parameters of a Gaussian phonetic model was introduced by Hunt \cite{hunt1981speaker} and further developed by Cox and Bridle \cite{cox1989unsupervised}.  

These probabilistic spectral modeling approaches can be viewed as precursors to maximum likelihood linear regression (MLLR) introduced by Leggetter and Woodland \cite{leggetter1995maximum} and generalized by Gales \cite{gales1998maximum}.  MLLR applies to continuous probability density HMM systems, composed of Gaussian probability density functions.  In MLLR, linear transforms are estimated to adapt the mean vectors and -- in \cite{gales1998maximum} -- covariance matrices of the Gaussian components.  If $\mu$ and $\Sigma$ are the mean vector and covariance matrix of a particular Gaussian, then MLLR adapts the parameters as follows:
\begin{align}
      \hat{\mu}_s &= A_s \, \mu - b_s \\
      \hat{\Sigma}_s &= H_s \, \Sigma \, H_{s}^\intercal \, .
\end{align}
The speaker-specific parameters $b_s, A_s$ and $H_s$ are estimated using maximum likelihood.  
MLLR is a compact adaptation technique since the transforms are shared across Gaussians: for instance all Gaussians corresponding to the same monophone  might share mean and covariance transforms.  Very often, especially when target data is sparse, a greater degree of sharing is employed -- for instance two shared adaptation transforms, one for Gaussians in  speech models and one for Gaussians in non-speech models.

Constrained MLLR \cite{neumeyer1995comparative,gales1998maximum}, is an important variant of MLLR, in which the same transform is used for both the mean and covariance:
\begin{align}
      \hat{\mu}_s &= A_s \mu - b_s \\
      \hat{\Sigma}_s &= A_s \, \Sigma \, A_s^\intercal
\end{align}
In this case, the log likelihood for a single Gaussian is given by
\begin{gather}
L^{\text{c}\textsc{MLLR}}(x; \hat{\mu}_s, \hat{\Sigma}_s) 
= 
\log \mathcal{N}(x; A_s \mu - b_s, A_s \, \Sigma \, A_s^\intercal)\\
= \log \mathcal{N}(A_s^{-1}x + A^{-1}_s b_s; \mu, \Sigma) - \log|A_s|
\end{gather}
It can be seen that this transform of the model parameters is equivalent to applying an affine transform to the data -- hence constrained MLLR is often referred to as feature-space MLLR (fMLLR),
although it is not strictly feature-space adaptation unless a single transform is shared across all Gaussians in the system, in which case the Jacobian term $-\log|A_s|$ can be ignored.  MLLR and its variants have been used extensively in the adaptation of Gaussian mixture model (GMM)-based HMM speech recognition systems \cite{woodland2001speaker,gales2008application}.

\subsection{Bayesian methods}

The above model-based adaptation approaches have aimed to estimate transforms between a speaker independent model and a model adapted to a target speaker.  An alternative Bayesian approach attempts to perform the adaptation by using the speaker independent model to inform the prior of a speaker-adapted model.  If the set of parameters of a speech recognition model are denoted by $\theta$, then maximum likelihood estimation sets $\theta$ to maximize the likelihood $p(X \,|\, \theta)$.  In MAP training, the estimation procedure maximizes the posterior of the parameters given the data $X = \left \{ x_1 \dots x_T \right \}$:
\begin{equation}
    P(\theta \mid X) \propto p(X \mid \theta) \, p(\theta)^r \, ,
\end{equation}
where $p(\theta)$ is the prior distribution of the parameters, which can be based on speaker independent models, and $r$ is an empirically determined weighting factor.  Gauvain and Lee \cite{gauvain1994maximum} presented an approach using MAP estimation as an adaptation approach for HMM/GMM systems.  A convenient choice of function for $p(\theta)$ is the  conjugate  to the likelihood -- the function which ensures the posterior has the same form as the prior.  For a GMM, if it is assumed that the mixture weights $c_i$ and the Gaussian parameters ($\mu_i$, $\Sigma_i$) are independent, then the conjugate prior may take the form of a mixture model $p_D(c_i) \prod_i p_W(\mu_i, \Sigma_i)$, where $p_D()$ is a Dirichlet distribution (conjugate to the multinomial) and $p_W()$ is the normal-Wishart density (conjugate to the Gaussian).  This results in the following intuitively understandable parameter estimate for the adapted mean of a Gaussian $\hat{\mu} \in \mathbb{R}^d$:
\begin{equation}
    \hat{\mu} = \frac{\tau \mu_0 + \sum_t \gamma (t) x_t} {\tau + \sum_t \gamma (t)} \, ,
\end{equation}
where $\mu_0 \in \mathbb{R}^d$ is the unadapted (speaker-independent) mean, $x_t \in \mathbb{R}^d$ is the adaptation vector at time $t$, $\gamma (t) \in \mathbb{R}$ is the component occupation probability (responsibility) for the Gaussian component at time $t$ (estimated by the forward-backward algorithm), and $\tau$ is a positive scalar-valued parameter of the normal-Wishart density, which is typically set to a constant empirically (although Gauvain and Lee \cite{gauvain1994maximum} also discuss an empirical Bayes estimation approach for this parameter). The re-estimated means of the Gaussian components take the form of a weighted interpolation between the speaker independent mean and data from the target speaker.  When there is no target speaker data for a Gaussian component, the parameters remain speaker-independent;  as the amount of target speaker data increases, so the Gaussian parameters approach the target speaker maximum likelihood estimate.

\subsection{Speaker adaptive training}

In the model-based approaches discussed above (MLLR and MAP), we have implicitly assumed that adaptation takes place at test time:  speaker independent models are trained using recordings of multiple speakers in the usual way, with only the test speakers used for adaptation.  In contrast to this, it is possible to employ a model-based adaptive training approach.
In speaker adaptive training \cite{anastasakos1996compact}, a transform is estimated for each speaker in the training set, as well as for each speaker in the test set.  During training, speaker-specific transforms and a speaker-independent canonical model are updated in an iterative fashion.




Speaker space approaches represent a speaker-adapted model as a weighted sum of a set of individual models which may represent individual speakers or, more commonly, speaker clusters.  In cluster-adaptive training (CAT)  \cite{gales2000cluster}, the mean for a Gaussian component for a specific speaker $s$ is given by:
\begin{equation}
    \hat{\mu}_s = \sum_{c=1}^C w_c \mu_c \,
\end{equation}
where $\mu_c \in \mathbb{R}^d$ is the mean of the particular Gaussian component for speaker cluster $c$, and $w_c \in \mathbb{R}$ is the cluster weight. This expresses the speaker-adapted mean vector as a point in a speaker space.  Given a set of canonical speaker cluster models, CAT is efficient in terms of parameters, since only the set of cluster weights need to be estimated for a new speaker.  Eigenvoices \cite{kuhn2000rapid} are alternative way of constructing speaker spaces, with a speaker model again represented as a weighted sum of canonical models.  In the Eigenvoices technique, principal component analysis of ``supervectors'' (concatenated mean vectors from the set of speaker-specific models) is used to create a basis of the speaker space.

A number of variants of cluster-adaptive training have been presented, including representing a speaker by combining MLLR transforms from the canonical models \cite{gales2000cluster}, and using sequence discriminative objective functions such as minimum phone error (MPE) \cite{yu2006discriminative}.  Techniques closely related to CAT have been used for the adaptation of neural network based systems (\secref{sec:structured_xforms}).

In contrast to model-based methods, in feature-based adaptation it is usual to adapt or normalize the acoustic features for each speaker in both the training and test sets-- this may be viewed as a form of speaker adaptive training.  For example, in the case of cepstral mean and variance normalization (CMVN), statistics are computed for each speaker and the features normalized accordingly, during both training and test.  Likewise, VTLN is also carried out for all speakers, transforming the acoustic features to a canonical form, with the variation from changes in vocal tract length being normalized away.    


\section{Adaptation algorithms for NN-based ASR}
\label{sec:adaptNN}

The literature describing methods for adaptation of NNs has tended to inherit terminology from the algorithms used to adapt HMM-GMM systems, for which there is an important distinction between \textit{feature space} and \textit{model space} formulations of MLLR-type approaches \cite{gales1998maximum}, as discussed in the previous section.  In a 2017 review of NN adaptation, Sim et al.~\cite{sim2017adaptation} divide adaptation algorithms into \textit{feature normalisation}, \textit{feature augmentation} and \textit{structured parameterization}.  (They also use a further category termed \textit{constrained adaptation}, discussed further below.)  

The task of an ASR model is to map a sequence of acoustic feature vectors, $X=(x_1, \dotsc x_t, \dotsc,  x_T)$, \mbox{$x_t \in \Rd$} to a sequence of words $W$.   
Although -- as we discuss below -- most techniques described in this paper apply equally to end-to-end models and hybrid HMM-NN models, we generally treat the model to be adapted as an acoustic model.  That is, we ignore aspects of adaptation that affect only $P(W)$, independently of the acoustics $X$ (LM adaptation is discussed in \secref{sec:lmadapt}).
Further, with only a small loss of generality, in what follows we will assume that the model operates in a framewise manner, thus we can define the model as:
\begin{equation}
y_t = f(x_t; \theta)
\end{equation}
where $f(x;\theta)$ is the NN model with parameters $\theta$ and $y_t$ is the output label at frame $t$.  In a hybrid HMM-NN system, for example, $y_t$ is taken to be a vector of posterior probabilities over a senone set.  In a CTC model, $y_t$ would be a vector of posterior probabilities over the output symbol set, plus blank symbol.  Note that NN models often operate on a wider windowed set of input features, $x_t(w) = [ x_{t-c}, x_{t-c+1}, \dotsc, x_{t+c-1}, x_{t+c} ]$ with the total window size $w = 2c+1$.  For reasons of notational clarity, we generally ignore the distinction between $x_t$ and $x_t(w)$, unless it is specifically relevant to a particular topic.    

In this framework, we can define \textit{feature normalisation} approaches as acting to transform the features in a speaker-dependent manner, on which the speaker-independent model operates.  For each speaker $s$, a transformation function \mbox{$g: \Rd \rightarrow \bbR^{d^\prime}$} computes:
\begin{equation}
x^\prime_t = g(x_t;\phi_s)  \label{eqn:feature_normalisation}
\end{equation}
where $\phi_s$ is a set of speaker-dependent parameters.  Commonly the dimension of the normalised features is the identical to the original (\ie $d=d^\prime$) but this is not required.  This family is closely related to feature space methods used in GMM systems described above in \secref{sec:early}, including fMLLR (when only a single affine transform is used), VTLN, and CMVN.   

\textit{Structured parameterization} approaches, in contrast, introduce a speaker-dependent transformation of the acoustic model parameters:
\begin{equation}
\theta^\prime_s = h(\theta; \varphi_s) \label{eqn:structured_parametrisation}
\end{equation}
In this case, the function $h$ would typically be structured so as to ensure that the number of speaker-dependent parameters $\varphi_s$ is sufficiently smaller than the number of parameters of the original model.   Such methods are closely related to model-based adaptation of GMMs such as MLLR.

Finally, \textit{feature-augmentation} approaches extend the feature vector $x_t$ with a speaker-dependent embedding $\lambda_s$, which we can write as
\begin{equation}
x^\prime_t = \begin{pmatrix} x_t \\ \lambda_s  \end{pmatrix} \label{eqn:feature_augmentation}
\end{equation}
Close variants of this approach use the embedding to augment the input to higher layers of the network.  Note that the incorporation of an embedding requires the addition of further parameters to the acoustic model controlling the manner in which the embedding acts to adapt the model, which can be written $f(x_t;\theta, \theta^E)$.  The embedding parameters $\theta^E$ are themselves speaker-independent.    

We suggest that the distinctions described above may not always be helpful when considering NN adaptation specifically, because all three approaches can be seen to be closely related or even special cases of each other.  As we saw in Section~\ref{sec:early} this is not the case in HMM-GMM systems, where the distinction between feature-space and model adaptation is important (as noted by Gales~\cite{gales1998maximum}) because in the former case, different feature space transformations can be carried out per senone class if the appropriate scaling by a Jacobian is performed; whilst in the latter case, it is necessary for the adapted probability density functions to be re-normalized.

As an example of the equivalence of the close relationship between the three approaches to NN adaptation, the normalisation function $g$ can generally be formulated as shallow NN, possibly without a non-linearity.  If there is a set of ``identity transform'' parameters $\phi^I$ such that 
\begin{equation}
g(x_t; \phi^I) = x_t,  \; \forall x_t
\end{equation}
then we have
\begin{equation}
y_t = f(x_t;\theta) = f(g(x_t;\phi^I);\theta) = f^\prime(x_t; \theta, \phi^I) 
\end{equation}
where $f^\prime$ is a new network comprising of a copy of the original network $f$ with the layers of $g$ prepended.  Applying  feature normalization \eqref{eqn:feature_normalisation} leads to:
\begin{equation}
y_t = f(x_t^\prime;\theta) = f(g(x_t;\phi_s);\theta) = f^\prime(x_t; \theta, \phi_s) 
\end{equation}
which we can write as a structured parameter transformation of $f^\prime$, as defined in \eqref{eqn:structured_parametrisation}:
\begin{equation}
\theta^\prime_s = \{ \theta, \phi_s \} = h(\{\theta, \phi^I\}; \varphi_s)
\end{equation}   
where the transformation $h(\,\cdot\,;\varphi_s)$ is simply set to replace the parameters pertaining to $g$ with the original normalisation parameters,  $\phi_s = \varphi_s$, leaving the other parameters unchanged.

Similarly, feature augmentation approaches may be readily seen to be a further special case of structured adaptation.  In the simple case of input feature augmentation \eqref{eqn:feature_augmentation}, we see that the output of the first layer, prior to the non-linearity, can be written as
\begin{equation}
z = Wx^\prime + b = W \begin{pmatrix}x \\ \lambda_s \end{pmatrix} + b
\end{equation}
where $W$ and $b$ are the weight and bias of the first layer respectively.  By introducing a decomposition of $W$, $W= \begin{pmatrix} U & V \end{pmatrix}$ we write this as
\begin{equation}
z = \begin{pmatrix} U & V \end{pmatrix} \begin{pmatrix}x \\ \lambda_s \end{pmatrix} + b = U x + b + V \lambda_s 
\end{equation}
with $U \in \theta$ and $V \in \theta^E$ being weight matrices pertaining to the input features and speaker embedding, respectively.

This can be expressed as a structured transformation of the bias:
\begin{equation}
\theta^\prime_s = \{U^\prime,b^\prime\} = h(\{U,b\}; \varphi_s) = \{U, b+V\lambda_s\}   
\end{equation}
with $\varphi_s = V\lambda_s$.  Similar arguments apply to embeddings used in other network layers.  

Certain types of
feature normalisation approaches can be expressed as feature augmentation.  For example, cepstral mean normalisation given by
\begin{equation}
x_t^\prime = g(x_t;\phi_s) = x_t - \mu_s
\end{equation}
can be expressed as
\begin{equation}
z = W(x - \mu_s) + b = \begin{pmatrix} W & W \end{pmatrix} \begin{pmatrix}x \\ - \mu_s \end{pmatrix} + b  
\end{equation}
with augmented features $\lambda_s = -\mu_s$.

As we have seen, approaches to NN adaptation under the traditional categorization of feature augmentation, structured parameterization and feature normalization can usually be seen as special cases of one another. Therefore, in the remainder of this paper, we adopt an alternative categorization:
\begin{itemize}
    \item \textbf{Embedding-based} approaches in which any speaker-dependent parameters are estimated independently of the model, with the model $f(x_t; \theta)$ itself being unchanged between speakers, other than the possible need for additional embedding parameters $\theta^E$;
    \item \textbf{Model-based} approaches in which the model parameters $\theta$ are directly adapted to data from the target speaker according to the primary objective function;
    \item \textbf{Data augmentation} approaches which attempt to synthetically generate additional training data with a close match to the target speaker, by transforming the existing training data. 
\end{itemize}

This distinction is, we believe, particularly important in  \textit{speaker} adaptation of NNs because in ASR it has become standard to perform adaptation in a semi-supervised manner, with no transcribed adaptation data for the target speaker.  In this setting, as we will discuss, standard objective functions such as cross-entropy, which may be very effective in supervised training or adaptation, are particularly susceptible to transcription errors in semi-supervised settings.  

We describe the model-independent approaches as embedding-based  because any set of speaker-dependent parameters can be viewed as an embedding.  Embedding-based approaches are discussed in \secref{sec:embeddings}.  Well-known examples of speaker embeddings include i-vectors \cite{dehak2011front,saon2013speaker}, and x-vectors \cite{snyder2018x}, but can also include parameter sets more classically viewed as normalizing transforms such as CMVN statistics and global fMLLR transforms (see \secref{sec:early} above). However, for the reasons mentioned above,
we exclude from this category methods where the embedding is simply a subset of the primary model parameters and estimated according to the model's objective function.  Note that methods using a one-hot encoding for each speaker are also excluded, since it would be impossible to use these with a speaker-independent model, without each test speaker having been present in training data; such methods might however be useful for closely related tasks such as domain adaptation, discussed in \secref{sec:domain_adaptation}.

The primary benefit of speaker adaptive approaches over simply using speaker-dependent models is the prevention of over-fitting to the adaptation data (and  its possibly errorful transcript). A large number of model-based adaptation techniques have been proposed to achieve this; in this paper, we sub-divide them into: 
\begin{itemize}
    \item \textbf{Structured transforms}: Methods in which a subset of the parameters are adapted, with many instances structuring the model so as to permit a reduced number of speaker-dependent parameters, as in the Learning Hidden Unit Contributions (LHUC) scheme \cite{swietojanski2014learning,swietojanski2016learning}. The can be viewed as an analogy to MLLR transforms for GMMs.  They are discussed in \secref{sec:structured_xforms}.
    \item \textbf{Regularization}: Methods with explicit regularization of the objective function to prevent over-fitting to the adaptation data, examples including the use of L2 loss or KL divergence terms to penalize the divergence from the speaker-independent parameters \cite{liao2013speaker,yu2013kl}.  Such methods can be viewed as related to the MAP approach for GMM adaptation.  They are discussed in \secref{sec:regularization}.
    \item \textbf{Variant objective functions}: Methods which  adopt variants of the primary objective function to overcome the problems of noise in the target labels, with examples including the use of lattice supervision \cite{klejch2019lattice} or multi-task learning \cite{huang2015rapid}.  They are discussed in \secref{sec:objective_functions}.
\end{itemize}
The second two categories above are collectively termed \textit{constrained adaptation} in the review by Sim et al.~\cite{sim2017adaptation}.  Within this, multi-task learning is labeled by Sim et al. as attribute aware training; however,
we do not believe that all multi-task learning approaches to adaptation can be labeled in this way.

Data augmentation methods have proved very successful in adaptation to other sources of variability, particularly those -- such as background noise conditions -- where the required model transformations are hard to explicitly estimate, but where it is easy to generate realistic data.  In the case of speaker adaptation, it is significantly harder to generate sufficiently good-quality synthetic data for a target speaker, given only limited data from the speaker in question.  However, there is a growing body of work in this area using, for example, techniques from the field of speech synthesis \cite{huang2020using}.  Approaches in this area are discussed in \secref{sec:data_augmentation}.

Most works suitable for adapting hybrid acoustic models can be leveraged to adapt acoustic encoders in E2E models. Both Kullback-Leibler divergence (KLD) regularization (\secref{sec:regularization}) and multi-task learning (MTL) methods (\secref{sec:objective_functions}) have been used for speaker adaptation for CTC and AED models \cite{li2018speaker, meng2019speaker}.

Sim et al.~\cite{sim2019investigation} updated  the  acoustic encoder of RNN-T models using speaker-specific adaptation data. Furthermore, by generating text-to-speech (TTS) audio from the target speaker, more data can be used to adapt the acoustic encoder. Such data augmentation adaptation (discussed in \secref{sec:data_augmentation}) was shown to be an effective way for the speaker adaptation of E2E models \cite{Huang2020Rapid} even with very limited raw  data from the target speaker.  Embeddings have also been used to train a speaker-aware  AED model \cite{fan2019speaker, sari2020unsupervised, zhao2020speech}. 

Because AED and RNN-T also have components corresponding to the language model, there are also techniques specific to adapting the language modeling aspect of  E2E models, for instance using a text embedding instead of an  acoustic embedding  to bias an E2E model in order to produce outputs relevant to the particular recognition context \cite{pundak2018deep, chen2019end, jain2020contextual}. If the new domain differs from the source domain mainly in content instead of acoustics, domain adaptation on E2E models can be performed by either interpolating the E2E model with an external language model or updating language model related components inside the E2E model with the text-to-speech audio generated from the text in the new domain \cite{sim2019personalization, Li2020RNNT}, discussed in \secref{sec:lmadapt}. 

\section{Speaker embeddings}
\label{sec:embeddings}

Speaker embeddings map speakers to a continuous space. In this section we consider embeddings that may be extracted in a manner independent of the model, and which are also typically unsupervised with respect to the transcript. They can therefore also be useful in a standalone manner for other tasks such as speaker recognition. When used with an acoustic model, the model learns how to incorporate the embedding information by, in effect, speaker-aware training. Speaker embeddings may encode speaker-level variations that are otherwise difficult for the AM to learn from short-term features \cite{huang2015investigation}, and may be included as auxiliary features to the network. Specifically, let $x\in\mathbb{R}^d$ denote the acoustic features, and $\lambda_s\in\mathbb{R}^k$ a k-dimensional speaker embedding. The speaker embeddings may be  concatenated with the acoustic input features, as previously seen in \eqref{eqn:feature_augmentation}:

\begin{equation}
x^\prime_t = \begin{pmatrix} x_t \\ \lambda_s  \end{pmatrix} \label{eqn:feature_augmentation_2}
\end{equation}

Alternatively they may be concatenated with the activations of a hidden layer. In either case the result is bias adaptation of the next hidden layer as  discussed in \secref{sec:structured_xforms}. As noted by Delcroix et al.~\cite{delcroix2018auxiliary} the auxiliary features may equivalently be added directly to the features using a learned projection matrix $P$, with the benefit that the downstream architecture can remain unchanged:

\begin{equation}
    x^\prime_t = x_t + P\lambda_s
\end{equation}

There are many other ways to incorporate embeddings into the AM: for example, they may be used to scale neuron activations as in LHUC \cite{swietojanski2016learning}. More generally we may consider embeddings applied to either biases or activations through context-adaptive \cite{delcroix2018context} or control networks \cite{rownicka2019embeddings}. It is possible to limit connectivity from the auxiliary features to the rest of the network in order to improve robustness at test time or to better incorporate static features \cite{garimella2015robust,tan2016speaker,parthasarathi2015fmllr}. We will further consider transformations of the features as speaker embeddings, such as with fMLLR \cite{neumeyer1995comparative,gales1998maximum}, and they may also be used as label targets \cite{meng2020vector}.

\ONDREJ{There is a considerable overlap between this paragraph and the penultimate paragraph of Sec. VI}
\PETER{I read both carefully and think it's ok}

\subsection{Feature transformations}
We may consider speaker-level transformations of the acoustic features as speaker embeddings. These include methods traditionally viewed as normalisation, such as CMVN and fMLLR, which produce affine transformations of the features:
\begin{equation}
    x_s' = A_s x + b_s
\end{equation}

CMVN derives its name from the application to cepstral features, but corresponds to the standardization of the features to zero mean and unit variance (z-score):
\begin{equation}
    x_s' = \frac{x-\mu}{\sqrt{\sigma^2 + \epsilon}}
\end{equation}
where $\mu \in \mathbb{R}^d$ is the cepstral mean, $\sigma^2 \in \mathbb{R}^d$ is the cepstral variance, and $\epsilon$ is a small constant for numerical stability.

fMLLR \cite{gales1998maximum} belongs to a family of speaker adaptation methods originally developed for HMM-GMM models, as discussed in Section~\ref{sec:early}. The technique has, however, later been used with success to transform features for hybrid models as well \cite{seide2011feature,rath2013improved}. While the fMLLR transforms were traditionally estimated using maximum likelihood and HMM-GMM models, the transforms may also be estimated using a neural network trained to estimate fMLLR features \cite{joy2016dnns} (in \secref{sec:structured_xforms} we will further discuss structurally similar transforms estimated using the main objective function). Instead of transforming the input features, some work has explored fMLLR features as an additional, auxiliary, feature stream to the standard features in order to improve robustness to mismatched transforms \cite{parthasarathi2015fmllr}, or to obtain speaker-adapted features derived from GMM log-likelihoods \cite{tomashenko2015gmm}, otherwise known as GMM-derived features.

Another technique with a long history is VTLN \cite{wakita1977normalization,andreou1994experiments,lee1996speaker,uebel1999investigation}, which was briefly introduced in Section~\ref{sec:early}. To control for varying vocal tract lengths between speakers, VTLN typically uses a piecewise linear warping function to adjust the filterbank in feature extraction. This requires only a single warping factor parameter that can be estimated using any AM with a line search. Alternatively, linear-VTLN (\eg \cite{kim2004using}) obtains a corresponding affine transform similar to fMLLR, but chooses from a fixed set of transforms at test time . A related idea is that of the exponential transform \cite{povey2011speaker}, which forgoes any notion of vocal tract length, but akin to VTLN is controlled by a single parameter. More recently, adaptation of learnable filterbanks, operating as the first layer in a deep network, has resulted in updates which compensate for vocal tract length differences between speakers \cite{fainberg2019acoustic}. 

\subsection{i-vectors}
Many types of embeddings stem from research in speaker verification and speaker recognition. One such approach is identity vectors, or i-vectors \cite{dehak2011front,karafiat2011ivector,saon2013speaker}, which are estimated using means from GMMs trained on the acoustic features. 
Specifically, the extraction of a speaker i-vector, $\lambda_s \in \mathbb{R}^k$, assumes a linear relationship between the global means from a background GMM (or universal background model, UBM), $m_{g}\in \mathbb{R}^m$, and the speaker-specific means, $m_s\in \mathbb{R}^m$
\begin{equation}
m_{s} = m_{g} + T\lambda_s
\end{equation}
where $T\in \mathbb{R}^{m\times k}$ is a matrix that is shared across all speakers which is sometimes called the total variability matrix from its relation to joint factor analysis \cite{kenny2007speaker}. An i-vector thus corresponds to coordinates in the column space of $T$. $T$ is estimated iteratively using the EM algorithm. It is possible to replace the GMM means with posteriors or alignments from the AM 
\cite{lei2014novel,kenny2014deep,garimella2015robust}
although this is no longer independent of the AM and requires transcriptions. The i-vectors are usually concatenated with the acoustic features as discussed above, but have also been used in more elaborate architectures to produce a feature mapping of the input features themselves \cite{miao2014towardssa,miao2014improvements}.

\subsection{Neural network embeddings}

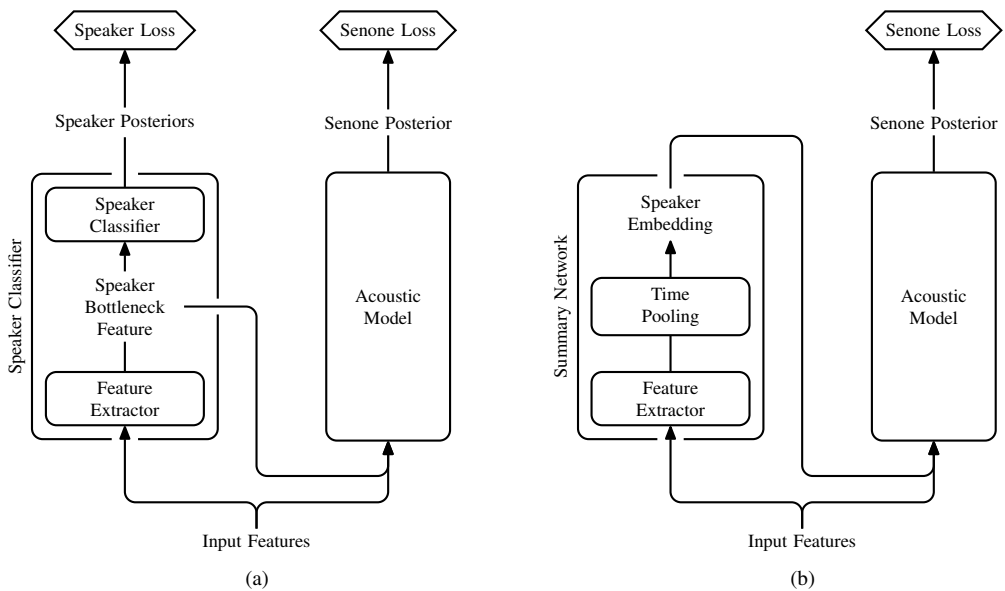
\begin{figure*}

\centering

\begin{minipage}[c]{.4\textwidth}
\centering
\subfloat[\label{diag:bottleneck}]{
\begin{tikzpicture}[atipstyle, node distance=1.75cm, scale=0.7, every node/.style={transform shape}]

    \tikzstyle{loss} = [ thick, chamfered rectangle, chamfered rectangle xsep=2cm, text centered, draw=black]
    \tikzstyle{data} = [text centered, text width=2.5cm]
 
\node (input) [data, text width=3.5cm] {Input Features};

\node (feature_extractor) [model, above of=input, xshift=-2.5cm, yshift=1cm] {Feature Extractor}; %
\node (speaker_bottleneck_feature) [data, above of=feature_extractor, text width=2cm] {Speaker Bottleneck Feature};
\node (speaker_classifier) [model, above of=speaker_bottleneck_feature] {Speaker Classifier};
\node (speaker_posterior) [data, above of=speaker_classifier, text width=3cm] {Speaker Posteriors};
\node (speaker_loss) [loss, above of=speaker_posterior, text depth=0.1em,] {Speaker Loss};

\node (acoustic_model) [model, minimum height=5.095cm, above of=input, xshift=2.5cm, yshift=2.75cm, minimum width=2cm] {Acoustic Model};
\node (senone_posterior) [data, text width=2.5cm, above of=acoustic_model, yshift=1.75cm] {Senone Posterior};
\node (senone_loss) [loss, above of=senone_posterior, text depth=0.1em,] {Senone Loss};

\draw[linestyle]     ($(speaker_classifier.north west)+(-0.25,0.25)$) rectangle ($(feature_extractor.south east)+(0.25,-0.25)$);
\draw [line width=2pt, white] ($(feature_extractor.east)+(0.25,1.6)$) -- ($(speaker_classifier.east)+(0.25,-1.6)$);
\draw [line width=2pt, white] ($(speaker_classifier.north)+(-0.25,0.25)$) -- ($(speaker_classifier.north)+(0.25,0.25)$);
\draw [line width=2pt, white] ($(feature_extractor.south)+(-0.25,-0.25)$) -- ($(feature_extractor.south)+(0.25,-0.25)$);

\node (bn_label) [left of=speaker_bottleneck_feature, rotate=90, yshift=0.3cm] {Speaker Classifier};

\draw [linestyle, ->,] ($(input.north)$) -- ($(input.north)+(0,0.5)$) -| (feature_extractor.south);
\draw [linestyle,] (feature_extractor) -- (speaker_bottleneck_feature);
\draw [linestyle, ->,] (speaker_bottleneck_feature) -- (speaker_classifier);
\draw [linestyle] (speaker_classifier) -- (speaker_posterior);
\draw [linestyle, ->] (speaker_posterior) -- (speaker_loss);

\draw [linestyle, ->] (input.north) -- ($(input.north)+(0,0.5)$) -| (acoustic_model.south);
\draw [linestyle,] (acoustic_model) -- (senone_posterior);
\draw [linestyle, ->,] (senone_posterior) -- (senone_loss);

\draw [linestyle, ->] (speaker_bottleneck_feature.east) -| ($(input.north)+(0,1)$) -| (acoustic_model.south);

\clip (current bounding box.south west) rectangle (4.85, 10.2);

\end{tikzpicture}
}
\end{minipage}%
\begin{minipage}[c]{.4\textwidth}
\centering
\subfloat[\label{diag:summary_network}]{
\begin{tikzpicture}[atipstyle, node distance=1.75cm,scale=0.7, every node/.style={transform shape}]

    \tikzstyle{loss} = [ thick, chamfered rectangle, chamfered rectangle xsep=2cm, text centered, draw=black]
    \tikzstyle{data} = [text centered, text width=2.5cm]
 
\node (input) [data, text width=3.5cm] {Input Features};

\node (feature_extractor) [model, above of=input, xshift=-2.5cm, yshift=1cm] {Feature Extractor}; %
\node (speaker_bottleneck_feature) [model, above of=feature_extractor, text width=2cm] {Time\\Pooling};
\node (speaker_embedding) [data, above of=speaker_bottleneck_feature,minimum width=3cm] {Speaker Embedding};

\node (acoustic_model) [model, minimum height=5.095cm, above of=input, xshift=2.5cm, yshift=2.75cm, minimum width=2cm] {Acoustic Model};
\node (senone_posterior) [data, text width=2.5cm, above of=acoustic_model, yshift=1.75cm] {Senone Posterior};
\node (senone_loss) [loss, above of=senone_posterior, text depth=0.1em,] {Senone Loss};

\draw[linestyle]     ($(speaker_embedding.north west)+(-0.25,0.25)$) rectangle ($(feature_extractor.south east)+(0.25,-0.25)$);
\draw [line width=2pt, white] ($(speaker_embedding.north)+(-0.25,0.25)$) -- ($(speaker_embedding.north)+(0.25,0.25)$);
\draw [line width=2pt, white] ($(feature_extractor.south)+(-0.25,-0.25)$) -- ($(feature_extractor.south)+(0.25,-0.25)$);

\node (bn_label) [left of=speaker_bottleneck_feature, rotate=90, yshift=0.3cm] {Summary Network};

\draw [linestyle, ->,] ($(input.north)$) -- ($(input.north)+(0,0.5)$) -| (feature_extractor.south);
\draw [linestyle,] (feature_extractor) -- (speaker_bottleneck_feature);
\draw [linestyle, ->,] (speaker_bottleneck_feature) -- (speaker_embedding);

\draw [linestyle, ->] (input.north) -- ($(input.north)+(0,0.5)$) -| (acoustic_model.south);
\draw [linestyle,] (acoustic_model) -- (senone_posterior);
\draw [linestyle, ->,] (senone_posterior) -- (senone_loss);

\draw [linestyle, ->] (speaker_embedding.north) -- ($(speaker_embedding.north)+(0,1)$) -| ($(input.north)+(0,1)$) -| (acoustic_model.south);

\clip (current bounding box.south west) rectangle (4.85, 10.2);

\end{tikzpicture}
}
\end{minipage}%

\caption{(a) Bottleneck feature extraction that uses a pretrained speaker classifier. (b) Summary network extracting speaker embeddings which is trained jointly with the acoustic model.}

\end{figure*}

A number of works proposed to extract low-dimensional embeddings from bottleneck layers in neural network models trained to distinguish between speakers \cite{huang2015investigation,tan2016speaker} or across multiple layers followed by dimensionality reduction in a separate AM (\eg CNN embeddings \cite{rownicka2018analyzing}). One such approach, using Bottleneck Speaker Vector (BSV) embeddings \cite{huang2015investigation}, trains a feed-forward network to predict speaker labels (and silence) from spliced MFCCs (\figref{diag:bottleneck}). Tan et al.~\cite{tan2016speaker} proposed to add a second objective to predict monophones in a multi-task setup. The bottleneck layer dimension is typically set to values commonly used for i-vectors. In fact, Huang and Sim~\cite{huang2015investigation} note that if the speaker label targets are replaced with speaker deviations from a UBM, then the bottleneck-features may be considered frame-level i-vectors. The extracted features are averaged across all speech frames of a given speaker to produce speaker-level i-vectors.

There are several more recent approaches that we may collectively refer to as $\star$-vectors. Like bottleneck features, these approaches typically extract embeddings from neural networks trained to discriminate between speakers, but not necessarily using a low-dimensional layer. For instance, deep vectors, or d-vectors~\cite{variani2014deep,li2015modeling}, extract embeddings from feed-forward or LSTM networks  trained on filterbank features to predict speaker labels. The activations from the last hidden layer are averaged over time.  X-vectors~\cite{snyder2018x,rownicka2019embeddings} use TDNNs  with a pooling layer that collects statistics over time and the embeddings are extracted following a subsequent affine layer. A related approach called r-vectors \cite{khokhlov2019r} uses the architecture of x-vectors, but predicts room impulse response (RIR) labels rather than speaker labels. In contrast to the above approaches, label embeddings, or l-vectors \cite{meng2020vector}, are designed to be used as soft output targets for the training of an AM. Each label embedding represents the output distribution for a particular senone target. In this way they are, in effect, uncoupled from the individual data points and can be used for domain adaptation without a requirement of parallel data. We will discuss this idea further in \secref{sec:domain_adaptation}. For completeness we also mention h-vectors \cite{shi2020h} which use a hierarchical attention mechanism to produce utterance-level embeddings, but has only been applied to speaker recognition tasks.

X-vector embeddings are not widely used for adaptating ASR algorithms in practice -- especially in comparison to commonly used i-vectors -- as experiments have not shown consistent improvements in recognition accuracy.  One reason for this is related to the speaker identification training objective for the x-vector network which implicitly factors out channel information, which might be beneficial for adaptation. The optimal objective for speaker embeddings used in ASR differs from the objective used in speaker verification.

Summary networks \cite{vesely2016sequence,delcroix2018auxiliary} produce sequence level summaries of the input features and are closely related to \mbox{$\star$-vectors} (\cf \figref{diag:summary_network}).  Auxiliary features are produced by a neural network that takes as input the same features as the AM, and produces embeddings by taking the time-average of the output. By incorporating the averaging into the graph, the network can be trained jointly with the AM in an end-to-end fashion~\cite{delcroix2018auxiliary}. A related approach is to produce LHUC feature vectors (\secref{sec:structured_xforms}) from an independent network with embedded averaging \cite{xie2019}. 

\subsection{Embeddings for E2E systems}
The embedding method is also helpful to the adaptation of E2E systems. Fan et al.~\cite{fan2019speaker} and Sari et al.~ \cite{sari2020unsupervised} generated a soft embedding vector by combining a set of i-vectors from multiple speakers with the combination weight calculated from the attention mechanism. The soft embedding vector is appended to the acoustic encoder output of the E2E model, helping the model to normalize speaker variations.  While the soft embedding vectors in \cite{fan2019speaker, sari2020unsupervised} are different at each frame, the speaker i-vectors are concatenated with the speech utterance as the input of every encoder layer in \cite{zhao2020speech} to form  a persistent memory through the depth of encoder, hence learning utterance-level speaker knowledge.  

In addition to acoustic embedding, E2E models can also leverage text embedding to improve their modeling accuracy. For example, E2E models can be optimized to produce outputs relevant to the particular recognition context, for instance user contacts or device location. One solution is to add a context bias encoder in addition to the original audio encoder into E2E models \cite{pundak2018deep, chen2019end, jain2020contextual}.  This bias encoder takes a list of biasing phrases as the input. The context vector of the biasing list is generated by using the attention mechanism, and is then  concatenated with the context vector of acoustic encoder and is fed into the decoder. 
\section{Structured transforms}
\label{sec:structured_xforms}

Methods to adapt the parameters $\theta$ of a neural network-based acoustic model $f(x; \theta)$ can be split into two groups.
The first group adapts the whole acoustic model or some of its layers~\cite{liao2013speaker,yu2013kl,huang2015regularized}.
The second group employs structured  transformations~\cite{sim2017adaptation} to transform input features $x$, hidden activations $h$ or outputs $y$ of the acoustic model.
Such transformations include the linear input network (LIN)~\cite{neto1995speaker}, linear hidden network (LHN)~\cite{gemello2007linear} and the linear output network (LON)~\cite{li2010comparison}. 
These transforms are parameterized with a transformation matrix $A_s \in \mathbb{R}^{n \times n}$ and a bias $b_s \in \mathbb{R}^n$.
The transformation matrix $A_s$ is initialized as an identity matrix and the bias $b_s$ is initialized as a zero vector prior to speaker adaptation.
The adapted hidden activations then become
\begin{equation}
    h' = A_s h + b_s.
\end{equation}
However, even a single transformation matrix $A_s$ can contain many speaker dependent parameters, making adaptation susceptible to overfitting to the adaptation data.
It also limits its practical usage in real world deployment because of memory requirements related to storing speaker dependent parameters for each speaker.
Therefore there has been considerable research into how to structure the matrix $A_s$ and the bias $b_s$ to reduce the number of speaker dependent parameters.

The first set of approaches restricts the adaptation matrix $A_s$ to be diagonal.  If we denote the diagonal elements as $r_s = diag(A_s)$, then
the adapted hidden activations become
\begin{equation}
    h' = r_s \odot h + b_s.
\end{equation}
There are several methods that belong to this set of adaptation methods.
LHUC \cite{swietojanski2014learning,swietojanski2016learning}  adapts only the parameters $r_s$: 
\begin{equation}
    h' = r_s \odot h.
    \label{eq:lhuc}
\end{equation}
Speaker Codes~\cite{bridle1990recnorm,abdel2013fast} prepend an adaptation neural network to an existing SI model in place of the input features. The adaptation network -- which operates somewhat similarly to control networks, described below -- uses the acoustic features as inputs, as well as an auxiliary low-dimensional speaker code which essentially adapts speaker dependent biases within the adaptation network:
\begin{equation}
    h' = h + b_s.
\end{equation}
The network and speaker codes are learned by back-propagating through the frozen SI network with transcribed training data. At test time the speaker codes are derived by freezing all but the speaker code parameters and back-propagating on a small amount of adaptation data. 

Similarly, Wang and Wang~\cite{wang2017unsupervised} proposed a method that adapts both $r_s$ and $b_s$ as parameters $\beta_s \in \mathbb{R}^n$ and $\gamma_s \in \mathbb{R}^n$ of a batch normalization layer, adapting both the scale and the offset of the hidden layer activations with mean $\mu \in \mathbb{R}^n$ and variance $\sigma^2 \in \mathbb{R}^n$:
\begin{equation}
    h' = \gamma_s \frac{h - \mu}{\sqrt{\sigma^2 + \epsilon}} + \beta_s.
\end{equation}
Mana et al.~\cite{mana2019online} showed that batch normalization layers can be also updated by recomputing the statistics $\mu$ and $\sigma^2$ in online fashion.

A similar approach with a low-memory footprint adapts the activation functions instead of the scale $r_s$ and offset $b_s$.
Zhang and Woodland~\cite{zhang2015parameterised} proposed the use of parameterised sigmoid and ReLU activation functions. 
With the parameterised sigmoid function,
  hidden activations $h$ are computed from hidden pre-activations $z$ as
\begin{equation}
    h = \eta_s \frac{1}{1 + e^{-\gamma_s z + \zeta_s}},
\end{equation}
where $\eta_s \in \mathbb{R}^n$, $\gamma_s \in \mathbb{R}^n$ and $\zeta_s \in \mathbb{R}^n$ are speaker dependent parameters.
$|\eta_s|$ controls the scale of the hidden  activations,
  $\gamma_s$ controls the slope of the sigmoid function
  and $\zeta_s$ controls the midpoint of the sigmoid function.
Similarly, parameterised ReLU activations were defined as
\begin{equation}
    h = \begin{cases}
        \alpha_s z & \text{if } z > 0 \\
        \beta_s z & \text{if } z \leq 0 
    \end{cases},
\end{equation}
where $\alpha_s \in \mathbb{R}^n$ and $\beta_s \in \mathbb{R}^n$ are speaker dependent parameters that correspond to slopes for positive and negative pre-activations, respectively.

\begin{figure*}[tb]

\centering

\newcommand{\offset}{0.50cm}

\begin{minipage}[c]{.2\textwidth}
\centering
\subfloat[\label{diag:LHUC}]{
\begin{tikzpicture}[x=0.5cm, y=0.5cm, >=stealth]
    \tikzstyle{neuron}=[circle, draw=black, minimum size = 3mm, thick]
    \tikzstyle{state}=[shape=circle,draw=black,minimum size = 3mm, thick]
    \tikzstyle{mainedge}=[<-, thick]
    \tikzstyle{lightedge}=[<-,dotted,thick]
    \tikzstyle{selfloop}=[<-,loop above, thick]

	\draw [thick] (3.5,0) rectangle (10.5,1);
	\foreach \x in {1,...,3}
		\draw node at (3 + \x*1,0.5) [neuron] (h1_\x) {};
	\foreach \x in {1,...,3}
		\draw node at (11 - \x*1,0.5) [neuron] (h1_n-\x) {};
	\foreach \x in {1,...,3}
		\fill (7.0 -0.625 + \x*0.3,0.5) circle (1pt);

	\draw [thick] (3.5,3) rectangle (10.5, 4);
	\foreach \x in {1,...,3}
		\draw node at (3 + \x*1, 3.5) [neuron] (h2_\x) {};
	\foreach \x in {1,...,3}
		\draw node at (11 - \x*1, 3.5) [neuron] (h2_n-\x) {};
	\foreach \x in {1,...,3}
		\fill (7.0 -0.625 + \x*0.3, 3.5) circle (1pt);

    \foreach \i in {1,...,3}
        \draw [->,thick] (h1_\i) to (h2_\i);
        \foreach \i in {1,...,3}
        \draw [->,thick] (h1_n-\i) to (h2_n-\i);

    \draw [thick,fill=white] (5.5,1.5) rectangle (8.5,2.5);
    \draw node at (7,2) (P_S) {$r_s \in \mathbb{R}^{n}$};
    
    \clip ($(current bounding box.south west)+(0,-3)$) rectangle ($(current bounding box.north east)+(0,3)$);
\end{tikzpicture}
}
\end{minipage}
\begin{minipage}[c]{.35\textwidth}
\centering
\subfloat[\label{diag:LRPD}]{
\begin{tikzpicture}[x=0.5cm, y=0.5cm, >=stealth]
    \tikzstyle{neuron}=[circle, draw=black, minimum size = 3mm, thick]
    \tikzstyle{state}=[shape=circle,draw=black,minimum size = 3mm, thick]
    \tikzstyle{mainedge}=[<-, thick]
    \tikzstyle{lightedge}=[<-,dotted,thick]
    \tikzstyle{selfloop}=[<-,loop above, thick]

	\draw [thick] (3.5,0) rectangle (10.5,1);
	\foreach \x in {1,...,3}
		\draw node at (3 + \x*1,0.5) [neuron] (h1_\x) {};
	\foreach \x in {1,...,3}
		\draw node at (11 - \x*1,0.5) [neuron] (h1_n-\x) {};
	\foreach \x in {1,...,3}
		\fill (7.0 -0.625 + \x*0.3,0.5) circle (1pt);
	
	\draw [thick] (5,3) rectangle (9,4);
	\foreach \x in {1,...,4}
		\draw node at (4.5 + \x*1, 3.5) [neuron] (b_\x) {};

	\draw [thick] (3.5,6) rectangle (10.5, 7);
	\foreach \x in {1,...,3}
		\draw node at (3 + \x*1, 6.5) [neuron] (h2_\x) {};
	\foreach \x in {1,...,3}
		\draw node at (11 - \x*1, 6.5) [neuron] (h2_n-\x) {};
	\foreach \x in {1,...,3}
		\fill (7.0 -0.625 + \x*0.3, 6.5) circle (1pt);

    \foreach \i in {1,...,3}
        \foreach \j in {1,...,4}
		    \draw [->,thick] (h1_\i) to (b_\j);
	\foreach \i in {1,...,3}
        \foreach \j in {1,...,4}
		    \draw [->,thick] (h1_n-\i) to (b_\j);
		    
    \foreach \i in {1,...,3}
        \foreach \j in {1,...,4}
		    \draw [->,thick] (b_\j) to (h2_\i);
	\foreach \i in {1,...,3}
        \foreach \j in {1,...,4}
		    \draw [->,thick] (b_\j) to (h2_n-\i);

    \foreach \i in {1,...,3}
        \draw [->,thick, bend left=75,looseness=1.3] (h1_\i) to (h2_\i);
        \foreach \i in {1,...,3}
        \draw [->,thick, bend right=75,looseness=1.3] (h1_n-\i) to (h2_n-\i);

    \draw [thick,fill=white] (5,1.5) rectangle (9,2.5);
    \draw node at (7,2) (P_S) {$P_s \in \mathbb{R}^{n \times k}$};
    \draw [thick,fill=white] (5,4.5) rectangle (9,5.5);
    \draw node at (7,5) (Q_S) {$Q_s \in \mathbb{R}^{k \times n}$};
    \draw [thick,fill=white] (0.5,3) rectangle (4.5,4);
    \draw node at (2.5,3.5) (D_S) {$D_s \in \mathbb{R}^{n \times n}$};
    
    \draw [thick, opacity=0.0] (9.5,3) rectangle (13.5,4);
    
    \clip ($(current bounding box.south west)+(0,-1.5)$) rectangle ($(current bounding box.north east)+(0,1.5)$);
\end{tikzpicture}
}
\end{minipage}
\begin{minipage}[c]{.35\textwidth}
\centering
\subfloat[\label{diag:eLRPD}]{
\begin{tikzpicture}[x=0.5cm, y=0.5cm, >=stealth]
    \tikzstyle{neuron}=[circle, draw=black, minimum size = 3mm, thick]
	
	\draw [thick] (3.5,0) rectangle (10.5,1);
	\foreach \x in {1,...,3}
		\draw node at (3 + \x*1,0.5) [neuron] (h1_\x) {};
	\foreach \x in {1,...,3}
		\draw node at (11 - \x*1,0.5) [neuron] (h1_n-\x) {};
	\foreach \x in {1,...,3}
		\fill (7.0 -0.625 + \x*0.3,0.5) circle (1pt);
	
	\draw [thick] (5,3) rectangle (9,4);
	\foreach \x in {1,...,4}
		\draw node at (4.5 + \x*1, 3.5) [neuron] (b1_\x) {};

	\draw [thick] (5,6) rectangle (9,7);
	\foreach \x in {1,...,4}
		\draw node at (4.5 + \x*1, 6.5) [neuron] (b2_\x) {};

	\draw [thick] (3.5,9) rectangle (10.5, 10);
	\foreach \x in {1,...,3}
		\draw node at (3 + \x*1, 9.5) [neuron] (h2_\x) {};
	\foreach \x in {1,...,3}
		\draw node at (11 - \x*1, 9.5) [neuron] (h2_n-\x) {};
	\foreach \x in {1,...,3}
		\fill (7.0 -0.625 + \x*0.3, 9.5) circle (1pt);

    \foreach \i in {1,...,3}
        \foreach \j in {1,...,4}
		    \draw [->,thick] (h1_\i) to (b1_\j);
	\foreach \i in {1,...,3}
        \foreach \j in {1,...,4}
		    \draw [->,thick] (h1_n-\i) to (b1_\j);
		    
    \foreach \i in {1,...,4}
        \foreach \j in {1,...,4}
	        \draw [->,thick] (b1_\i) to (b2_\j);
    
    \foreach \i in {1,...,3}
        \foreach \j in {1,...,4}
		    \draw [->,thick] (b2_\j) to (h2_\i);
	\foreach \i in {1,...,3}
        \foreach \j in {1,...,4}
		    \draw [->,thick] (b2_\j) to (h2_n-\i);
		    
    \foreach \i in {1,...,3}
        \draw [->,thick, bend left=75,looseness=1.3] (h1_\i) to (h2_\i);
        \foreach \i in {1,...,3}
        \draw [->,thick, bend right=75,looseness=1.3] (h1_n-\i) to (h2_n-\i);

    \draw [thick,fill=white] (5,1.5) rectangle (9,2.5);
    \draw node at (7,2) (U) {$P \in \mathbb{R}^{n \times k}$};
    \draw [thick,fill=white] (5,4.5) rectangle (9,5.5);
    \draw node at (7,5) (E) {$T_s \in \mathbb{R}^{k \times k}$};
    \draw [thick,fill=white] (5,7.5) rectangle (9,8.5);
    \draw node at (7,8) (V) {$Q \in \mathbb{R}^{k \times n}$};
    \draw [thick,fill=white] (-0.5,4.5) rectangle (3.5,5.5);
    \draw node at (1.5,5) (D_S) {$D_s \in \mathbb{R}^{n \times n}$};
    
    \draw [thick, opacity=0.0] (10.5,4.5) rectangle (14.5,5.5);

\end{tikzpicture}
}
\end{minipage}

\caption{Structured transforms of an adaptation matrix $A_s$: (a) Learning Hidden Unit Contributions (LHUC) adapts only diagonal elements of the transformation matrix $r_s = diag(A_s)$; (b) Low-Rank Plus Diagonal factorizes the adaptation matrix as $A_s \approx D_s + P_s Q_s$; (c) Extended LRPD factorizes the adaptation matrix as $A_s \approx D_s + P T_s Q$.}
\label{fig:structured_transforms}
\end{figure*}
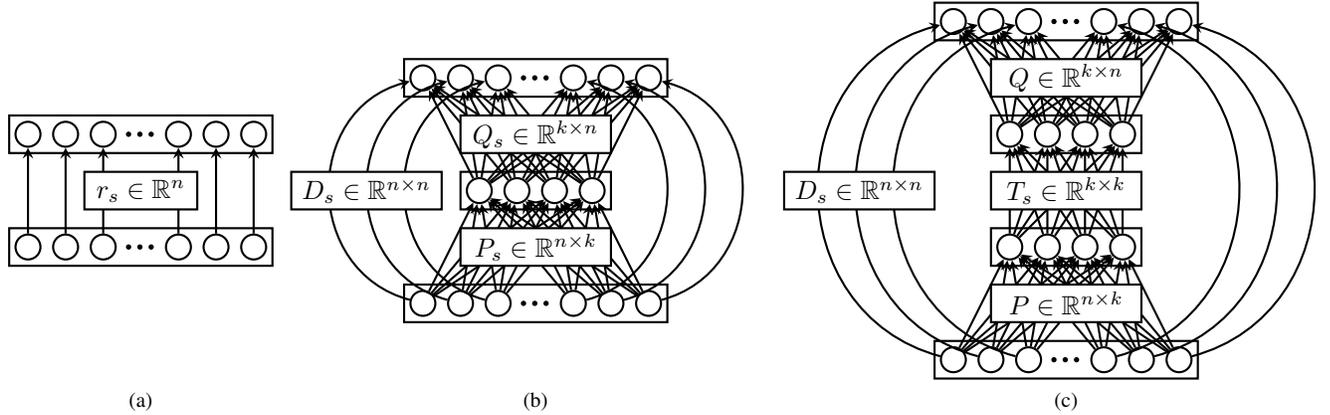

Other approaches factorize the transformation matrix $A_s$ into a product of low-rank matrices to obtain a compact set of speaker dependent parameters.
Zhao et al.~\cite{zhao2016low} proposed the \textit{Low-Rank Plus Diagonal} (LRPD) method, which reduces the number of speaker dependent parameters by approximating the linear transformation matrix $A_s \in \mathbb{R}^{n \times n}$ as
\begin{equation}
    A_s \approx D_s + P_s Q_s,
\end{equation}
where the $D_s \in \mathbb{R}^{n \times n}$, $P_s \in \mathbb{R}^{n \times k}$ and $Q_s \in \mathbb{R}^{k \times n}$ are treated as speaker dependent matrices ($k < n$) and $D_s$ is a diagonal matrix.
This approximation was motivated by the assumption that the adapted hidden activations should not be very different from the unadapted hidden activations  when only a limited amount of adaptation data is available; hence the adaptation linear transformation should be close to a diagonal matrix.
In fact, for $k = 0$ LRPD reduces to LHUC adaptation.
LRPD adaptation can be implemented by inserting two hidden linear layers and a skip connection as illustrated in \figref{diag:LRPD}.

Zhao et al.~\cite{zhao2017extended} later presented an extension to LRPD called \textit{Extended LRPD} (eLRPD),  which removed the dependency of the number of speaker dependent parameters on the hidden layer size by performing a different approximation of the linear transformation matrix $A_s$,
\begin{equation}
    A_s \approx D_s + P T_s Q,
\end{equation}
where matrices $D_s \in \mathbb{R}^{n \times n}$ and $T_s \in \mathbb{R}^{k \times k}$ are treated as speaker dependent, and matrices $P \in \mathbb{R}^{n \times k}$ and $Q \in \mathbb{R}^{k \times n}$ are treated as speaker independent.
Thus the number of speaker dependent parameters is mostly dependent on $k$, which can be chosen arbitrarily.

Instead of factorizing the transformation matrix, a technique typically known as feature-space discriminative linear regression (fDLR) \cite{abrash1995connectionist,seide2011feature,yao2012adaptation} imposes a block-diagonal structure such that each input frame shares the same linear transform. This is, in effect, a tied variation of LIN with a reduction in the number of speaker dependent parameters.

Another set of approaches uses the speaker dependent parameters as mixing coefficients $\theta_s = \left \{ \alpha_0 \dots \alpha_k \right \}$
  for a set of $k$ speaker independent bases $\left \{ B_0 \dots B_k \right \}$
  which factorize the transformation matrix $A_s$.
Samarakoon and Sim~\cite{samarakoon2015learning,samarakoon2016factorized} proposed to use factorized hidden layers (FHL) that allow both speaker-independent and speaker dependent modelling.
With this approach, activations of a hidden layer $h$ with an activation function $\sigma$ are computed as
\begin{equation}
    h = \sigma\left( (W + \sum_{i=0}^{k} \alpha_i B_i) x + b_s + b \right ).
    \label{eq:fhl1}
\end{equation}
Note, that  when $\alpha_s = 0$ and $b_s = 0$,
  the activations correspond to a standard speaker independent model.
If the bases $B_i$ are rank-1 matrices, $B_i = \gamma_i \psi_i^T$, then this allows the reparameterization of \eqref{eq:fhl1} as \cite{samarakoon2016factorized}:
\begin{equation}
    h = \sigma\left( (W + \Gamma D \Psi^T) x + b_s + b \right ),
\end{equation}
where vectors $\gamma_i$ and $\psi_i$ are $i$-th columns of matrices $\Gamma$ and $\Psi$, respectively, and the mixing coefficients $\alpha_s$ correspond to the diagonal of matrix $D$. 
This approach is very similar to the factorization of hidden layers used for Cluster Adaptive Training of DNN networks (CAT-DNN) \cite{tan2016cluster} that uses full rank bases instead of rank-1 bases.

Similarly, Delcroix et al.~\cite{delcroix2018context} proposed to adapt the activations of a hidden layer using a mixture of experts~\cite{jacobs1991adaptive}.
The adapted hidden unit activations are then
\begin{equation}
    h' = \sum_{i = 0}^{k} \alpha_i B_i h.
\end{equation}

There have also been approaches, that further reduce the number of speaker dependent parameters by removing the dependency on the hidden layer width by using control networks that predict the speaker-dependent parameters
\begin{equation}
    \theta_s = c(\lambda_s; \phi),
\end{equation}
In contrast to the adaptation network used in the Speaker Codes scheme, the control networks themselves are speaker-independent,
taking as input some lower dimensional speaker embedding $\lambda_s \in \mathbb{R}^k$.
As such, they form a link between structured transforms and the embedding-based approaches of Sec. \ref{sec:embeddings}.
The control networks $c(\lambda_s, \phi)$ can be implemented as a single linear transformation or as a multi-layer neural network.
These control networks are similar to the conditional affine transformations referred to as Feature-wise Linear Modulation (FiLM) \cite{perez2018film}.
For example, Subspace LHUC~\cite{samarakoon2016subspace} uses a control network to predict LHUC parameters $r_s$ from i-vectors $\lambda_s$, resulting in a $94\%$ memory footprint reduction compared to standard LHUC adaptation.
Cui et al. \cite{cui2017embedding} used auxiliary features to adapt both the scale $r_s$ and offset $b_s$.
Other approaches adapted the scale $r_s$ or the offset $b_s$ by leveraging the information extracted with summary networks instead of auxiliary features \cite{kim2017dynamic,sari2019speaker,xie2019fast}.

Finally, the number of speaker dependent parameters in all the aforementioned linear transformations can be reduced by applying them to bottleneck layers that have much lower dimensionality than the standard hidden layers.
These bottleneck layers can be obtained directly by training a neural network with bottleneck-layers or by applying Singular Value Decomposition (SVD) to the hidden layers~\cite{xue2013restructuring,xue2014singular}.

\section{Regularization methods}
\label{sec:regularisation}
\label{sec:regularization}

Even with the small number of speaker dependent parameters required by structured transformations, speaker adaptation can still overfit to the adaptation data.
One way to prevent this overfitting is through the use of regularization methods that prevent the adapted model from diverging too far from the original model.
This can be achieved by using early stopping and appropriate learning rates,
which can be obtained with a hyper-parameter grid-search or by meta-learning~\cite{klejch2018learning,klejch2019speaker}.
Another way to prevent the adapted model from diverging too far from the original can be achieved by limiting the distance between the original and the adapted model.
Liao~\cite{liao2013speaker} proposed to use the L2 regularization loss of the distance between the original speaker dependent parameters $\theta_s$ and the adapted speaker dependent parameters $\theta_s'$
\begin{equation}
    \mathcal{L}_{L2} = \lvert\theta_s - \theta_s'\rvert^2_2.
\end{equation}
Yu et al.~\cite{yu2013kl} proposed to use Kullback-Leibler (KL) divergence to measure the distance between the senone distributions of the adapted model and the original model
\begin{equation}
    \mathcal{L}_{KL} = D_{KL}(f(x; \theta) \,||\, f(x; \theta_s')).
\end{equation}
If we consider the overall adaptation loss using cross-entropy:
\begin{equation}
    \mathcal{L} = (1 - \lambda) \mathcal{L}_{xent} + \lambda \mathcal{L}_{KL},
\end{equation}
we can show that this loss equals to cross-entropy with the target distribution for a label $y$ given the input frame $x_t$ 
\begin{equation}
    (1 - \lambda) \hat{P}(y \mid x_t) + \lambda f(x_t; \theta),
\end{equation}
where $\hat{P}(y \,\vert\, x_t)$  is a distribution corresponding to the provided labels $y^\text{adapt}$.
Although initially proposed for adapting hybrid models, the KLD regularization method may also be used for speaker adaptation of E2E models \cite{li2018speaker, meng2019speaker, weninger2019listen}.

\usetikzlibrary{arrows.meta}

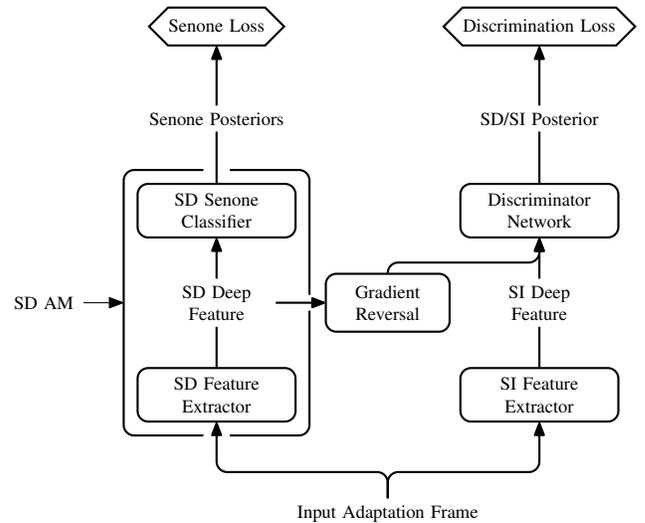
\begin{figure}[tb]

\centering

\begin{tikzpicture}[atipstyle, node distance=1.75cm,scale=0.7, every node/.style={transform shape}]

    \tikzstyle{loss} = [ thick, chamfered rectangle, chamfered rectangle xsep=2cm, text centered, draw=black]
    \tikzstyle{data} = [text centered, text width=2.5cm]
    
\node (senone_loss) [loss, text depth=0.1em,] {Senone Loss}; 

\node (student_senone) [data, below of=senone_loss, text width=3cm] {Senone Posteriors};
\node (senone_classifier) [model, below of=student_senone] {SD Senone Classifier}; 
\node (deep_feature) [data, below of=senone_classifier, text width=2cm] {SD Deep Feature}; 
\node (feature_extractor) [model, below of=deep_feature] {SD Feature Extractor}; 
\node (input) [data, right of=feature_extractor, xshift=1.5cm, yshift=-2.25cm, text width=3.5cm] {Input Adaptation Frame}; 

\node (condition_loss) [loss, right=3cm of senone_loss, text depth=0.1em,] {Discrimination Loss}; 

\node (condition_posterior) [data, text width=2.5cm, below of=condition_loss] {SD/SI Posterior};
\node (condition_classifier) [model, below of=condition_posterior] {Discriminator Network}; 
\node (si_deep_feature) [data, below of=condition_classifier, text width=2cm] {SI Deep Feature}; 
\node (si_feature_extractor) [model, below of=si_deep_feature] {SI Feature Extractor}; 

\node (student_am) [text width=1.2cm, left of=deep_feature, xshift=-1.5cm] {SD AM};

\draw[linestyle]     ($(senone_classifier.north west)+(-0.25,0.25)$) rectangle ($(feature_extractor.south east)+(0.25,-0.25)$);
\draw [line width=2pt, white] ($(feature_extractor.east)+(0.25,1.6)$) -- ($(senone_classifier.east)+(0.25,-1.6)$);
\draw [line width=2pt, white] ($(senone_classifier.north)+(-0.25,0.25)$) -- ($(senone_classifier.north)+(0.25,0.25)$);
\draw [line width=2pt, white] ($(feature_extractor.south)+(-0.25,-0.25)$) -- ($(feature_extractor.south)+(0.25,-0.25)$);

\node (grl) [model, right of=deep_feature, xshift=1.5cm, minimum width=2cm, minimum height=1.1cm] {Gradient Reversal}; 

\draw [linestyle, ->,] ($(input.north)$) -- ($(input.north)+(0,0.5)$) -| (feature_extractor.south);
\draw [linestyle,] (feature_extractor) -- (deep_feature);
\draw [linestyle, ->,] (deep_feature) -- (senone_classifier);
\draw [linestyle] (senone_classifier) -- (student_senone);
\draw [linestyle, ->] (student_senone) -- (senone_loss);

\draw [linestyle, ->,] (input.north) -- ($(input.north)+(0,0.5)$) -| (si_feature_extractor.south);
\draw [linestyle,] (si_feature_extractor) -- (si_deep_feature);
\draw [linestyle, ->,] (si_deep_feature) -- (condition_classifier);
\draw [linestyle,] (condition_classifier) -- (condition_posterior);
\draw [linestyle, ->,] (condition_posterior) -- (condition_loss);

\draw [linestyle, ->,] (deep_feature.east) -- (grl.west);
\draw [linestyle, ->] (grl.north) -- ($(grl.north)+(0,0.2)$) -| ($(condition_classifier.south)$);

\draw[draw=black] (student_am) edge [->] ($(deep_feature.west -| senone_classifier.west)+(-0.25,0)$);

\end{tikzpicture}

\caption{Adversarial speaker adaptation.}
\label{diag:adversarial_speaker_adaptation}
\end{figure}

Meng et al.~\cite{meng2019adversarial} noted that KL divergence is not a 
distance metric between distributions because it is asymmetric, and therefore proposed to use adversarial learning which guarantees that the local minimum of the regularization term is reached only 
if the senone distributions of the speaker independent and the speaker dependent models are identical.
They achieve this by adversarially training a discriminator $d(x; \phi)$ whose task is to discriminate between the speaker dependent deep features $h'$ and speaker independent deep features $h$ that are obtained by passing the input adaptation frames through speaker dependent and speaker independent feature extractor respectively.
This process is illustrated in \figref{diag:adversarial_speaker_adaptation}.
The regularization loss of the discriminator is
\begin{equation}
    \mathcal{L}_{disc} = - \log d(h; \phi) - \log \left [ 1 - d(h'; \phi)  \right ],
\end{equation}
where $h$ are hidden layer activations of the speaker independent model and $h'$ are hidden layer activations of the adapted model.
The discriminator is trained in a minimax fashion during adaptation by minimizing $\mathcal{L}_{disc}$ with respect to $\phi$ and maximizing $\mathcal{L}_{disc}$ with respect to $\theta_s$.
Consequently, the distribution of activations of the i-th hidden layer of the speaker dependent model will be indistinguishable from the distribution of activations of the i-th hidden layer of the speaker independent model,
  which ought to result in more robust performance of speaker adaptation.

Other approaches aim to prevent overfitting by leveraging the uncertainty of the speaker-dependent parameter space. 
Huang et al.~\cite{huang2015maximum} proposed Maximum A Posteriori (MAP) adaptation of neural networks, inspired by MAP adaptation of GMM-HMM models~\cite{gauvain1994maximum} (\secref{sec:early}).
MAP adaptation estimates speaker dependent parameters as a mode of the distribution
\begin{equation}
    \hat{\theta}_s = \arg \max_{\theta_s} P(Y \mid X, \theta_s) p(\theta_s),   
\end{equation}
where $p(\theta_s)$ is a prior density of the speaker dependent parameters.
In order to obtain this prior density, Huang et al.~\cite{huang2015maximum} employed an empirical Bayes approach (following Gauvain and Lee~\cite{gauvain1994maximum}) and treated each speaker in the training data as a data point.
They performed speaker adaptation for each speaker and observed that the speaker parameters across speakers resemble Gaussians.
Therefore they decided to parameterise the prior density $p(\theta_s)$ as
\begin{equation}
    p(\theta_s) = \mathcal{N}(\theta_s ; \mu, \Sigma),
\end{equation}
where $\mu$ is the mean of adapted speaker dependent parameters across different speakers,
  and $\Sigma$ is the corresponding diagonal covariance matrix. 
With this parameterisation the regularization term of the prior density $p(\theta_s)$ is
\begin{equation}
    \mathcal{L}_{MAP} = \frac{1}{2}(\theta_s - \mu)^T \Sigma^{-1} (\theta_s - \mu), 
\end{equation}
which for the prior density $p(\theta_s) = \mathcal{N}(\theta_s ; 0, I)$ degenerates to the 
L2
regularization loss.  Huang et al. investigated their proposed MAP approach with LHN structured transforms, but noted that it may be
used in combination with other schemes.

Xie at al~\cite{xie2019blhuc} proposed a fully Bayesian way of dealing with uncertainty inherent in speaker dependent parameters $\theta_s$, in the context of estimating the LHUC parameters $r_s$ (see Sec. \ref{sec:structured_xforms}).  In this method, known as BLHUC, the
posterior distribution of the adapted model is approximated as:
\begin{equation}
    P(Y \mid X, \mathcal{D}^\text{adapt}) \approx P(Y \mid X, \mathbb{E}[r_s \mid \mathcal{D}^\text{adapt}]),
    \label{eq:blhuc_posterior}
\end{equation}
Xie at al propose to use a distribution $q(r_s)$ as a variational approximation of the posterior distribution of the LHUC parameters, $p(r_s | \mathcal{D}^\text{adapt})$.
For simplicity, they assume that both $q(r_s)$ and $p(r_s)$ are normal, such that $q(r_s) = \mathcal{N}(r_s ; \mu_s, \gamma_s)$ and $p(r_s) = \mathcal{N}(r_s ; \mu_0, \gamma_0)$, which results in the expectation for the speaker dependent parameters in (\ref{eq:blhuc_posterior}) being given by :
\begin{equation}
    \mathbb{E}[r_s \mid \mathcal{D}^\text{adapt} ] = \mu_s.
\end{equation}
The parameters are computed using gradient descent with a Monte Carlo approximation.
Similarly to MAP adaptation, the effect is to force the adaptation to stay close to the speaker independent model when we perform adaptation with a small amount of adaptation data. 
\section{Variant objective functions}
\label{sec:objective_functions}

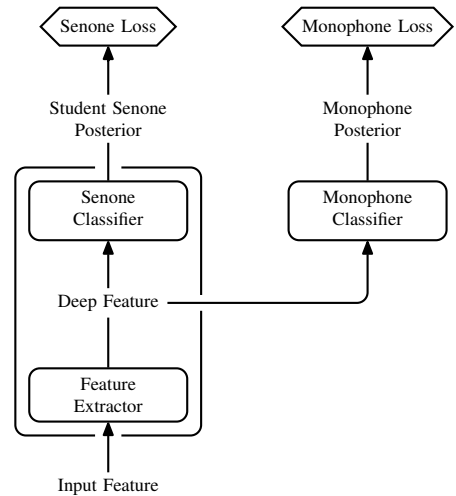
\begin{figure}

\centering

\begin{tikzpicture}[atipstyle, node distance=1.75cm, scale=0.7, every node/.style={transform shape}]
    \tikzstyle{loss} = [chamfered rectangle, chamfered rectangle xsep=2cm, text centered, draw=black, thick,
    				text depth=0.1em,]
    \tikzstyle{data} = [text centered, text width=2.5cm]

\node (senone_loss) [loss, text depth=0.1em] {Senone Loss}; 

\node (student_senone) [data, below of=senone_loss] {Student Senone Posterior};
\node (senone_classifier) [model, below of=student_senone, text width=2.2cm, text depth=1.5em] {Senone\\Classifier}; 
\node (deep_feature) [data, below of=senone_classifier, text width=2cm] {Deep Feature}; 
\node (feature_extractor) [model, below of=deep_feature, text width=2.2cm] {Feature\\Extractor}; 
\node (student_input) [data, below of=feature_extractor] {Input Feature}; 

\node (monophone_loss) [loss, text depth=0.1em, right=2cm of senone_loss] {Monophone Loss}; 

\node (monophone_posterior) [data, text width=2cm, below of= monophone_loss] {Monophone Posterior};
\node (monophone_classifier) [model, below of=monophone_posterior, text width=2.5cm,  text depth=1.5em] {Monophone Classifier}; 

\draw[linestyle] ($(senone_classifier.north west)+(-0.25,0.25)$) rectangle ($(feature_extractor.south east)+(0.25,-0.25)$);
\draw [line width=2pt, white] ($(feature_extractor.east)+(0.25,1.6)$) -- ($(senone_classifier.east)+(0.25,-1.6)$);
\draw [line width=2pt, white] ($(senone_classifier.north)+(-0.25,0.25)$) -- ($(senone_classifier.north)+(0.25,0.25)$);
\draw [line width=2pt, white] ($(feature_extractor.south)+(-0.25,-0.25)$) -- ($(feature_extractor.south)+(0.25,-0.25)$);

\draw [linestyle, ->] (student_input) -- (feature_extractor);
\draw [linestyle] (feature_extractor) -- (deep_feature);
\draw [linestyle, ->] (deep_feature) -- (senone_classifier);
\draw [linestyle] (senone_classifier) -- (student_senone);
\draw [linestyle, ->] (student_senone.north) -- +(0,0.5) -| (senone_loss.south);

\draw [linestyle, ->] (deep_feature) -| (monophone_classifier);
\draw [linestyle] (monophone_classifier) -- (monophone_posterior);
\draw [linestyle, ->] (monophone_posterior.north) -- (monophone_loss.south);
\end{tikzpicture}

\caption{ Multi-task learning speaker adaptation.}
\label{diag:multitask_adaptation}

\end{figure}

Another challenge in speaker adaptation is overfitting to targets seen in the adaptation data and to errors in semi-supervised transcriptions.
This issue can be mitigated by an appropriate choice of objective function.

Gemello et al.~\cite{gemello2007linear} proposed \textit{Conservative Training},
  which modifies the target distribution to ensure that labels not seen in the adaptation data will not be catastrophically forgotten. 
Given a set of labels not seen in the adaptation data  $U$ and the reference label $\hat{y}_t$ at a time-step $t$ the adjusted target distribution $\hat{P}$ is defined as
\begin{equation}
    \hat{P}(y \mid x_t) = \begin{cases}
        P(y \mid x_t)                              & \text{if } y \in U \\
        1 - \sum\limits_{y' \in U} P(y' | x_t)   & \text{if } y = \hat{y}_t \\
        0                                           & \text{otherwise}.  
    \end{cases}
\end{equation}

To mitigate errors in semi-supervised transcriptions we can replace the transcriptions with a lattice of supervision,
  which encodes the uncertainty arising from the first pass decoding. 
Lattice supervision has previously been used in work on unsupervised adaptation~\cite{padmanabhan2000lattice} and training~\cite{fraga2011lattice} of GMMs,
as well as discriminative \cite{povey2005discriminative} and semi-supervised training \cite{manohar2018semi}, and adaptation \cite{klejch2019lattice}, of neural network models.
For instance, lattice supervision can be used with the MMI criterion where for a single utterance we have:
\begin{equation}
    \mathcal{F}_{MMI}(\theta) = \log \frac{p (X \mid \mathbb{M}^{num}; \theta)}{p (X \mid \mathbb{M}^{den}; \theta)},
\end{equation}
where the $\mathbb{M}_r^{num}$ is a numerator lattice containing multiple hypotheses from a first pass decoding and $\mathbb{M}_r^{den}$ is a denominator lattice containing all possible sequences of words.


Another family of methods prevents overfitting to adaptation targets by performing adaptation through the use of a lower entropy task such as monophone or senone cluster targets.
This has the advantage that the unsupervised targets might be less noisy
  and also that the targets have higher coverage even with small amounts of adaptation data.
Price et al.~\cite{price2014speaker} proposed to append a new output layer predicting monophone targets on top of the original output layer predicting senones.
The layer can be either full rank or sparse -- leveraging knowledge of relationships between monophones and senones.
Its parameters are trained on the training data with a fixed speaker independent model.
Only the monophone targets are used for the adaptation of the speaker dependent parameters.

Huang et al.~\cite{huang2015rapid} presented an approach that used  multi-task learning~\cite{caruana1997multitask} to leverage both senone and monophone/senone clusters targets.
It worked by having multiple output layers, each on top of the last hidden layer, that predicted the corresponding targets.
These additional output layers were also trained after a complete training pass of the speaker independent model with its parameters fixed.
Thus, the adaptation loss was a weighted sum of individual losses, 
  for example monophone and senone losses (\figref{diag:multitask_adaptation}).
Swietojanski et al.~\cite{swietojanski2015structured} combined these two approaches and used multi-task learning for speaker adaptation through a structured output layer, which predicts both monophone targets and senone targets.  Unlike the approach by Price et al.~\cite{price2014speaker},  the monophone predictions are used for the prediction of senones.

Li et al.~\cite{li2018speaker} and Meng et al.~\cite{meng2019speaker} applied multi-task learning to speaker adaptation of CTC and AED models. These E2E models typically use subword units, such as word piece units, as the output target in order to achieve high recognition accuracy. The number of subword units is usually at the scale of thousands or even more. Given very limited speaker-specific adaptation data, these units may not be fully covered.  Multi-task learning using both character and subword units can significantly alleviate such sparseness issues. 
\section{Data augmentation}
\label{sec:data_augmentation}

Data augmentation has been proven to be an effective way to decrease the acoustic mismatch between training and testing conditions.  Data augmentation approaches supplement the training data with distorted or synthetic variants of speech with characteristics resembling the target acoustic environment, for instance with reverberation  or interfering sound sources. Thanks to realistic room acoustic simulators~\cite{allen1979image} one can generate large numbers of room impulse responses and reuse clean corpora to create multiple copies of the same sentence under different acoustic conditions~\cite{ko2017study, kim2017generation, li2014noise}. 

Similar approaches have been proposed for increasing robustness in speaker space by augmenting training data with, typically label-preserving, speaker-related distortions or transforms. Examples include creating multiple copies of clean utterances with perturbed VTL warp factors~\cite{jaitly2013vocal, cui2015data}, augmenting related properties such as volume or speaking rate~\cite{peddinti2015time, ko2015audio, huang2020acoustic}, or voice-conversion~\cite{stylianou1998continuous} inspired transformations of speech uttered by one speaker into another speaker using stochastic feature mapping~\cite{cui2015data2,cui2015data,fainberg2016improving}. 

While voice conversion does not create any new data with respect to unseen acoustic / linguistic complexity (just replicas of the utterances with different voices, often from the same dataset), recent advances in text-to-speech (TTS) allows the rapid building of new multi-speaker TTS voices~\cite{ye2018transfer} from small amounts of data.  TTS may then be used to arbitrarily expand the adaptation set for a given speaker, possibly to cover unseen acoustic domains~\cite{huang2020using, Huang2020Rapid}. If TTS is coupled with a related natural language generation module, it is possible to generate speech for domain-related texts. In this way, the speaker adaptation uses more data, not only from the speaker's original speech but also from the TTS speech.  Because the transcription used for TTS generation is also used for model adaptation, this approach also circumvents the obstacle of the hypothesis error in unsupervised adaptation. Moreover, TTS generated data can also help to adapt E2E models to a new domain which has more discrepancy in contents from the source domain, which will be discussed in \secref{sec:lmadapt}.

Finally, for unbalanced data sets the acoustic models may under-perform for certain demographics that are not sufficiently represented in training data. There is an ongoing effort to address this using generative adversarial networks (GANs). For example, Hosseini-Asl et al.~\cite{hosseini2018augmented} used GANs with a cycle-consistency constraint~\cite{zhu2017unpaired} to balance the speaker ratios with respect to gender representation in training set.  

\section{Accent adaptation}
\label{sec:accent_adaptation}

Although there is significant literature on automatic dialect identification from speech (\eg \cite{ali2016automatic}), there has been less work on accent and dialect adaptive speech recognition systems.  The MGB--3~\cite{ali2017mgb3} and MGB--5~\cite{ali2019mgb5} evaluation challenges have used dialectal Arabic test sets, with a modern standard Arabic (MSA) training set, using broadcast and internet video data.  The best results reported on these challenges have used a straightforward model-based transfer learning approach in an lattice-free maximum mutual information (LF-MMI) framework \cite{povey2016lfmmi},  adapting MSA trained baseline systems to specific Arabic dialects \cite{smit2017aalto,khurana2019darts}. 

Much of the reported work on accent adaptation has taken approaches for speaker adaptation, and applied them using an adaptation set of utterances from the target accent.  For instance, Vergyri et al.~\cite{vergyri2010automatic} used MAP adaptation of a GMM/HMM system. Zheng et al.~\cite{zheng2005accent} used both MAP and MLLR adaptation, together with features selected to be discriminative towards accent, with the accent adaptation controlled using hard decisions made by an accent classifier.  

Earlier work on accent adaptation focused on automatic adaptation of the pronunciation dictionary  \cite{kat1999fast,liu2000mandarin}.  These approaches resemble approaches for acoustic adaptation of VQ codebooks (discussed in section \ref{sec:early}), in that they learn an accent-specific transition matrix between the phonemic symbols in the dictionary.  Selection of utterances for accent adaptation has been explored, with Nallasamy et al.~\cite{nallasamy2012active} proposing an active learning approach.

Approaches to accent adaptation of neural network-based systems have typically employed accent-dependent output layers and shared hidden layers \cite{huang2014multi,chen2015improving}, based on a similar approach to the multilingual training of deep neural networks \cite{ghoshal2013multilingual,heigold2013multilingual,huang2013cross}.  Huang et al.~\cite{huang2014multi} combined this with KL regularization (\secref{sec:regularization}), and Chen et al.~\cite{chen2015improving} used accent-dependent i-vectors (\secref{sec:embeddings});  Yi et al.~\cite{yi2016improving} used accent-dependent bottleneck features in place of i-vectors; and Turan et al.~\cite{turan2020achieving} used x-vector accent embeddings in a semi-supervised setting.  

Multi-task learning approaches, where the secondary task is accent/dialect identification has been explored by a number of researchers \cite{elfeky2016towards,yang2018joint,jain2018improved,li2018speaker_adapt,viglino2019end} in the context of both hybrid and end-to-end models.   Improvements with multi-task training were observed in some instances, but the evidence indicates that it gives a small adaptation gain.  Sun et al.~\cite{sun2018domain} replaced multi-task learning with domain adversarial learning (\secref{sec:objective_functions}), in which the objective function treated accent identification as an adversarial task, finding that this improved accented speech recognition over multi-task learning.

More successfully, Li et al.~\cite{li2018multi} explored learning multi-dialect sequence-to-sequence models using one-hot dialect information as input. Grace et al.~\cite{grace2018occam} also used one-hot dialect codes and also explored a family of cluster adaptive training and hidden layer factorization approaches. In both cases using one-hot dialect codes as an input augmentation (corresponding to bias adaptation) proved to be the best approach, and cluster-adaptive approaches did not result in a consistent gain.  These approaches were extended by Yoo et al.~\cite{yoo2019highly} and Viglino et al.~\cite{viglino2019end} who both explored the use of dialect embeddings for multi-accent end-to-end speech recognition. Ghorbani et al.~\cite{ghorbani2018advancing} used accent-specific teacher-student learning, and Jain et al.~\cite{jain2019multi} explored a mixture of experts (MoE) approach, using mixtures of experts both at the phonetic and accent levels.  

Yoo et al.~\cite{yoo2019highly} also applied a method of feature-wise affine transformations on the hidden layers (FiLM), that are dependent both on the network’s internal state and the dialect/accent code (discussed in \secref{sec:structured_xforms}).  This approach, which can be viewed as a conditioned normalization, differs from the previous use of one-hot dialect codes and multi-task learning in that it has the goal of learning a single normalized model rather than an implicit combination of specialist models.  A related approach is gated accent adaptation \cite{zhu2019multi}, although this focused on a single transformation conditioned on accent.

Winata et al.~\cite{winata2020learning} experimented with a meta-learning approach for few-shot adaptation to accented speech, where the meta-learning algorithm learns a good initialization and hyperparameters for the adaptation.
\section{Domain adaptation}

\label{sec:domain_adaptation}

The performance of automatic speech recognition (ASR) always drops significantly when the recognition model is evaluated in a mismatched new domain. Domain adaptation is the technology used to adapt the well-trained source domain model to the new domain. The most straightforward way is to collect and label data in the new domain to fine-tune the model. Most adaptation technologies discussed in this paper can also be applied to domain adaptation \cite{huang2015regularized, long2017domain, fainberg2017factorised, sim2018domain, huang2020cross}. When the amount of adaptation data is limited, a common practice is adapting only partial layers of the network \cite{ueno2018encoder}. To let the adapted model still perform well on the source domain, Moriya et al.~\cite{moriya2018progressive} proposed progressive neural networks by adding an additional model column to the original model for each new domain and only update the new model column with the new domain data. In the following, we focus on technologies more specific to domain adaptation. 

\subsection{Teacher-student learning}
While  conventional adaptation techniques require large amounts of labeled data in the target domain, the teacher-student (T/S) paradigm \cite{li2014learning, hinton2015distilling} can better take advantage of large amounts of unlabeled data and has been widely used for industrial scale tasks \cite{Li2018Developing, movsner2019improving}. 

The most popular T/S learning strategy was proposed in 2014 by Li et al. \cite{li2014learning} to minimize the KL divergence between the output posterior distributions of the teacher network and the student network. This can also be considered as learning soft targets generated by a teacher model instead of 1-hot hard targets
\begin{equation}
    -\sum_{t=1}^T\sum_{y=1}^N P_T(y\mid x_t) \log P_S(y\mid x_t),
\end{equation}
where $P_T$ and $P_S$ are posteriors of teacher and student networks, $x_t$ and $y_t$ are the input speech and output senone at time $t$, respectively. $T$ is the number of speech frames in an utterance, and $N$ is the number of senones in the network output layer. 

Later, Hinton et al.~\cite{hinton2015distilling} proposed knowledge distillation by introducing a temperature parameter (like chemical distillation) to scale the posteriors. This has been applied to speech by \eg Asami et al.~\cite{asami2017domain}. There are also variations such as learning the interpolation of soft and hard targets \cite{hinton2015distilling} and conditional T/S learning \cite{meng2019conditional}. Although initially proposed for model compression, T/S learning is also widely used for model adaptation if source and target signals are frame-synchronized, which can be realized by simulation. The loss function is   \cite{Li17TS, watanabe2017student}
\begin{equation}
    -\sum_{t=1}^T\sum_{y=1}^N P_T(y\mid x_t) \log P_S(y\mid \hat{x}_t),
\end{equation}
where $x_t$ is the source speech signal while $\hat{x}_t$ is the frame-synchronized target signal. It can be further improved with sequence-level loss function as the speech signal is a sequence signal \cite{wong2016sequence, manohar2018teacher}.

The biggest advantage of T/S learning is that it can leverage large amounts of unlabeled data by using soft labels $P_T(y_t=y|x_t)$. This is particularly useful in industrial setups where effectively unlimited unlabeled data is available \cite{Li2018Developing,movsner2019improving}. Furthermore, soft labels produced by the teacher network carry knowledge learned by the teacher on the difficulty of classifying each sample, while the hard labels do not contain such information. Such knowledge helps the student to generalize better, especially when adaptation data size is small. 

E2E models tend to memorize the training data well, and therefore may not generalize well to a new domain. Meng et al.~\cite{meng2019domain} proposed T/S learning for the domain adaptation of E2E models. The loss function is
\begin{equation}
    -\sum_{l=1}^L\sum_{y=1}^N P_T(y \mid Y_{1:u-1}, X) \log P_S(y \mid Y_{1:u-1}, \hat{X}),
\end{equation}
where $X$ and $\hat{X}$ are the source and target domain speech sequence, $Y$ is the label sequence of length $L$ which is either the ground truth in the supervised adaptation setup or the hypothesis generated by the decoding of the teacher model with $X$ in the unsupervised adaptation setup.  Note that in the unsupervised case, there are two levels of knowledge transfer: the teacher’s token posteriors (used as soft labels) and one-best predictions as decoder guidance. 

One constraint to T/S adaptation is that it requires paired source and target domain data. While the paired data can be obtained with simulation in most cases, there are scenarios in which it is hard to simulate the target domain data from the source domain data. For example, simulation of children's speech  or accented speech remains challenging. In \cite{meng2020vector}, a neural label embedding scheme was proposed for domain adaptation with unpaired data. A  label embedding, l-vector, represents the output distribution of the deep network trained in the source domain for each output token, \eg, senone. To adapt the deep network model to the target domain,  the l-vectors learned from the source domain are used as the soft targets in the cross entropy criterion.

\subsection{Adversarial learning}
It is usually hard to obtain the transcription in the target domain, therefore unsupervised adaptation is critical. Although the transcription can be generated by decoding the target domain data using the source domain model, the generated hypothesis quality is often poor given the domain mismatch. 
Recently, adversarial training was applied to the area of unsupervised domain adaptation in a form of multi-task learning \cite{grl_ganin} without the need for transcription in the target domain. Unsupervised adaptation is achieved by learning deep intermediate representations that are both discriminative for the main task on the source domain and invariant with respect to mismatch between source and target domains. Domain invariance is achieved by adversarial training of the domain classification objective functions using a \emph{gradient reversal layer (GRL)}~\cite{grl_ganin}. This GRL approach has been applied to acoustic models for unsupervised adaptation in \cite{grl_sun, meng2017unsupervised, denisov2018unsupervised}. Meng et al.~\cite{meng2018adversarial} further combine adversarial learning and T/S learning as adversarial T/S learning to improve the robustness against condition variability during adaptation.

There is also increasing interest in the use of GANs with cycle consistency constraints for domain adaptation~\cite{mimura2017cross,hosseini_asl2018,meng2018cycle}. This enables the use of non-parallel data without labels in the target domain by learning to map the acoustic features into the style of the target domain for training. The cycle-consistency constraint also provides the possibility of mapping features from the target to the source style for, in effect, test-time adaptation or speech enhancement.




Unsupervised domain adaptation is more attractive than the supervised one because there is usually large amount of unlabeled data in the new domain while transcribing the new domain data usually is time consuming with large cost. T/S learning and adversarial learning both can utilize  unlabeled data well. Specifically, T/S learning has been very successful in industry-scale tasks. In contrast, adversarial learning was reported successful in relatively smaller tasks. Therefore, T/S learning is more promising if the parallel data is available. However, if there is no prior knowledge about the new domain, adversarial learning can be a good choice. There are also other works on unsupervised domain adaptation. For example, Hsu et al.~\cite{hsu2017unsupervised} use a variational autoencoder instead of adversarial learning to obtain a latent representation robust to domains. However, similar to adversarial learning, such method is pending examination when large amount of unlabeled training data is available.  
\section{Language model adaptation}
\label{sec:lmadapt}

LM adaptation typically involves updating an LM estimated from a large general corpus, with data from a target domain.  Many approaches to LM adaptation were developed in the context of n-gram models, and are reviewed by Bellegarda~\cite{bellegarda2004statistical}. Hybrid NN/HMM speech recognition systems still make use of n-gram language models and a finite state structure, at least in the first pass; it is difficult to use neural network LMs (with infinite context) directly in  first pass decoding in such systems.  Neural network LMs are typically used to rescore lattices in hybrid systems, or may be combined (in a variety of ways) in end-to-end systems.

The main techniques for n-gram language model adaptation include interpolation of multiple language models \cite{clarkson1997language,tur2007unsupervised,liu2008context}, updating the model using a cache of recently observed (decoded) text  \cite{kuhn1990cache,kuhn1992corrections,federico1996bayesian,clarkson1997language}, or merging or interpolating n-gram counts from decoded transcripts \cite{bacchiani2003unsupervised}.  There is also a large body of work incorporating longer scale context, for instance modelling the topic and style of the recorded speech  \cite{seymore1998nonlinear,chen2001using,hsu2006style,huang2008unsupervised}. LM adaptation approaches making use of wider context have often built on approaches using unigram statistics or bag-of-words models, and a number of approaches for combination with n-gram models have been proposed, for example dynamic marginals \cite{kneser1997language}.

Neural network language modelling \cite{bengio2003neural} has become state-of-the-art, in particular recurrent neural network language models (RNNLMs) \cite{mikolov2011extensions}.  
There has been a range of work on adaptation of RNNLMs, including the use of topic or genre information as auxiliary features  \cite{chen2015recurrent,deena2018recurrent} or combined as marginal distributions \cite{li2018recurrent},  domain specific embeddings \cite{moriokal2018language}, and the use of curriculum learning and fine-tuning to take account of shifting contexts  \cite{shi2015recurrent,gangireddy2016unsupervised}.  Approaches based on acoustic model adaptation, such as LHUC \cite{gangireddy2016unsupervised} and LHN \cite{deena2018recurrent}, have also been explored.  

There have a been a number of approaches to apply the ideas of cache language model adaptation to neural network language models \cite{li2018recurrent,grave2017improving,merity2016pointer}, along with so-called dynamic evaluation approaches in which the recent context is used for fine tuning \cite{li2018recurrent,krause2018dynamic}.

E2E models are trained with paired speech and text data. The amount of text data in such a paired setup is much smaller than the amount of text data used in training a separate external LM. Therefore, it is popular to adjust E2E models by fusing the external LM trained with a large amount of text data. The simplest and most popular approach is shallow fusion \cite{gulcehre2015Shallow, hannun2014first, kannan2018shallowfusion,zeyer2018shallowfusion}, in which the external LM is interpolated log-linearly with the E2E model at inference time only. 

However, shallow fusion does not have  a clear probabilistic interpretation. McDermott et al.~\cite{mcdermott2019density} proposed a density ratio approach based on Bayes' rule. An LM is built on text transcripts from  the training set which has paired speech and text data, and a second LM is built on the target domain. When decoding on the target domain, the output of the E2E model is modified by the ratio of target/training LMs. While it is well grounded with Bayes' rule, the density ratio method requires the training of two separate LMs, from the training and target data respectively. Variani et al.~ \cite{variani2020hybrid} proposed a hybrid autoregressive transducer (HAT) model to improve the RNN-T model.  The HAT model builds a training set LM internally and the label distribution is derived by normalizing the score functions across all labels excluding blank. Therefore, it is mathematically justified to integrate the HAT model with an external or target LM using the density ratio formulation. 

In \cite{sim2019personalization, Li2020RNNT}, RNN-T models were adapted to a new domain  with the TTS data generated from the domain-specific text.  Because the prediction network in RNN-T works similarly to a LM, adapting it without updating the acoustic encoder  is shown to be more effective than interpolating the RNN-T model with an external LM trained from the domain-specific text \cite{Li2020RNNT}. 

\begin{table*}[ht!]
\caption{Adaptation studies used in the meta-analysis categorized on the level they operate at, and system architecture.}
\label{tab:ma_refs_partition}
\centering
    \begin{tabular}{l|c|l}
        \toprule
        Level & System &  References  \\
        \midrule
        \multirow{2}{1.3cm}{Model} & Hybrid & \cite{Li2018Developing,chen2015improving,fainberg2019acoustic,huang2014multi,huang2015regularized,huang2020acoustic,kim2017dysarthric,kitza2018comparison,klejch2018learning,klejch2019speaker,liao2013speaker,liu2016investigations,miao2015speaker,samarakoon2016factorized,seki2018rapid,serizel2014deep,swietojanski2016differentiable,swietojanski2016learning,weninger2019listen,xie2019,zhang2016dnn,zhu2019multi} \\
        & E2E & \cite{ghorbani2018advancing,ghorbani2018leveraging,li2018speaker,meng2019domain,meng2019speaker,sim2019investigation,weninger2019listen,winata2020learning} \\
        \midrule
        \multirow{2}{1.3cm}{Embedding}    & Hybrid & \cite{abdel2013fast,chen2015improving,gupta2014vector,li2015modeling,mana2019online,miao2015speaker,pan2020memory,rownicka2018analyzing,rownicka2019embeddings,samarakoon2016factorized,saon2013speaker,senior2014improving,tan2016speaker,tomashenko2015gmm,xie2019}  \\
        & E2E & \cite{delcroix2018auxiliary,sari2020unsupervised,yi2016improving} \\
        \midrule
        Feature     & Hybrid & \cite{kitza2018comparison,miao2015speaker,samarakoon2016factorized,saon2013speaker,seide2011feature,swietojanski2016differentiable,swietojanski2016learning,tomashenko2015gmm,woodland2015cambridge,zhang2016dnn,zhu2019multi}  \\
        \midrule
        Data        &  Hybrid & \cite{cui2015data,huang2020using} \\
        \bottomrule
    \end{tabular}
\end{table*}

\section{Meta Analysis}
\label{sec:experimental}
\label{sec:meta_analysis}

In this section we present an aggregated review of published results in experiments applying adaptation algorithms to speech recognition.  This differs from typical experimental reporting that focuses on one-to-one system comparisons typically using a small fixed set of systems and benchmark tasks and data.  The proposed meta-analysis approach offers insights into the performance of adaptation algorithms that are difficult to capture from individual experiments. 

We divide this section into four main parts. The first, \secref{ssec:init}, explains the protocol and overall assumptions of the meta-analysis, followed by a top-level summary of findings in \secref{ssec:findings_generic}, with a more detailed analysis  in \secref{ssec:findings_details}.  The final part, \secref{ssec:findings_corpora}, aims to quantify the adaptation performance across languages, speaking styles and datasets.

\subsection{Protocol and Literature} \label{ssec:init}

The meta-analysis is based on \nstudies~peer-reviewed studies selected such that they cover a wide range of systems, architectures, and adaptation tasks. Each study was required to compare adaptation results versus a baseline, enabling the configurations of interest to be compared quantitatively. There was no fixed target for the total number of papers included, due to our aim to cover as many different methods as possible. Note that the meta-analysis spans several model architectures, languages, and domains; although most studies use word error rate (WER) as the evaluation metric, some studies used character error rate (CER) or phone error rate (PER). Since we are interested in the relative improvement brought by adaptation, we  report Relative Error Rate Reductions (RERR).

 The meta-analysis is based on the studies shown in Table~\ref{tab:ma_refs_partition}, with additional splits into level of operation and top-level system architecture. The positions were selected such that they cover most of the topics mentioned in the review. For an adaptation of end-to-end systems we included all peer-reviewed works we could find (their number is relatively limited). For the hybrid approach, the studies are shortlisted such that they enable the quantification of the gains for the categories outlined in the preceding theoretical sections. As a generic rule, when choosing papers for the analysis we first included works that introduced a specific adaptation method in the context of neural models, or that offered some additional experiments allowing the comparison of different areas of interest - such as the impact of objective functions, the complementarity of adaptation transforms or that show behavior under different operating regimes.  In the case of certain more commonly-used techniques, due to the laborious nature of the analysis, it was not always possible to include an exhaustive set of somewhat similar papers.  In this situation, the papers selected were those with higher citation counts.
 
The analysis spans \ndsets~datasets (more than 50 unique \{train, test\} pairings), 28 of which are public and 10 are proprietary. These cover different speaking styles, domains, acoustic conditions, applications and languages (though the study is strongly biased towards English resources). The public corpora used include the following:
AISHELL2~\cite{aishell2}, AMI~\cite{carletta2007unleashing}, APASCI~\cite{angelini1994speaker}, Aurora4~\cite{parihar2004aurora}, CASIA~\cite{casia}, ChildIt~\cite{giuliani2003investigating}, Chime4~\cite{vincent2017analysis}, CSJ~\cite{maekawa2003corpus}, ETAPE~\cite{gravier2012etape}, HKUST~\cite{liu2006hkust}, MGB\cite{bell2015mgb}, RASC863~\cite{rasc}, SWBD~\cite{Godfrey92:switchboard}, TED~\cite{cettolo2012wit3}, TED-LIUM~\cite{rousseau2012ted}, TED-LIUM2~\cite{rousseau2014ted}, TIMIT~\cite{garofolo1993timit}, WSJ~\cite{paul1992design}, PF-STAR~\cite{batliner2005pf_star}, Librispeech~\cite{panayotov2015librispeech}, Intel Accented Mandarin Speech Recognition Corpus~\cite{chen2015improving}, UTCRSS-4EnglishAccent~\cite{ghorbani2018leveraging}. To save space we do not provide detailed corpora statistics in this paper, but make them available via a corresponding repository\footnote{\repourl.} alongside raw data and scripts used to perform the analysis.

Overall, the meta-analysis is based on ASR systems trained on datasets with a combined duration of over 30,000 hours, while the baseline acoustic models were estimated from as little as 5 hours to around 10,000 hours of speech. Adaptation data varies from a few seconds per speaker to over 25,000 hours of acoustic material used for domain adaptation.

\subsection{Overall findings} \label{ssec:findings_generic}

\begin{figure}
    \centering
    \begin{tikzpicture}
    \begin{axis}[
        height=3.0cm, width=9cm,
        xmin=-10, xmax=93.0, 
        ytick={1},
        yticklabels={Total},
        grid=major, grid style={dashed,gray!30},
        y=1.7cm, ymax=1.7, ymin=0.4,
        xtick={0, 10, 20, 40, 60, 80},
        xticklabels={},
        title = {All Datapoints}
    ]
	\desboxplottotal{-4.55}{3.68}{7.00}{12.70}{74.74}{9.72}{356}{38}{47}
   \end{axis}
\end{tikzpicture}
           \begin{tikzpicture}
        \begin{axis}[
          height=2cm, width=9cm,
          xmin=-10, xmax=93.0, 
          ytick={1},
          yticklabels={Total},
          grid=major, grid style={dashed,gray!30},
          xtick={0, 10, 20, 40, 60, 80},
          xticklabels={},
          title={Single Adaptation Method}
        ]
	\desboxplot{-4.08}{3.57}{7.0}{11.28}{74.74}{9.50}{283}{35}{46}
        \end{axis}
     \end{tikzpicture}
           \begin{tikzpicture}
        \begin{axis}[
          height=2cm, width=9cm,
          xmin=-10, xmax=93.0, 
          ytick={1},
          yticklabels={Total},
          grid=major, grid style={dashed,gray!30},
          xlabel={Relative Error Rate Reduction [\%]},
          xtick={0, 10, 20, 40, 60, 80},
          xticklabels={0, 10, 20, 40, 60, 80},
          title = {Two (or more) Adaptation Methods}
        ]
	\desboxplot{-4.55}{4.55}{11.35}{14.21}{72.03}{10.58}{73}{12}{13}

        \end{axis}
     \end{tikzpicture}
    \caption{Aggregated summary of adaptation RERR from all studies (top), considering single method only (middle) and two or more methods stacked (bottom). The top graph is annotated to explain the information presented in each of the boxplot graphs in this section.}
    \label{fig:ma_total}
\end{figure}
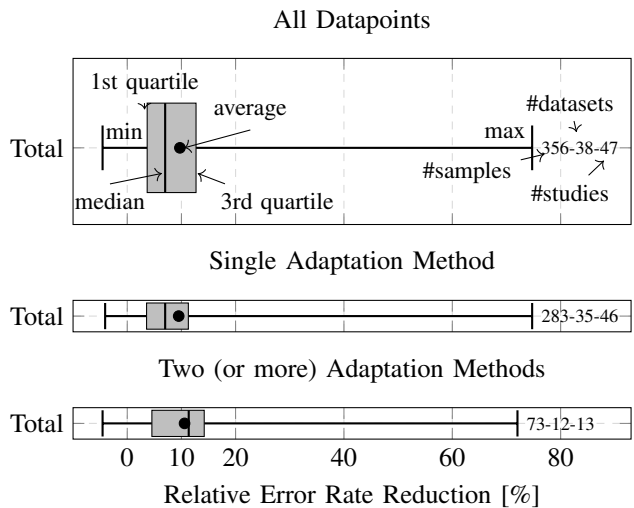

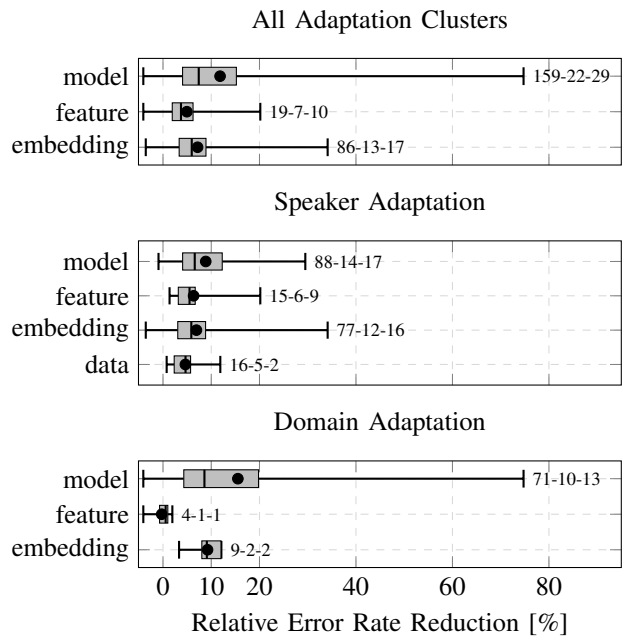
\begin{figure}
    \centering
           \begin{tikzpicture}
        \begin{axis}[
          height=3.0cm, width=8cm,
          xmin=-5.08, xmax=95.0, 
          ytick={1,2,3},
          yticklabels={embedding,feature,model},
          grid=major, grid style={dashed,gray!30},
          xticklabels={},
          xtick={0, 10, 20, 40, 60, 80},
          title={All Adaptation Clusters}
        ]
	\desboxplot{-3.57}{3.35}{5.96}{8.91}{34.13}{7.17}{86}{13}{17}
	\desboxplot{-4.08}{1.92}{3.73}{6.23}{20.17}{4.95}{19}{7}{10}
	\desboxplot{-4.08}{4.06}{7.41}{15.21}{74.74}{11.83}{159}{22}{29}
        \end{axis}
     \end{tikzpicture}
           \begin{tikzpicture}
        \begin{axis}[
          height=3.5cm, width=8cm,
          xmin=-5.08, xmax=95.0, 
          ytick={1,2,3,4},
          yticklabels={data,embedding,feature,model},
          grid=major, grid style={dashed,gray!30},
          xticklabels={},
          xtick={0, 10, 20, 40, 60, 80},
          title={Speaker Adaptation}
        ]
	\desboxplot{0.76}{2.32}{4.67}{5.72}{11.88}{4.64}{16}{5}{2}
	\desboxplot{-3.57}{3.06}{5.87}{8.83}{34.13}{6.93}{77}{12}{16}
	\desboxplot{1.36}{3.17}{5.50}{6.68}{20.17}{6.34}{15}{6}{9}
	\desboxplot{-0.93}{4.07}{6.58}{12.28}{29.50}{8.86}{88}{14}{17}
        \end{axis}
     \end{tikzpicture}
           \begin{tikzpicture}
        \begin{axis}[
          height=3.0cm, width=8cm,
          xmin=-5.08, xmax=95.0, 
          ytick={1,2,3},
          yticklabels={embedding,feature,model},
          grid=major, grid style={dashed,gray!30},
          xlabel={Relative Error Rate Reduction [\%]},
          title = {Domain Adaptation},
          xtick={0, 10, 20, 40, 60, 80},
          xticklabels={0, 10, 20, 40, 60, 80}
        ]
	\desboxplot{3.33}{8.05}{9.09}{12.02}{12.10}{9.23}{9}{2}{2}
	\desboxplot{-4.08}{-0.71}{0.55}{1.00}{1.94}{-0.26}{4}{1}{1}
	\desboxplot{-4.08}{4.31}{8.59}{19.80}{74.74}{15.52}{71}{10}{13}

        \end{axis}
     \end{tikzpicture}
    \caption{Comparison of feature, embedding, and model-level adaptation approaches. Speaker (middle) and domain (bottom) adaptations are based on \{utterance, speaker\} and \{accent, child, domain, disordered\} clusters, respectively. 
    }
    \label{fig:ma_levels}
\end{figure}

\figref{fig:ma_total} (Top) presents the average adaptation gains for all considered systems, adaptation methods, and adaptation classes. The overall RERR is 9.72\%%
\footnote{We do not report exact numbers in tabular form due to space limitations, but they are available in the GitHub repository}. 
Since grouping data across attributes of interest may result in unbalanced (or very sparse) sample sizes, we also report additional statistics such as the number of samples, datasets and studies the given statistic is based on. As can be seen in the right part of  \figref{fig:ma_total} (Top), the results in this review were derived from \npoints~samples produced using \ndsets~datasets reported in \nstudies~studies. A single sample is defined as a 1:1 system comparison for which one can unambiguously state the RERR. Likewise, a dataset refers to a particular training corpus configuration. Note that there may be some data-level overlap between different corpora originating from the same source (\eg TED talks) and we make a distinction for the acoustic condition (\eg AMI close-talking and distant channels are counted as two different datasets when they are used to estimate separate acoustic models). A study refers to a single peer-reviewed publication.

Depending on which property we want to measure the analysis set can be split into smaller subsets, as the ones shown in the lower part of \figref{fig:ma_total}. The majority of analyses in this review are reported for models adapted using a single method 
with some additional groupings used to better capture further details such as complementarity of adaptation methods or their performance in different operating regimes.  

As mentioned in \secref{sec:adaptNN}, adaptation methods were historically categorized based on the level they operated at in the speech processing pipeline. \figref{fig:ma_levels} (top) quantifies the ASR performance along this attribute, showing that model-based adaptation obtains the best average improvements of 11.8\%, followed by embedding and feature levels at 7.2\% and 5.0\% RERR, respectively. This is not surprising, as model level adaptation allows large amounts of adaptation data to be leveraged by allowing the update of large portions of the model (including re-training the whole model). In more data-constrained regimes, such as utterance or speaker-level adaptation, where only a limited amount of adaptation data is typically available, differences are less pronounced and model-based speaker adaptation obtains 8.9\% RERR while adapting to domains gives 15.5\% RERR (\cf middle and bottom plots in Fig~\ref{fig:ma_levels}). Embedding approaches stay at a similar level for speaker adaptation,  improving to 9.2\% RERR for domain adaptation (although based on only two studies). Feature-space domain adaptation was used in only one study, which reported a small deterioration of -0.3\% RERR. 

\figref{fig:ma_levels} (middle) additionally shows results for speaker-oriented data augmentation as described in \secref{sec:data_augmentation}. These were found to increase accuracy by 4.6\% RERR on average, or by 3.3\%, 3.6\% and 8.2\% RERR for VTL perturbations (VTLP)~\cite{jaitly2013vocal,cui2015data}, stochastic feature mapping (SFM)~\cite{cui2015data2, cui2015data} and when using synthetically generated TTS utterances~\cite{huang2020using}, respectively. Note that the TTS method was used to augment the adaptation set to better estimate additional adaptation transforms while VTLP and SFM were used to directly expand the training data, and were found particularly effective for low-resource training conditions. Data augmentations are beyond the scope of this meta-analysis and will not be further investigated in this review.

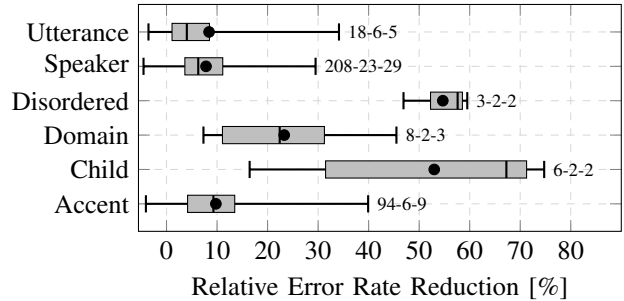
\begin{figure}
    \centering
           \begin{tikzpicture}
        \begin{axis}[
          height=4.6cm, width=8cm,
          xmin=-5.55, xmax=90., 
          ytick={1,2,3,4,5,6},
          yticklabels={Accent,Child,Domain,Disordered,Speaker,Utterance},
          grid=major, grid style={dashed,gray!30},
          xlabel={Relative Error Rate Reduction [\%]},
          xtick={0, 10, 20, 30, 40, 50, 60, 70, 80},
          xticklabels={0, 10, 20, 30, 40, 50, 60, 70, 80}
        ]
	\desboxplot{-4.08}{4.17}{9.28}{13.50}{39.90}{9.77}{94}{6}{9}
	\desboxplot{16.45}{31.48}{67.26}{71.26}{74.74}{52.98}{6}{2}{2}
	\desboxplot{7.31}{11.07}{22.40}{31.25}{45.50}{23.29}{8}{2}{3}
	\desboxplot{46.90}{52.25}{57.61}{58.56}{59.52}{54.68}{3}{2}{2}
	\desboxplot{-4.55}{3.63}{6.28}{11.13}{29.50}{7.82}{208}{23}{29}
	\desboxplot{-3.57}{1.09}{4.02}{8.50}{34.13}{8.42}{18}{6}{5}

        \end{axis}
     \end{tikzpicture}
    \caption{Adaptation results for different adaptation clusters.}
    \label{fig:ma_clusters}
\end{figure}

The results for different adaptation clusters, introduced in \secref{sec:adaptation_targets},  are shown in \figref{fig:ma_clusters}. Models benefit more when adapting to accent, from adult to child speech, to the domain, and to disordered speech conditions (such as arising from speech motor disorders), as opposed to speaker or utterance adaptation. This is expected, since domain adaptation usually has more adaptation data, and the acoustic mismatch introduced by unseen domains is greater than the mismatch caused by unseen speakers -- unless these are substantially mismatched to the training data as is often the case for child or disordered speech recognition. But in the latter case the adaptation is typically not carried out at the speaker level, but at the domain level (\ie tailoring the acoustic model to better handle dysarthric speech, not a single dysarthric speaker).

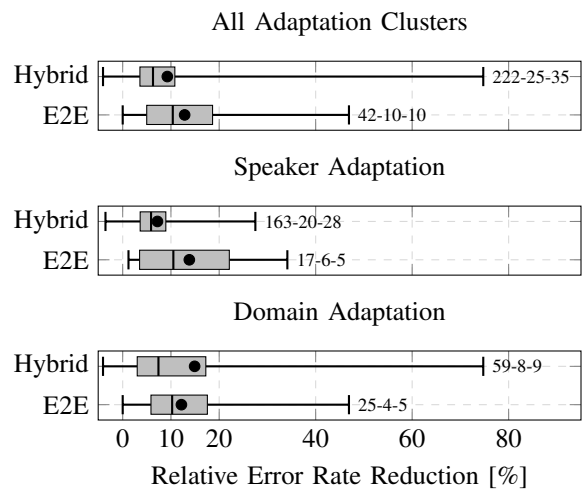
\begin{figure}
    \centering
           \begin{tikzpicture}
        \begin{axis}[
          height=2.5cm, width=8cm,
          xmin=-5.08, xmax=95.0, 
          ytick={1,2},
          yticklabels={E2E,Hybrid},
          grid=major, grid style={dashed,gray!30},
          xtick={0,10,20, 40, 60, 80},
          xticklabels={},
          title = {All Adaptation Clusters}
        ]
	\desboxplot{0.00}{4.98}{10.38}{18.63}{46.90}{12.84}{42}{10}{10}
	\desboxplot{-4.08}{3.57}{6.28}{10.81}{74.74}{9.25}{222}{25}{35}

        \end{axis}
     \end{tikzpicture}
           \begin{tikzpicture}
        \begin{axis}[
          height=2.5cm, width=8cm,
          xmin=-5.08, xmax=95.0, 
          ytick={1,2},
          yticklabels={E2E,Hybrid},
          grid=major, grid style={dashed,gray!30},
          xticklabels={},
          xtick={0,10,20, 40, 60, 80},
          title = {Speaker Adaptation}
        ]
	\desboxplot{1.20}{3.50}{10.50}{22.10}{34.13}{13.81}{17}{6}{5}
	\desboxplot{-3.57}{3.62}{5.88}{8.94}{27.50}{7.20}{163}{20}{28}

        \end{axis}
     \end{tikzpicture}
           \begin{tikzpicture}
        \begin{axis}[
          height=2.5cm, width=8cm,
          xmin=-5.08, xmax=95.0, 
          ytick={1,2},
          yticklabels={E2E,Hybrid},
          grid=major, grid style={dashed,gray!30},
          xlabel={Relative Error Rate Reduction [\%]},
          xtick={0,10,20, 40, 60, 80},
          xticklabels={0,10,20, 40, 60, 80},
          title = {Domain Adaptation}
        ]
	\desboxplot{0.00}{5.86}{10.25}{17.55}{46.90}{12.17}{25}{4}{5}
	\desboxplot{-4.08}{3.01}{7.41}{17.23}{74.74}{14.91}{59}{8}{9}

        \end{axis}
     \end{tikzpicture}
    \caption{Comparison of adaptation results for hybrid and E2E systems.}
    \label{fig:ma_hyb_e2e}
\end{figure}

\figref{fig:ma_hyb_e2e} aggregates the adaptation along the two main neural network-based ASR approaches - hybrid and E2E. It is interesting to observe that E2E systems gain more from adaptation (12.8\% RERR) than hybrid systems (9.2\% RERR) in both the overall and speaker-based regimes. This is somewhat expected, as hybrid systems benefit from strong inductive biases -- such as access to pronunciation dictionaries and hand engineered modeling constraints -- whereas E2E models must learn these from data. Given limited amounts of training data one may expect that E2E may struggle to learn these as well as hybrid models, as such adaptation may bring greater gains. This reverses for domain adaptation, with E2E and hybrid improving by 12.2 and 14.9\% RERR, respectively. Note that for domain adaptation, the hybrid approach was studied more often for child and disordered speech applications, which makes adaptation gains bigger (see also~\figref{fig:ma_clusters}).
Table~\ref{tab:ma_hours} further reports average amounts of training data used to estimate hybrid and E2E models. It is interesting to notice that E2E systems on average leverage twice as much acoustic material when compared to hybrid setups but still seem to substantially benefit from adaptation.
These results suggest that adaptation for E2E is a promising direction for future investigations, that remains under-investigated as of now based on the relatively few works published to date. 
 
\begin{table}
    \caption{Amounts of data used to estimate Hybrid and E2E models for \texttt{Speaker} and \texttt{Domain} adaptation clusters.}
    \label{tab:ma_hours}
    \centering
    \begin{tabular}{c|c|c}
        \toprule
        \multirow{2}{*}{Cluster} & \multicolumn{2}{c}{Avg. Training Data [hours]} \\ \cmidrule(lr){2-3} 
               & Hybrid & E2E  \\
         \midrule
         All & 862 & 1747 \\
        Speaker & 874 & 2640  \\
        Domain & 824 & 1033 \\
        \bottomrule
    \end{tabular}
\end{table}

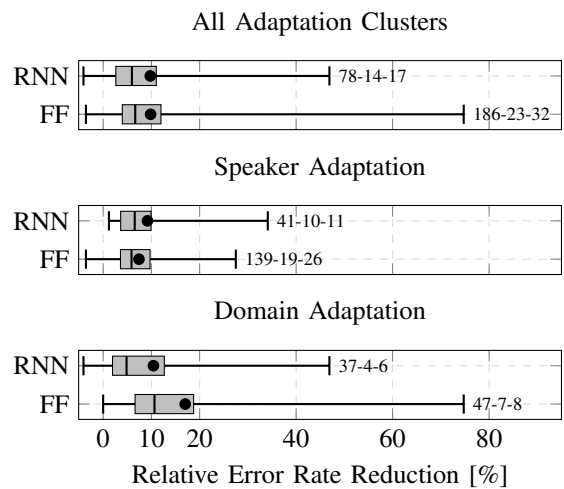
\begin{figure}
    \centering
           \begin{tikzpicture}
        \begin{axis}[
          height=2.5cm, width=8cm,
          xmin=-5.08, xmax=95.0, 
          ytick={1,2},
          yticklabels={FF,RNN},
          grid=major, grid style={dashed,gray!30},
          xtick = {0, 10, 20, 40, 60, 80},
          xticklabels={},
         title = {All Adaptation Clusters}
        ]
	\desboxplot{-3.57}{3.97}{6.65}{11.98}{74.74}{9.84}{186}{23}{32}
	\desboxplot{-4.08}{2.63}{5.96}{11.02}{46.90}{9.77}{78}{14}{17}

        \end{axis}
     \end{tikzpicture}
           \begin{tikzpicture}
        \begin{axis}[
          height=2.5cm, width=8cm,
          xmin=-5.08, xmax=95.0, 
          ytick={1,2},
          yticklabels={FF,RNN},
          grid=major, grid style={dashed,gray!30},
          xticklabels = {},
          xtick = {0, 10, 20, 40, 60, 80},
          title = {Speaker Adaptation}
        ]
	\desboxplot{-3.57}{3.59}{5.88}{9.72}{27.50}{7.42}{139}{19}{26}
	\desboxplot{1.20}{3.63}{6.54}{9.93}{34.13}{9.19}{41}{10}{11}

        \end{axis}
     \end{tikzpicture}
           \begin{tikzpicture}
        \begin{axis}[
          height=2.5cm, width=8cm,
          xmin=-5.08, xmax=95.0, 
          ytick={1,2},
          yticklabels={FF,RNN},
          grid=major, grid style={dashed,gray!30},
          xlabel={Relative Error Rate Reduction [\%]},
          title = {Domain Adaptation},
          xtick = {0, 10, 20, 40, 60, 80},
          xticklabels = {0, 10, 20, 40, 60, 80}
        ]
	\desboxplot{0.00}{6.62}{10.64}{18.76}{74.74}{17.00}{47}{7}{8}
	\desboxplot{-4.08}{1.94}{4.85}{12.68}{46.90}{10.42}{37}{4}{6}

        \end{axis}
     \end{tikzpicture}
    \caption{Comparison of adaptation results for FF and RNN architectures.}
    \label{fig:ma_am_type}
\end{figure}

Next we compare feed-forward (FF) and recurrent neural network (RNN) architectures in both hybrid and E2E models. Hybrid models can leverage either FF or RNN architectures while most E2E systems use some form of RNN. (Note, transformer-based E2E models~\cite{vaswani2017attention} are built from FF (CNN) modules, however, due to their relative novelty in ASR there is only one accent adaptation study included in our meta-analysis~\cite{winata2020learning}). \figref{fig:ma_am_type} reports similar adaptation gains of 9.8\% RERR for both FF and RNN architectures. RNNs seem to benefit more when adapting to speakers (9.2\% vs 7.4\% RERR for RNN and FF, respectively), and less when adapting to domain (10.4\% vs 17.0\% RERR for RNN and FF, respectively). When controlling for the system paradigm (E2E vs. Hybrid), RNNs mostly benefit through adapting E2E models (\cf~Fig~\ref{fig:ma_am_type_sys} 6.6\% vs 15.7\% RERR for Hybrid (RNN) and E2E (RNN), respectively). We observed a similar trend for speaker and domain clusters separately (figure not shown).

\begin{figure}
    \centering
           \begin{tikzpicture}
        \begin{axis}[
          height=2.5cm, width=7.5cm,
          xmin=-5.08, xmax=95.0, 
          ytick={1,2},
          yticklabels={Supervised,Unsupervised},
          grid=major, grid style={dashed,gray!30},
          xlabel={Relative Error Rate Reduction [\%]},
          xtick={0,10,20,40,60,80},
          xticklabels={0,10,20,40,60,80}
        ]
	\desboxplot{-4.08}{4.02}{8.24}{15.68}{74.74}{12.78}{99}{16}{20}
	\desboxplot{-3.57}{3.55}{5.98}{9.54}{45.50}{8.04}{165}{21}{32}

        \end{axis}
     \end{tikzpicture}
    \caption{Comparison of adaptation results for supervision modes.}
    \label{fig:ma_unsup_sup}
\end{figure}
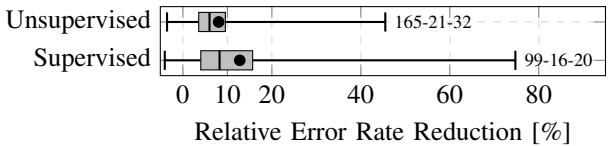

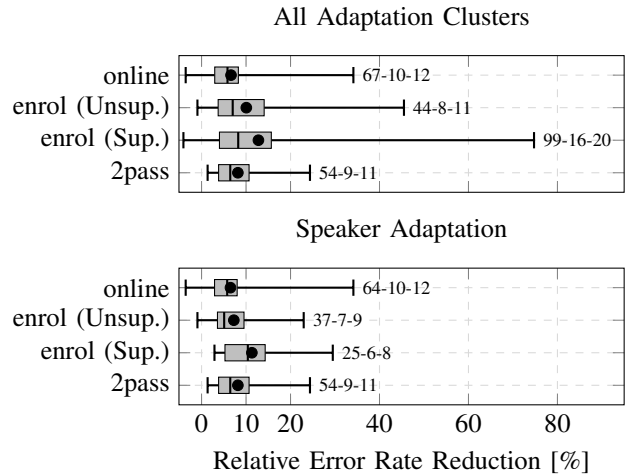
\begin{figure}
    \centering
           \begin{tikzpicture}
        \begin{axis}[
          height=3.4cm, width=7.5cm,
          xmin=-5.08, xmax=95.0, 
          ytick={1,2,3,4},
          yticklabels={2pass,enrol (Sup.), enrol (Unsup.), online},
          grid=major, grid style={dashed,gray!30},
          xtick={0,10,20, 40, 60, 80},
          xticklabels={},
          title = {All Adaptation Clusters}
        ]
	\desboxplot{1.36}{3.83}{6.45}{10.68}{24.37}{8.16}{54}{9}{11}
	\desboxplot{-4.08}{4.02}{8.24}{15.68}{74.74}{12.78}{99}{16}{20}
	\desboxplot{-0.93}{3.73}{7.00}{14.07}{45.50}{10.06}{44}{8}{11}
	\desboxplot{-3.57}{2.96}{5.83}{8.27}{34.13}{6.62}{67}{10}{12}

        \end{axis}
     \end{tikzpicture}
           \begin{tikzpicture}
        \begin{axis}[
          height=3.4cm, width=7.5cm,
          xmin=-5.08, xmax=95.0, 
          ytick={1,2,3,4},
          yticklabels={2pass,enrol (Sup.), enrol (Unsup.), online},
          grid=major, grid style={dashed,gray!30},
          xlabel={Relative Error Rate Reduction [\%]},
          xtick={0,10,20, 40, 60, 80},
          xticklabels={0,10,20, 40, 60, 80},
          title = {Speaker Adaptation}
        ]
	\desboxplot{1.36}{3.83}{6.45}{10.68}{24.37}{8.16}{54}{9}{11}
	\desboxplot{2.90}{5.26}{10.40}{14.30}{29.50}{11.31}{25}{6}{8}
	\desboxplot{-0.93}{3.57}{5.08}{9.54}{22.96}{7.25}{37}{7}{9}
	\desboxplot{-3.57}{2.92}{5.76}{8.00}{34.13}{6.50}{64}{10}{12}

        \end{axis}
     \end{tikzpicture}
    \caption{Comparison of adaptation results for different adaptation targets: online adaptation, supervised and unsupervised enrollment, and two-pass decoding.}
    \label{fig:ma_methods_sup}
\end{figure}

\figref{fig:ma_unsup_sup} compares the RERR for unsupervised and supervised modes of adaptation. Overall, deriving the adaptation transform with manually annotated targets results in an average 12.8\% RERR, whereas unsupervised methods result in 8\% RERR. \figref{fig:ma_unsup_sup} shows results specifically for semi-supervised adaptation, which are captured by the 2pass and enrol (Unsup.) conditions.
\figref{fig:ma_methods_sup} also shows further analysis on the modes of deriving adaptation statistics (\secref{sec:adaptation_targets}). Both online and two-pass adaptation are unsupervised, while the enrollment mode may be either supervised or unsupervised. The supervised approach offers most accurate adaptation, as expected.  Unsupervised enrollment  outperforms the other two unsupervised methods mainly due to the T/S domain adaptation study~\cite{meng2019domain} (\secref{sec:domain_adaptation}) that leverages large amounts of data. When considering speaker adaptation only, the two-pass approach obtains 8.2\% RERR and is more effective than enrol (Unsup.) (7.3\% RERR) and online adaptation (6.5\% RERR).

\begin{figure}
    \centering
           \begin{tikzpicture}
        \begin{axis}[
          height=3.4cm, width=7.5cm,
          xmin=-5.08, xmax=95.0, 
          ytick={1,2,3,4},
          yticklabels={E2E (FF) ,E2E (RNN), Hybrid (FF), Hybrid (RNN)},
          grid=major, grid style={dashed,gray!30},
          xlabel={Relative Error Rate Reduction [\%]},
          xtick={0,10,20,40,60,80},
          xticklabels={0,10,20,40,60,80}
        ]
	\desboxplot{0.00}{3.52}{6.67}{10.45}{20.18}{7.60}{15}{1}{1}
	\desboxplot{1.20}{5.89}{12.68}{23.40}{46.90}{15.75}{27}{9}{9}
	\desboxplot{-3.57}{3.99}{6.64}{12.07}{74.74}{10.03}{171}{22}{31}
	\desboxplot{-4.08}{2.05}{4.01}{7.76}{45.50}{6.61}{51}{7}{8}

        \end{axis}
     \end{tikzpicture}
    \caption{Comparison of adaptation results for FF and RNN architectures split by hybrid and E2E systems.}
    \label{fig:ma_am_type_sys}
\end{figure}
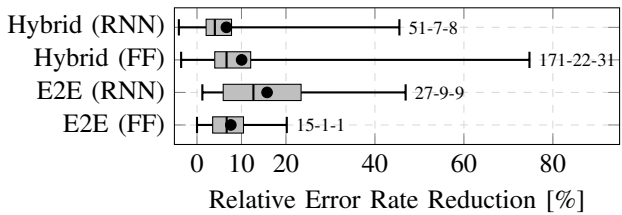

\begin{figure}
    \centering
           \begin{tikzpicture}
        \begin{axis}[
          height=5.2cm, width=8cm,
          xmin=-5.08, xmax=93.74, 
          ytick={1,2,3,4,5,6,7},
          yticklabels={0-1 min,1-10 mins,10-30 mins,30-60 mins,1-10 hours,10-100 hours,+100 hours},
          grid=major, grid style={dashed,gray!30},
          xlabel={Relative Error Rate Reduction [\%]},
          xtick={0,10,20,30,40,60,80}
        ]
	\desboxplot{-3.57}{3.30}{5.24}{7.97}{34.13}{6.87}{54}{11}{11}
	\desboxplot{-0.93}{3.73}{6.45}{9.51}{24.37}{7.55}{126}{17}{24}
	\desboxplot{2.10}{5.27}{9.32}{13.46}{19.50}{9.49}{12}{3}{3}
	\desboxplot{6.67}{9.65}{13.38}{16.59}{18.02}{12.86}{4}{2}{2}
	\desboxplot{-4.08}{2.07}{5.20}{13.43}{59.52}{10.66}{38}{7}{7}
	\desboxplot{2.00}{6.31}{12.02}{23.84}{74.74}{19.30}{27}{6}{7}
	\desboxplot{20.10}{25.57}{34.05}{40.27}{45.50}{33.13}{6}{1}{2}
        \end{axis}
     \end{tikzpicture}
    \caption{Comparison of adaptation results for different amount of adaptation data.}
    \label{fig:ma_adaptbins}
\end{figure}
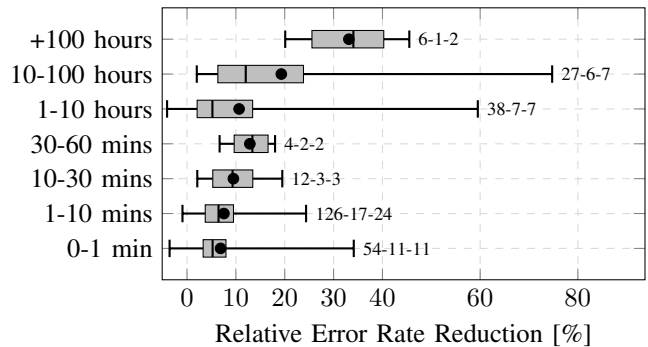

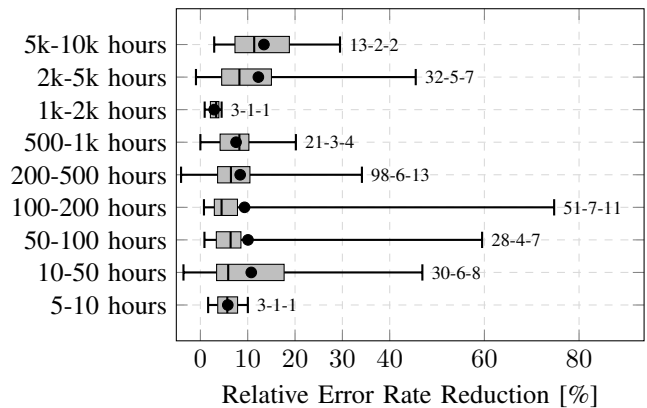
\begin{figure}
    \centering
           \begin{tikzpicture}
        \begin{axis}[
          height=6.0cm, width=7.8cm,
          xmin=-5.08, xmax=93.74, 
          ytick={1,2,3,4,5,6,7,8,9},
          yticklabels={5-10 hours,10-50 hours, 50-100 hours, 100-200 hours, 200-500 hours, 500-1k hours, 1k-2k hours,2k-5k hours, 5k-10k hours},
          grid=major, grid style={dashed,gray!30},
          xlabel={Relative Error Rate Reduction [\%]},
          xtick={0,10,20,30,40,60,80}
        ]
	\desboxplot{1.64}{3.67}{5.69}{7.86}{10.03}{5.79}{3}{1}{1}
	\desboxplot{-3.57}{3.43}{5.88}{17.69}{46.90}{10.75}{30}{6}{8}
	\desboxplot{0.86}{3.37}{6.39}{8.57}{59.52}{10.06}{28}{4}{7}
	\desboxplot{0.76}{2.95}{4.48}{7.84}{74.74}{9.33}{51}{7}{11}
	\desboxplot{-4.08}{3.64}{6.46}{10.47}{34.13}{8.41}{98}{6}{13}
	\desboxplot{0.00}{4.17}{8.24}{10.25}{20.18}{7.54}{21}{3}{4}
	\desboxplot{0.91}{2.12}{3.33}{3.91}{4.49}{2.91}{3}{1}{1}
	\desboxplot{-0.93}{4.50}{8.25}{15.01}{45.50}{12.25}{32}{5}{7}
	\desboxplot{2.92}{7.31}{11.37}{18.80}{29.50}{13.43}{13}{2}{2}
        \end{axis}
     \end{tikzpicture}
    \caption{Comparison of adaptation results for acoustic models estimated from different amounts of training data.}
    \label{fig:ma_trainbins}
\end{figure}

\begin{figure*}[!ht]
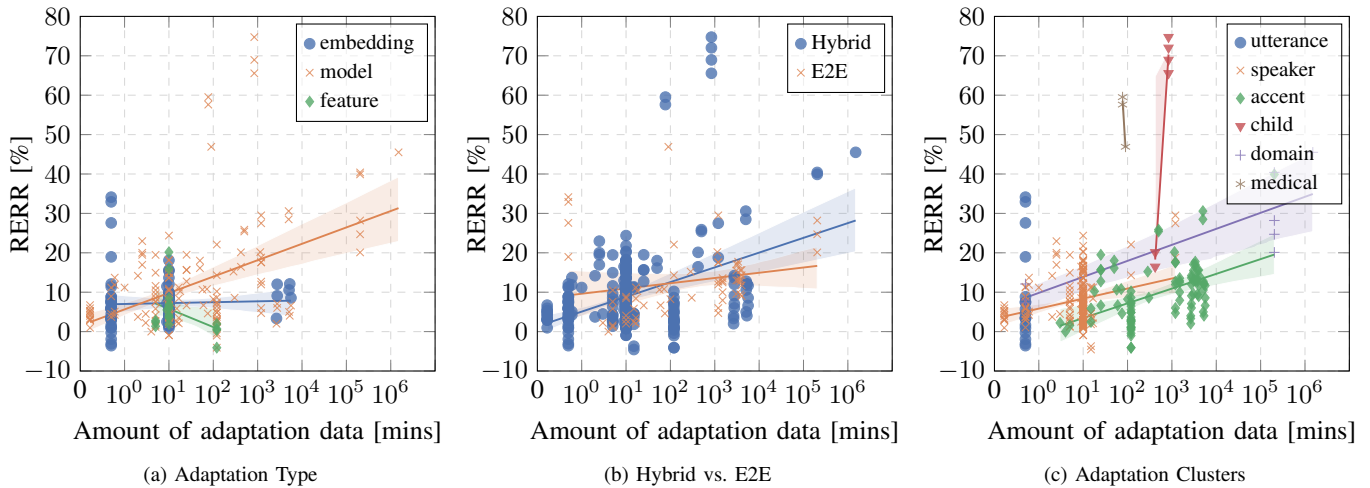

    \centering
    \subfloat[Adaptation Type]{
    \input{figs_meta/adaptminutes_level}
    }\hspace*{-1em}
    \subfloat[Hybrid vs. E2E]{
    \input{figs_meta/adaptminutes_system}
    }\hspace*{-1em}
    \subfloat[Adaptation Clusters]{
    \input{figs_meta/adaptminutes_clusters}
    }
    \caption{Regression analysis for the three major control variables.}
    \label{fig:ma_regressions}
\end{figure*}

Finally, we consider the overall trends for the considered systems and their operating regions. \figref{fig:ma_adaptbins} reports results obtained with different amounts of adaptation data. \figref{fig:ma_regressions} 
further shows regression trends when splitting by adaptation type, hybrid or E2E, and adaptation clusters. These are in line with the observations so far: i) more adaptation data brings (on average) larger improvements; ii) model-based adaptation is more powerful and gives better results than embedding or feature-based approaches; and iii) adaptation is particularly effective in scenarios with a large mismatch and where obtaining matched training data is difficult.

In \figref{fig:ma_trainbins} we further report adaptability of acoustic models estimated from different amounts of training material. Interestingly, models trained on small amounts of data (up to 50 hours) benefit from adaptation to a similar degree as models estimated from several thousands of hours. This is somewhat an unexpected result - if test sets are kept fixed, increasing the training material typically results in a less mismatched model, thus lowering gains from adaptation (and most experiments evaluating adaptation performance as a function of data are carried out in this way). However, when training from more data one should proportionally increase the complexity of the testing conditions. We hypothesize that this is what implicitly occurs across different datasets in the meta-analysis - someone who has access to a large training set may also sample a more diverse testing set. Note that the acoustic models in this work were trained from relatively limited amounts of data (up to 10k hours), and adaptation protocols between studies may not be exactly comparable. However, this does not change the conclusion that some form of adaptation is beneficial for most considered systems, regardless of how many hours of acoustic data was used to train it.

Since this meta-analysis combines results across many different studies with many reference systems,  the results should not necessarily be compared at the sample level, but rather in an aggregated form to outline dominant trends and typical data regimes that each category was tried in. Data amounts for some systems for the purpose of plotting were assumed approximately to be at a given level: \eg two-pass systems unless shown otherwise assumed 10 minutes per speaker, while embedding approaches 30 seconds.

\subsection {Detailed findings} \label{ssec:findings_details}

\begin{figure}
    \centering
           \begin{tikzpicture}
        \begin{axis}[
          height=4.6cm, width=7.0cm,
          xmin=-5.08, xmax=95.0, 
          ytick={1,2,3,4,5,6},
          yticklabels={Finetune, Activations,Linear Transform, NNTransformEmb, NNEmb, GMMEmb},
          grid=major, grid style={dashed,gray!30},
          xlabel={Relative Error Rate Reduction [\%]},
          xtick = {0, 10, 20, 40, 60, 80},
          xticklabels={0, 10, 20, 40, 60, 80}
        ]
        
	\desboxplot{-1.02}{4.84}{10.20}{18.02}{74.74}{13.87}{97}{17}{22}
	\desboxplot{0.97}{3.96}{6.36}{10.41}{68.96}{8.98}{38}{8}{10}
	\desboxplot{-4.08}{1.90}{4.00}{8.05}{57.61}{6.72}{43}{11}{14}
	\desboxplot{-0.27}{4.58}{7.78}{9.36}{34.13}{9.20}{18}{7}{5}
	\desboxplot{-3.08}{2.71}{4.80}{7.29}{12.08}{5.00}{32}{7}{8}
	\desboxplot{-3.57}{3.49}{7.13}{10.93}{32.93}{8.08}{36}{11}{14}

        \end{axis}
     \end{tikzpicture}
    \caption{Comparison of results for different adaptation approaches.\vspace{-0.2cm}}
    \label{fig:ma_families}
\end{figure}
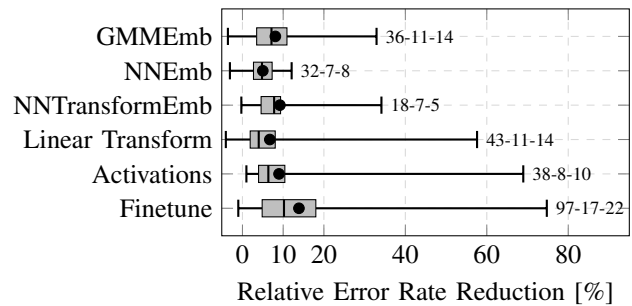

\begin{figure*}[!ht]
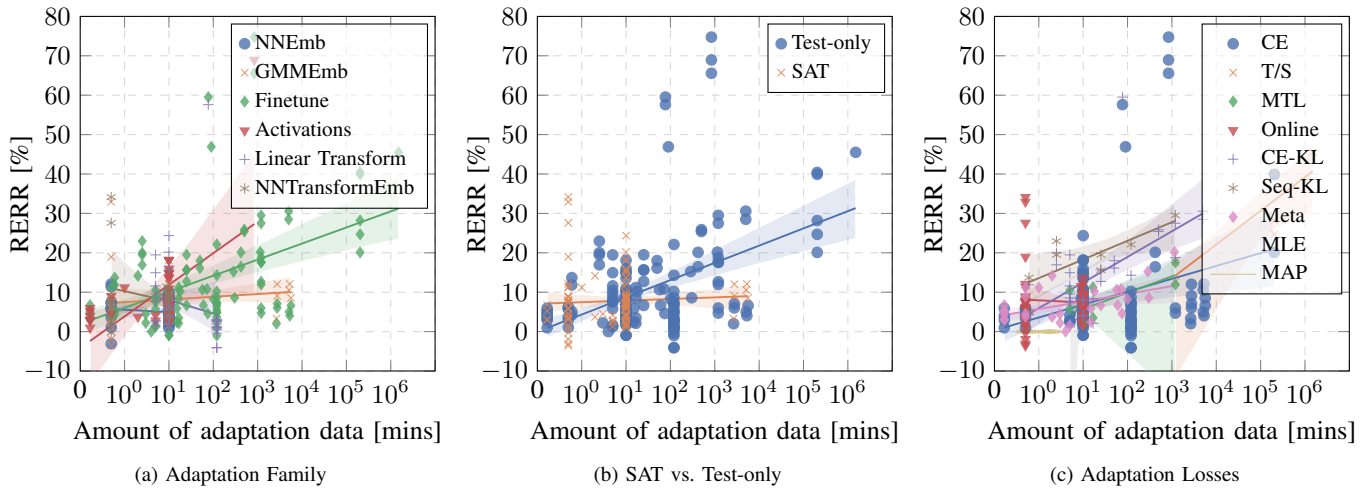

    \centering
    \subfloat[Adaptation Family]{
        \input{figs_meta/adaptminutes_family}
    }\hspace*{-1em}
    \subfloat[SAT vs. Test-only]{
        \input{figs_meta/adaptminutes_sat}
    }\hspace*{-1em}
    \subfloat[Adaptation Losses]{
        \input{figs_meta/adaptminutes_adapt_loss}
    }
    \caption{Regression analysis for adaptation families, speaker-adaptive training and adaptation losses.}
    \label{fig:ma_regressions_details}
\end{figure*}

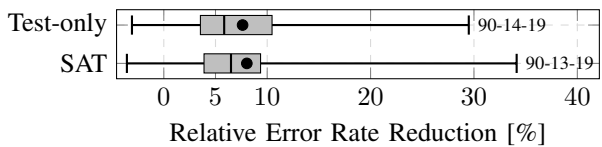
\begin{figure}
    \centering
           \begin{tikzpicture}
        \begin{axis}[
          height=2.5cm, width=8cm,
          xmin=-4.57, xmax=42.13, 
          ytick={1,2},
          yticklabels={SAT,Test-only},
          grid=major, grid style={dashed,gray!30},
          xlabel={Relative Error Rate Reduction [\%]},
          xtick={0, 5, 10, 20, 30, 40}
        ]
	\desboxplot{-3.57}{3.89}{6.50}{9.36}{34.13}{8.03}{90}{13}{19}
	\desboxplot{-3.08}{3.54}{5.84}{10.47}{29.50}{7.62}{90}{14}{19}

        \end{axis}
     \end{tikzpicture}
    \caption{Comparison of adaptation results for SAT vs Test-only modes.\vspace{-0.3cm}}
    \label{fig:ma_sat_vs_test_spk}
\end{figure}

In this subsection we investigate the effect of the specific approach to adaptation, beyond the broad categories discussed above. \figref{fig:ma_families} reorganizes the earlier split into feature, embedding, and model-level adaptation (\figref{fig:ma_levels}) into embedding (\cf \secref{sec:embeddings}) and model-based transformations (\cf \secref{sec:structured_xforms}). 

For the embeddings, we introduce three sub-categories referred to as GMMEmb, NNEmb and NNTransformEmb. 
GMMEmb comprises GMM-related embedding extractors primarily based on i-vectors~\cite{dehak2011front, saon2013speaker, senior2014improving} but also include adaptation results for other GMM-derived (GMMD) features~\cite{tomashenko2015gmm}. 
NNEmb are neural network-based embedding extractors that estimate speaker/utterance statistics from speaker-independent acoustic features. Examples of NNEmb approaches include $\star$-vector techniques, such as d-vectors~\cite{variani2014deep} and x-vectors~\cite{snyder2018x}, discussed in \secref{sec:embeddings}, sentence-level embeddings~\cite{delcroix2018auxiliary, vesely2016sequence}, and other bottleneck approaches~\cite{tan2016speaker, rownicka2019embeddings}.
NNTransformEmb are transformed embeddings which typically rely on i-vectors as input instead of acoustic features. These have been proposed to help alleviate issues related to inconsistent DNN adaptation performance when using raw i-vectors~\cite{senior2014improving, karanasou2014adaptation, karanasou2017vectors}. The NNTransformEmb group includes studies doing standard i-vector transformations with NNEmb~\cite{miao2014improvements, miao2015speaker, yi2016improving} but also more recent memory-based approaches in which an embedding is selected via attention from a fixed training stage embedding inventory~\cite{pan2020memory, sari2020unsupervised}. As shown in \figref{fig:ma_families} GMMEmb, NNEmb and NNTransformEmb obtain 8.1\%, 5.2\% and 9.2\% average RERR, respectively. 

The second group in \figref{fig:ma_families} comprises model-based approaches split into Linear Transform (LT), Activation, and Finetuning--based methods. LT methods introduce new speaker dependent affine transformations in the model, either in the form of new LIN/LHN/LON layers (\ie~\cite{neto1995speaker,li2010comparison,seide2011feature,kitza2018comparison}) or transforms estimated using a GMM system such as fMLLR~\cite{mohamed2011deep, seide2011feature, saon2013speaker, swietojanski2014learning}.  
Finetune refers to approaches which assume that the adaptation is carried out by altering a subset of the existing model parameters. This is often done in a similar manner to an LT approach by adapting an input, output and/or one or more hidden layers that are already present in the model~\cite{liao2013speaker,huang2014multi,chen2015improving,fainberg2019acoustic}. 
Finally, activation methods perform adaptation by introducing speaker-dependent parameters in the activation functions of the neural network~\cite{siniscalchi2013hermitian, abdel2013rapid, swietojanski2014learning, swietojanski2015differentiable, zhang2015parameterised}. Note that, as outlined in \secref{sec:structured_xforms}, some of activation-based methods can be expressed as constrained LT methods. The results obtained by LT, Activation and Finetune--based methods score 6.7\%, 9.0\% and 13.9\% average RERR, respectively.
\figref{fig:ma_regressions_details} (a) shows the regression trends for amounts of adaptation data for each of the six considered categories.

The use of embeddings implies that the acoustic model is trained in a speaker adaptive manner, whereas the majority of model-based techniques are carried out in a test-only manner -- meaning that  speaker-level information is not used during training --  though some methods offer SAT variants~\cite{swietojanski2016sat, samarakoon2015learning}. \figref{fig:ma_sat_vs_test_spk} shows that SAT trained systems offer a small advantage (8\%  vs. 7.6\% RERR) when adapted with limited amounts of data (up to around 10 minutes). When looking at the average performance across all data-points, however, test-only approaches obtain 10.8\% RERR, primarily because of greater adaptation gains for larger amounts of data. See also \figref{fig:ma_regressions_details} (b) for operating regions of SAT and non-SAT systems.

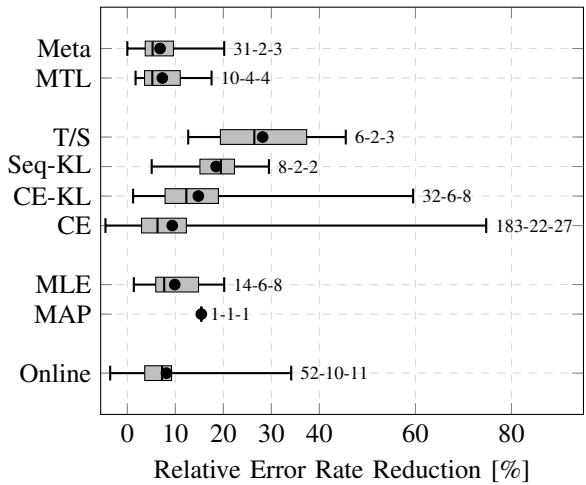
\begin{figure}
   \centering
           \begin{tikzpicture}
        \begin{axis}[
          height=7cm, width=8cm,
          xmin=-5.55, xmax=95.0, 
          ytick={1,3,4,6,7,8,9,11,12},
          yticklabels={Online,MAP,MLE,CE,CE-KL,Seq-KL,T/S,MTL,Meta},
          grid=major, grid style={dashed,gray!30},
          xlabel={Relative Error Rate Reduction [\%]},
          xtick={0, 10, 20, 30, 40, 60, 80},
          xticklabels={0, 10, 20, 30, 40, 60, 80},
        ]
	\desboxplotgrouped{-3.57}{3.63}{7.24}{9.22}{34.13}{8.14}{52-10-11}{1}
	\desboxplotgrouped{15.41}{15.41}{15.41}{15.41}{15.41}{15.41}{1-1-1}{3}
	\desboxplotgrouped{1.36}{5.92}{7.68}{14.83}{20.17}{9.88}{14-6-8}{4}
	\desboxplotgrouped{-4.55}{2.96}{6.29}{12.29}{74.74}{9.36}{183-22-27}{6}
	\desboxplotgrouped{1.20}{7.86}{12.29}{18.97}{59.52}{14.79}{32-6-8}{7}
	\desboxplotgrouped{5.08}{15.12}{19.49}{22.32}{29.50}{18.49}{8-2-2}{8}
	\desboxplotgrouped{12.68}{19.38}{26.45}{37.35}{45.50}{28.18}{6-2-3}{9}
	\desboxplotgrouped{1.72}{3.59}{5.21}{11.02}{17.55}{7.29}{10-4-4}{11}
	\desboxplotgrouped{0.00}{3.73}{5.26}{9.61}{20.18}{6.82}{31-2-3}{12}
        \end{axis}
     \end{tikzpicture}
    \caption{Comparison of results for different adaptation loss functions.}
    \label{fig:ma_adapt_loss}
\end{figure}

\figref{fig:ma_adapt_loss} quantifies gains for different adaptation objectives and regularization approaches -- results for  the online condition are given only for reference, as in this case adaptation information is obtained via an embedding extractor (which is usually not updated, although not always~\cite{yi2016improving}). The second group depicts approaches where the adaptation information is derived by adapting a GMM in model-space using an MLE or MAP criterion when extracting speaker-adapted auxiliary features for NN training~\cite{tomashenko2015gmm, tomashenko2016exploring} or by estimating fMLLR transforms with MLE under a GMM to obtain speaker adapted acoustic features~\cite{mohamed2011deep,seide2011feature,saon2013speaker}.

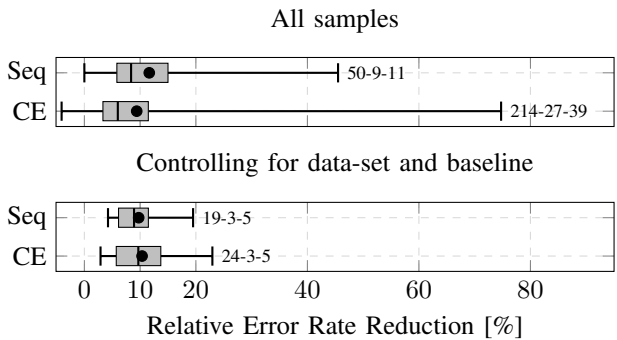
\begin{figure}
    \centering
           \begin{tikzpicture}
        \begin{axis}[
          height=2.5cm, width=9cm,
          xmin=-5.08, xmax=95., 
          ytick={1,2},
          yticklabels={CE,Seq},
          grid=major, grid style={dashed,gray!30},
          xtick={0, 10, 20, 40, 60, 80},
          xticklabels={},
          title={All samples}
        ]
	\desboxplot{-4.08}{3.33}{6.00}{11.47}{74.74}{9.39}{214}{27}{39}
	\desboxplot{0.00}{5.83}{8.38}{15.00}{45.50}{11.64}{50}{9}{11}

        \end{axis}
     \end{tikzpicture}
           \begin{tikzpicture}
        \begin{axis}[
          height=2.5cm, width=9cm,
          xmin=-5.08, xmax=95., 
          ytick={1,2},
          yticklabels={CE,Seq},
          grid=major, grid style={dashed,gray!30},
          xlabel={Relative Error Rate Reduction [\%]},
          xtick={0, 10, 20, 40, 60, 80},
          title={Controlling for data-set and baseline}
        ]
	\desboxplot{2.90}{5.72}{9.68}{13.71}{22.96}{10.35}{24}{3}{5}
	\desboxplot{4.24}{6.13}{8.91}{11.48}{19.50}{9.74}{19}{3}{5}

        \end{axis}
     \end{tikzpicture}
    \caption{Comparison of adaptation results for acoustic models trained with CE and Sequence-level objectives.}
    \label{fig:ma_am_loss}
\end{figure}

The third group comprises methods which aim to explicitly match the model's output distribution to the one found in the adaptation data. CE is a non-regularized frame-level cross-entropy baseline obtaining 8.7\% average RERR. This can be improved to 14.8\% average RERR by penalizing the adapted model's predictions such that they do not deviate too much from the speaker independent variant by KL regularization (CE-KL)~\cite{yu2013kl}. KL regularization can be applied to either CE or sequential objective functions~\cite{huang2015regularized}, although most models estimated in a sequential discriminative manner can successfully be adapted with a CE (or CE-KL) criterion~\cite{swietojanski2016learning, samarakoon2016factorized, woodland2015cambridge, huang2020acoustic} (see also \figref{fig:ma_am_loss}). Teacher-student (T/S)~\cite{li2014learning} is a special case (see \secref{sec:domain_adaptation}) where the adaptation is carried with the targets directly produced by a teacher model, rather than the ones obtained from first pass decodes (possibly KL-regularized with the SI model). T/S allows the use of large amounts of unsupervised data and in this analysis was found to offer an average 28.2\% RERR when adapting to domains~\cite{Li2018Developing, ghorbani2018advancing, meng2019domain}. 

The final group in \figref{fig:ma_adapt_loss} includes objectives that try to leverage auxiliary information at the objective function level.
Meta-learning~\cite{klejch2018learning,klejch2019speaker,winata2020learning} estimates the adaptation hyper-parameters jointly with the adaptation transform while multi-task learning~\cite{huang2015rapid, swietojanski2015structured, ghorbani2018leveraging, meng2019speaker} leverages  additional phonetic priors to circumvent the (potential) sparsity of senones when adapting with small amounts of data. Meta-learning and multi-task adaptation obtain 6.8\% and 7.6\% average RERR, respectively. See also \figref{fig:ma_regressions_details} (c).

\figref{fig:ma_am_loss} further summarizes the adaptability of acoustic models trained in a frame-based (CE) or a sequential (Seq) manner.  The results indicate that sequential models benefit more from adaptation when compared to frame-based systems (11.6\% vs. 9.8\% average RERR). However, when controlling for the same dataset and baseline (reference systems were expected to exist for both CE and Seq) the difference decreases to around 0.6\% RERR in favor of the frame-based systems.

\begin{figure}
    \centering
           \begin{tikzpicture}
        \begin{axis}[
          height=5.2cm, width=8cm,
          xmin=-5.55, xmax=90.0, 
          ytick={1,2,3,4,5,6,7},
          yticklabels={BLSTM,CNN,DNN,LSTM,ResNet,TDNN,Transformer},
          grid=major, grid style={dashed,gray!30},
          xlabel={Relative Error Rate Reduction [\%]},
          xtick={0, 10, 20, 30, 40, 60, 80},
          xticklabels={0, 10, 20, 30, 40, 60, 80},
        ]
	\desboxplot{-4.55}{2.63}{6.79}{11.90}{17.74}{7.11}{70}{5}{7}
	\desboxplot{1.64}{3.59}{6.27}{8.41}{18.10}{6.51}{15}{7}{4}
	\desboxplot{-0.93}{4.24}{7.84}{14.86}{59.52}{10.16}{157}{23}{27}
	\desboxplot{1.20}{4.01}{8.55}{24.70}{46.90}{15.13}{37}{10}{11}
	\desboxplot{8.83}{10.10}{11.37}{11.72}{12.08}{10.76}{3}{1}{1}
	\desboxplot{-3.57}{3.37}{5.54}{8.92}{74.74}{11.70}{40}{4}{5}
	\desboxplot{0.00}{3.52}{6.67}{10.45}{20.18}{7.60}{15}{1}{1}

        \end{axis}
     \end{tikzpicture}
    \caption{Comparison of adaptation results for different architectures.}
    \label{fig:ma_adapt_archs}
\end{figure}
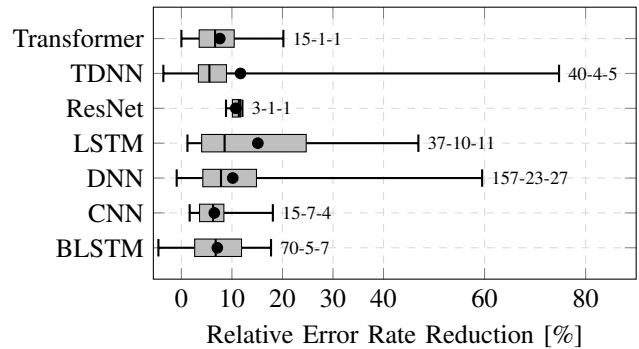

\figref{fig:ma_adapt_archs} compares the adaptation gains  obtained using various model architectures. LSTM benefits the most (15.4\% average RERR). The feed-forward TDNN, DNN, and ResNet architectures all improve by around 10.5\% RERR. Smaller gains were observed for Transformer, CNN and BLSTM, improving by 7.6, 6.5 and 4.9\% average RERR, respectively. This result is somewhat expected as the last three architectures either normalize some of the variability by design, or have access to a larger speech context during recognition. 

\begin{figure}
    \centering
           \begin{tikzpicture}
        \begin{axis}[
          height=10.0cm, width=6.2cm,
          xmin=-0.5, xmax=28, 
          ymin=0, ymax=13,
          ytick={3, 4, 5, 6, 7, 8, 9, 10, 11, 12, 13},
          yticklabels={BSV + ivectors~\cite{tan2016speaker}, DNNEmb + LHUC~\cite{miao2015speaker}, DNNEmb + fMLLR~\cite{miao2015speaker}, DiffPooling + fMLLR~\cite{swietojanski2016differentiable}, LHUC + fMLLR~\cite{swietojanski2016learning}, f-LHUC + LHUC~\cite{xie2019}, f-LHUC + ivectors~\cite{xie2019}, ivectors + LHUC~\cite{samarakoon2016factorized}, ivectors + fMLLR/VTLN~\cite{saon2013speaker}, pRELU + fMLLR~\cite{zhang2016dnn}},
          grid=major, grid style={dashed,gray!30},
          xlabel={Relative Error Rate Reduction [\%]},
          xtick={0, 5, 10, 15, 20, 25},
          xticklabels={0,5,10,15,25},
          xbar stacked,
          ticklabel style={font=\small}
        ]
	\addplot[fill=gray!50] coordinates
		{(2.16,3) (11.5,4) (5.5,5) (10.2,6) (11.8, 7) (5.4,8) (6.9,9) (5.88,10) (12.86,11) (2.61, 12)};
	\addplot[fill=white] coordinates
		{(4.63-2.16,3) (16.5-11.5,4) (13-5.5,5) (13.73-10.2,6) (16.8-11.8,7) (12.5-5.4,8) (6.1-6.9,9) (17.64-5.9, 10) (16.8-12.86, 11) (4.9-2.61, 12)};
        \end{axis}
        \begin{axis}[
             height=10.0cm, width=6.2cm,
             xmin=-0.5, xmax=28, 
             ymin=0, ymax=13,
             ytick={1,2},
             yticklabels={Method1 (Avg),Method1+Method2 (Avg)},
             xticklabels={},
             ticklabel style={font=\small}
          ]
          \desboxplotgrouped{1}{8.34}{12.44}{16.7}{21.8}{12}{22-8-6}{2}
          \desboxplotgrouped{2.16}{4.43}{6.4}{11.34}{18.1}{7.85}{22-8-6}{1}
        \end{axis}
     \end{tikzpicture}
    \caption{Complementarity of selected adaptation techniques.}
    \label{fig:ma_complementarity}
\end{figure}
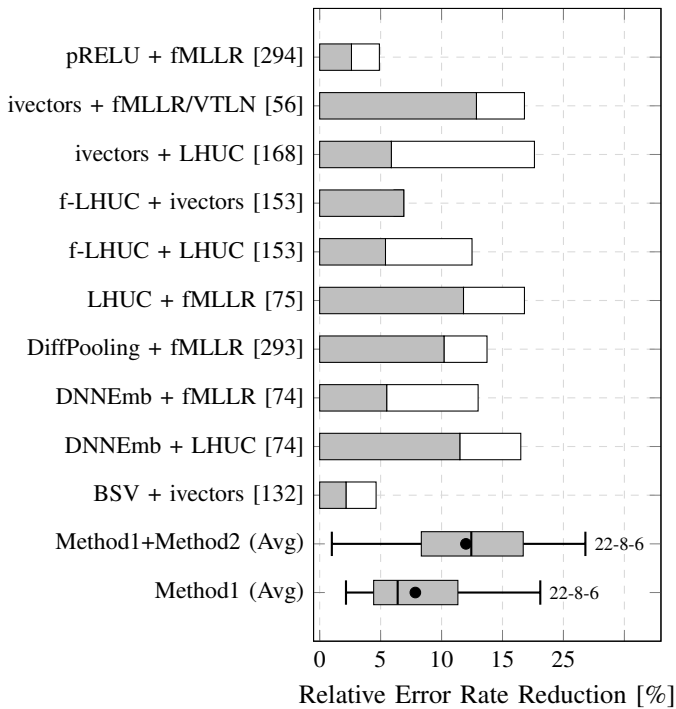

In \figref{fig:ma_complementarity} we study the complementarity of the different adaptation techniques. These results are based on 22 samples and 6 studies for which there were a complete set of baseline experiments allowing improvements to be quantified when adapting an SI model with Method1, and then measuring further gains when adding Method2. \figref{fig:ma_complementarity} shows that, on average, stacking adaptation techniques improved the adaptation performance by an additional 4\%, from 8\% to 12\% RERR.

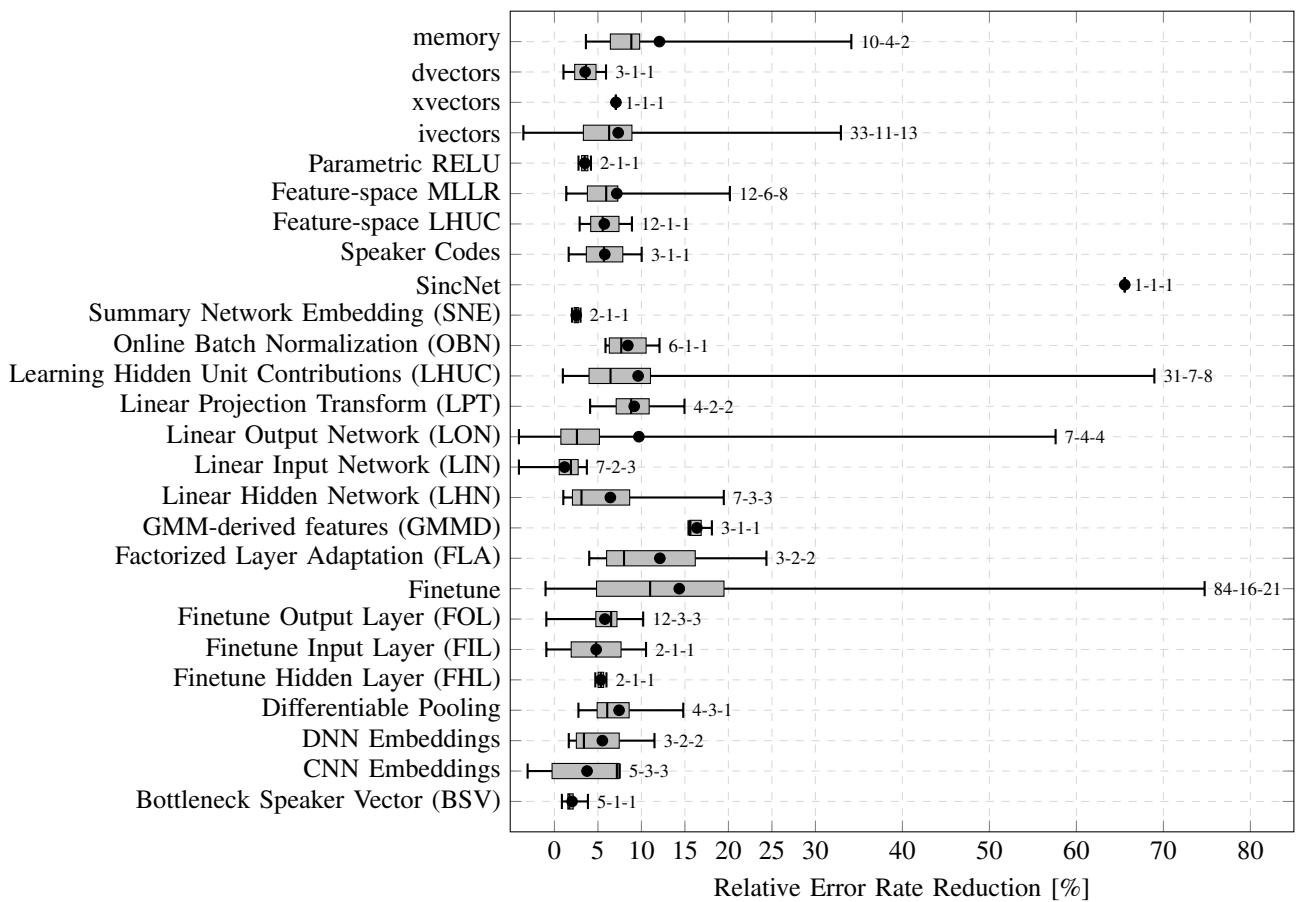
\begin{figure*}[htb]
    \centering
           \begin{tikzpicture}
        \begin{axis}[
          height=12.5cm, width=12cm,
          xmin=-5.08, xmax=85.00, 
          ymin=0, ymax=27.00, 
          ytick={1,2,3,4,5,6,7,8,9,10,11,12,13,14,15,16,17,18,19,20,21,22,23,24,25,26},
          yticklabels={Bottleneck Speaker Vector (BSV),
                       CNN Embeddings,
                       DNN Embeddings,
                       Differentiable Pooling,
                       Finetune Hidden Layer (FHL),
                       Finetune Input Layer (FIL),
                       Finetune Output Layer (FOL),
                       Finetune,
                       Factorized Layer Adaptation (FLA),
                       GMM-derived features (GMMD),
                       Linear Hidden Network (LHN),
                       Linear Input Network (LIN),
                       Linear Output Network (LON),
                       Linear Projection Transform (LPT),
                       Learning Hidden Unit Contributions (LHUC),
                       Online Batch Normalization (OBN),
                       Summary Network Embedding (SNE),
                       SincNet,
                       Speaker Codes,
                       Feature-space LHUC,
                       Feature-space MLLR,
                       Parametric RELU,
                       ivectors,
                       xvectors,
                       dvectors,
                       memory},
          grid=major, grid style={dashed,gray!30},
          xlabel={Relative Error Rate Reduction [\%]},
          xtick={0,5,10,15,20,25,30,40,50,60,70,80},
          xticklabels={0,5,10,15,20,25,30,40,50,60,70,80}
        ]
	\desboxplot{0.86}{1.54}{1.72}{2.16}{3.85}{2.03}{5}{1}{1}
	\desboxplot{-3.08}{-0.27}{7.18}{7.42}{7.51}{3.75}{5}{3}{3}
	\desboxplot{1.65}{2.52}{3.39}{7.45}{11.50}{5.51}{3}{2}{2}
	\desboxplot{2.76}{4.91}{6.07}{8.59}{14.81}{7.43}{4}{3}{1}
	\desboxplot{4.70}{5.02}{5.34}{5.66}{5.98}{5.34}{2}{1}{1}
	\desboxplot{-0.93}{1.93}{4.80}{7.66}{10.53}{4.80}{2}{1}{1}
	\desboxplot{-0.93}{4.76}{6.52}{7.19}{10.20}{5.80}{12}{3}{3}
	\desboxplot{-1.02}{4.84}{11.00}{19.49}{74.74}{14.35}{84}{16}{21}
	\desboxplot{4.00}{6.00}{8.00}{16.19}{24.37}{12.12}{3}{2}{2}
	\desboxplot{15.41}{15.52}{15.63}{16.87}{18.10}{16.38}{3}{1}{1}
	\desboxplot{1.02}{2.07}{3.11}{8.65}{19.47}{6.43}{7}{3}{3}
	\desboxplot{-4.08}{0.55}{1.90}{2.73}{3.73}{1.16}{7}{2}{3}
	\desboxplot{-4.08}{0.74}{2.59}{5.17}{57.61}{9.71}{7}{4}{4}
	\desboxplot{4.10}{7.10}{8.82}{10.89}{14.95}{9.17}{4}{2}{2}
	\desboxplot{0.97}{3.98}{6.45}{11.04}{68.96}{9.62}{31}{7}{8}
	\desboxplot{5.87}{6.31}{7.68}{10.54}{12.08}{8.44}{6}{1}{1}
	\desboxplot{2.02}{2.27}{2.51}{2.75}{3.00}{2.51}{2}{1}{1}
	\desboxplot{65.56}{65.56}{65.56}{65.56}{65.56}{65.56}{1}{1}{1}
	\desboxplot{1.64}{3.67}{5.69}{7.86}{10.03}{5.79}{3}{1}{1}
	\desboxplot{2.90}{4.17}{5.55}{7.42}{8.91}{5.73}{12}{1}{1}
	\desboxplot{1.36}{3.79}{5.95}{7.29}{20.17}{7.17}{12}{6}{8}
	\desboxplot{2.77}{3.13}{3.49}{3.84}{4.20}{3.49}{2}{1}{1}
	\desboxplot{-3.57}{3.33}{6.29}{8.91}{32.93}{7.33}{33}{11}{13}
	\desboxplot{7.07}{7.07}{7.07}{7.07}{7.07}{7.07}{1}{1}{1}
	\desboxplot{1.04}{2.33}{3.63}{4.78}{5.93}{3.53}{3}{1}{1}
	\desboxplot{3.62}{6.43}{8.84}{9.80}{34.13}{12.07}{10}{4}{2}
        \end{axis}
     \end{tikzpicture}
    \caption{Comparison of adaptation results for the standalone techniques.}
    \label{fig:ma_methods_single_method}
\end{figure*}

Finally, in~\figref{fig:ma_methods_single_method} we report results for all techniques included in the meta-analysis. These are based on samples where only a single method was used to adapt the acoustic model (\cf \figref{fig:ma_total} (middle)), spanning results for all adaptation clusters (\cf \figref{fig:ma_clusters}).  These should not be directly compared owing to differences in operating regions, but they offer an indication of the performance of the individual methods.

\subsection{Speech styles, applications, languages} \label{ssec:findings_corpora}

In this subsection, we analyze the efficacy of adaptation methods across acoustic and linguistic dimensions by reporting adaptation gains for different types of speech styles, applications (including ones with a large mismatch to the training conditions), and languages.

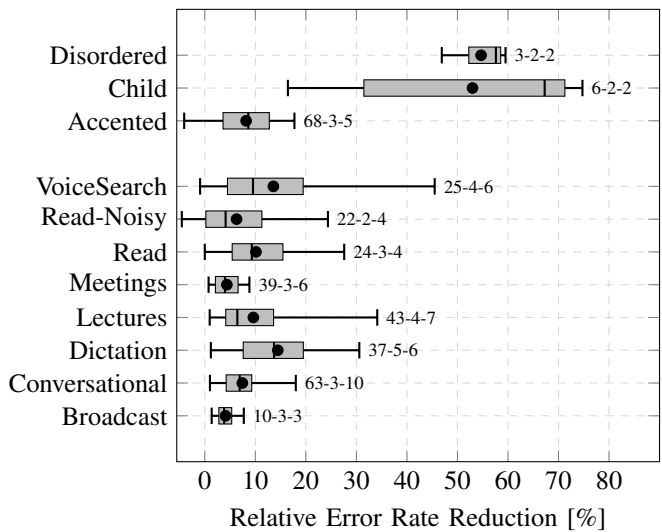
\begin{figure}
    \centering
           \begin{tikzpicture}
        \begin{axis}[
          height=7.6cm, width=8cm,
          xmin=-5.55, xmax=90.0, 
          ytick={1,2,3,4,5,6,7,8,10,11,12},
          yticklabels={Broadcast,Conversational,Dictation,Lectures,Meetings,Read,Read-Noisy,VoiceSearch, Accented, Child, Disordered},
          grid=major, grid style={dashed,gray!30},
          xlabel={Relative Error Rate Reduction [\%]},
          xtick={0,10,20,30,40,50,60,70,80},
          xticklabels={0,10,20,30,40,50,60,70,80},
        ]
	\desboxplot{1.36}{2.78}{3.83}{5.33}{7.72}{4.10}{10}{3}{3}
	\desboxplot{1.02}{4.23}{6.91}{9.29}{18.01}{7.43}{63}{3}{10}
	\desboxplot{1.20}{7.60}{13.70}{19.50}{30.58}{14.45}{37}{5}{6}
	\desboxplot{0.97}{4.15}{6.44}{13.61}{34.13}{9.61}{43}{4}{7}
	\desboxplot{0.75}{2.10}{4.00}{6.59}{8.83}{4.37}{39}{3}{6}
	\desboxplot{0.00}{5.42}{9.31}{15.47}{27.59}{10.12}{24}{3}{4}
	\desboxplot{-4.55}{0.21}{4.12}{11.30}{24.37}{6.29}{22}{2}{4}
	\desboxplot{-0.93}{4.49}{9.54}{19.47}{45.50}{13.57}{25}{4}{6}
	\desboxplot{100}{100}{100}{100}{100}{100}{0}{0}{0}
	\desboxplot{-4.08}{3.61}{8.61}{12.77}{17.74}{8.18}{68}{3}{5}
	\desboxplot{16.45}{31.48}{67.26}{71.26}{74.74}{52.98}{6}{2}{2}
	\desboxplot{46.90}{52.25}{57.61}{58.56}{59.52}{54.68}{3}{2}{2}
	
        \end{axis}
     \end{tikzpicture}
    \caption{Comparison of adaptation results for different speech styles}
    \label{fig:ma_styles}
\end{figure}

\figref{fig:ma_styles} compares gains as obtained for different speech styles. At the top we report  three special cases spanning disordered, children's, and accented speech (these are similar to the adaptation clusters from \figref{fig:ma_clusters}). As expected, acoustic models estimated largely from adult speech of healthy individuals perform poorly in these highly mismatched domains, especially for disordered and children's speech, and domain adaptation improves ASR by over 50\% average RERR.

\begin{figure}
    \centering
           \begin{tikzpicture}
        \begin{axis}[
          height=2.5cm, width=8cm,
          xmin=-5.08, xmax=90.0, 
          ytick={1,2},
          yticklabels={Proprietary,Public},
          grid=major, grid style={dashed,gray!30},
          xtick={0,10,20,40,60,80},
          xticklabels={},
          title={Speaker Adaptation}
        ]
	\desboxplot{-0.93}{4.00}{7.78}{12.31}{29.50}{9.17}{52}{9}{11}
	\desboxplot{-3.57}{3.57}{5.88}{8.91}{34.13}{7.27}{128}{13}{21}

        \end{axis}
     \end{tikzpicture} 
           \begin{tikzpicture}
        \begin{axis}[
          height=2.5cm, width=8cm,
          xmin=-5.08, xmax=90.0, 
          ytick={1,2},
          yticklabels={Proprietary,Public},
          grid=major, grid style={dashed,gray!30},
          xlabel={Relative Error Rate Reduction [\%]},
          xtick={0,10,20,40,60,80},
          xticklabels={0,10,20,40,60,80},
          title={Domain Adaptation}
        ]
	\desboxplot{2.00}{8.18}{15.64}{27.62}{59.52}{20.14}{34}{6}{8}
	\desboxplot{-4.08}{2.28}{5.70}{10.54}{74.74}{9.99}{50}{5}{6}

        \end{axis}
     \end{tikzpicture} 
    \caption{Performance of adaptation techniques as obtained on public and proprietary datasets.}
    \label{fig:ma_ownership}
\end{figure}
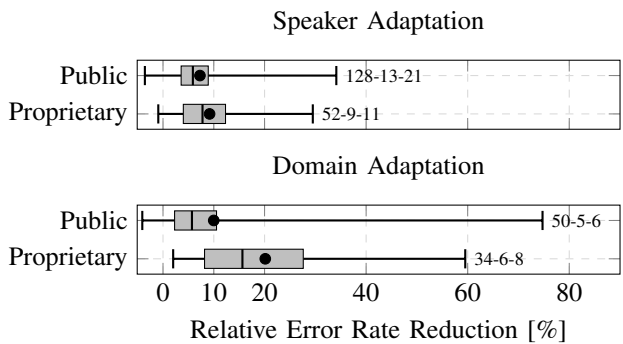

Performance gains from adapting models with accented speech are similar to that obtained on other speech tasks. Note that the presence of non-native speakers in (English) training corpora is fairly common, so the underlying acoustic models may learn to better normalize this variability at the training stage. Interestingly, adaptation brings relatively larger gains in commercial applications such as VoiceSearch and Dictation tasks (14\% RERR on average). This is also visible in \figref{fig:ma_ownership} comparing performance on public and proprietary data. We hypothesize that commercial data is more likely to contain a mix of speech from a diverse set of speakers (including non-native and child speech) and thus benefits more from adaptation. 
Another explanation could be that the public benchmarks have been around for some time, and systems built on these are likely to be more over-fitted in general. 

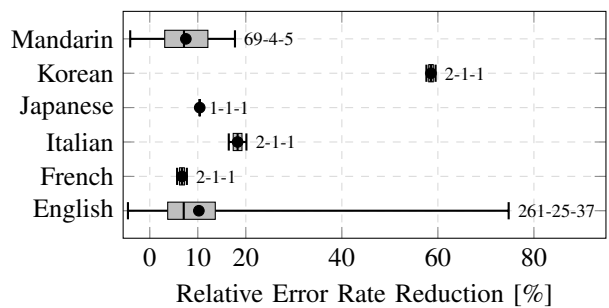
\begin{figure}
    \centering
           \begin{tikzpicture}
        \begin{axis}[
          height=4.6cm, width=8cm,
          xmin=-5.55, xmax=95.0, 
          ytick={1,2,3,4,5,6},
          yticklabels={English,French,Italian,Japanese,Korean,Mandarin},
          grid=major, grid style={dashed,gray!30},
          xlabel={Relative Error Rate Reduction [\%]},
          xtick={0, 10, 20, 40, 60, 80},
          xticklabels={0, 10, 20, 40, 60, 80},
        ]
	\desboxplot{-4.55}{3.73}{7.07}{13.64}{74.74}{10.22}{261}{25}{37}
	\desboxplot{5.69}{6.20}{6.71}{7.21}{7.72}{6.71}{2}{1}{1}
	\desboxplot{16.45}{17.37}{18.29}{19.20}{20.12}{18.29}{2}{1}{1}
	\desboxplot{10.40}{10.40}{10.40}{10.40}{10.40}{10.40}{1}{1}{1}
	\desboxplot{57.61}{58.09}{58.56}{59.04}{59.52}{58.56}{2}{1}{1}
	\desboxplot{-4.08}{3.11}{7.12}{12.10}{17.74}{7.51}{69}{4}{5}

        \end{axis}
     \end{tikzpicture}
    \caption{Adaptation gains for different languages.\vspace{-0.3cm}}
    \label{fig:ma_langs}
\end{figure}

Finally, \figref{fig:ma_langs} summarizes the adaptation performance for several languages. Note that speaker adaptation was performed on English, French, Japanese, and Mandarin while for Korean and Italian we only report adaptation gains for disordered and children's speech recognition. 
The overall improvements for non-English languages when adapting to speakers are similar to gains obtained for English when controlling for the adaptation method (\ie improvements are between 6 and 10\% average RERR), giving some evidence that adaptation helps to a similar degree for different languages, and that some of these primarily English-based findings generalize across languages.

\section{Summary and discussion}
\label{sec:summary}

The rapid developments in speech recognition over the past decade have been driven by deep neural network models of acoustics, deployed in both hybrid and E2E systems.  Compared to the previous state-of-the-art approaches based on GMMs, neural network-based systems have less constrained and more flexible models and are open to a richer set of adaptation algorithms, compared to previous approaches based on linear transforms of the model parameters and acoustic features.  

In this overview article we have surveyed approaches to the adaptation of neural network-based speech recognition systems.  We structured the field into embedding-based, model-based, and data augmentation adaptation approaches, arguing that this organization gives a more coherent understanding of the field compared with the usual split into feature-based and model-based approaches.  We presented these adaptation  algorithms in the context of speaker adaptation, with a discussion on their application to accent and domain adaptation.

A key aspect of this overview was a meta-analysis of recent published results for the adaptation of speech recognition systems.  The meta-analysis indicates that adaptation algorithms apply successfully to both hybrid and E2E systems, across different corpora and adaptation classes.

E2E modeling is less mature than the hybrid approach, and much of the research focus on E2E modeling is to improve the general modeling technology. Therefore, in this overview paper, many more adaptation methods were introduced in the context of hybrid systems. However, most adaptation technologies successfully applied to hybrid models by adapting acoustic model or language model should also work well for E2E models because E2E models usually contain sub-networks corresponding to the acoustic model and language model in hybrid models; this is supported by findings in our meta-analysis. 

Different from hybrid models in which components are optimized separately, E2E models are optimized using a single objective function. Therefore, E2E models tend to memorize the training data more and hence the generalization or robustness to unseen data \cite{li2014noise} is challenging to E2E models. Consequently, adaptation to new environment or new domain is very important to the large scale application of E2E models. We would expect more research toward this direction as E2E modeling becomes increasingly mainstream in ASR. 

Because the size of E2E models is much smaller than that of hybrid models,  E2E models have clear advantages when being deployed to device. Therefore, the personalization or adaptation of E2E models \cite{sim2019personalization, sim2019investigation, Li2020RNNT, Huang2020Rapid} is a rapidly growing area. While it is possible to adapt every user's model in the cloud and then push it back to each device, it is more reasonable to adapt the model on device, which requires adjusting the adaptation algorithm to overcome the challenge of limited memory and computation power \cite{sim2019investigation}. Another interesting direction for the adaptation of E2E models is how to leverage unpaired data especially text only data in a new domain. In \cite{Li2020RNNT}, several methods have been explored in this direction, but we are expecting more innovations there. 

Adaptation algorithms are often deployed for conditions in which there is very limited labeled data, or none at all.  In this case unsupervised and semi-supervised learning approaches are central, and indeed many current adaptation approaches strongly leverage such algorithms. However there are significant open research challenges in this area, particularly relating to unsupervised and semi-supervised training of E2E systems, using methods which are able to propagate uncertainty.  Current approaches often do this indirectly (\eg through T/S training), but more direct modeling of uncertainty would be desirable. 

Domain adaptation has become central to work in computer vision and image processing, as discussed in \secref{sec:intro}, with large scale base models (typically trained on ImageNet) being adapted to specific tasks.  The closest analogies to this in speech recognition are some of the domain recognition approaches discussed in \secref{sec:domain_adaptation} and for multilingual speech recognition.  The idea of shared multilingual representations and language-specific or language-adaptive output layers was proposed in 2013~\cite{ghoshal2013multilingual,heigold2013multilingual,huang2013cross} and has become a standard architectural pattern.  More recently several authors have proposed highly multilingual E2E systems, with a shared multilingual output layer \cite{toshniwal2018multilingual,kannan2019large,adams2019massively,pratap2020massively}, with the potential to be adapted to new languages.  

State-of-the-art NLP systems are characterized by an unsupervised, large-scale base model~\cite{vaswani2017attention, devlin2019bert} which may then be adapted to specific domains and tasks~\cite{raffel2020exploring}. An analogous approach for speech recognition would be based on the unsupervised learning of speech representations, from diverse and potentially multilingual speech recordings.  Initial work in this direction includes the unsupervised learning from large-scale multilingual speech data~\cite{conneau2020unsupervised,kawakami2020learning}.
More generally, deep probabilistic generative modeling has become a highly active research area, in particular through approaches such as normalizing flows~\cite{rezende2015variational,papamakarios2019normalizing,serra2019blow,prenger2019waveglow}.   Such deep generative models offer different ways of addressing the problem of adaptation including powerful approaches to data augmentation, and the development of rich adaptation algorithms building on a base model with a joint distribution over acoustics and symbols.   This offers the possibility of finetuning general encoders to specific acoustic domains, and adapting the decoder to specific tasks (such as speech recognition, speaker identification, language recognition, or emotion recognition), noting that classic adaptation to speakers can bring further gains~\cite{pascual2019learning,ravanelli2020mt}.

\bibliographystyle{IEEEtran}
\bibliography{adaptationRefs}


\vskip 0pt plus -1fil
\begin{IEEEbiography}[{\includegraphics[width=1.1in,height=2.25in,clip,keepaspectratio]{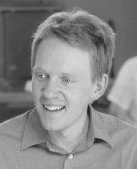}}] 
{Peter Bell} (M'09) received  a BA in mathematics (2002) and an MPhil in computer speech, text and internet technology (2005) from the University of Cambridge, and a PhD in automatic speech recognition (2010) from the University of Edinburgh.  He is a reader in speech technology at the School of Informatics, University of Edinburgh, with research  interests  that include  domain  adaptation,  regularization,  and  low-resource  methods  for  acoustic modeling.
\end{IEEEbiography}
\vskip -2\baselineskip plus -1fil
\begin{IEEEbiography}[{\includegraphics[width=1.1in,height=2.25in,clip,keepaspectratio]{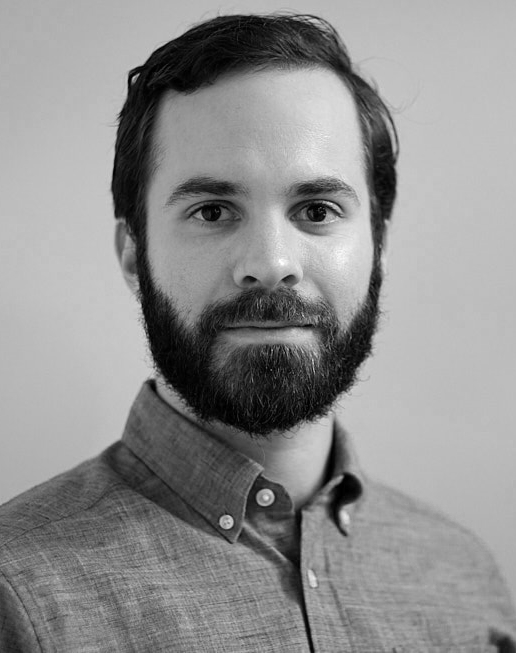}}]{Joachim Fainberg} (M'20) received a BMus in music and sound recording (Tonmeister) from the University of Surrey (2014), an MSc in artificial intelligence (2015) and a PhD in automatic speech recognition (2020) from the University of Edinburgh. He is currently at the Machine Learning Center of Excellence at JPMorgan Chase. His research interests include domain adaptation, and training methods for acoustic modeling.
\end{IEEEbiography}
\vskip -2\baselineskip plus -1fil
\begin{IEEEbiography}[{\includegraphics[width=1.1in,height=2.25in,clip,keepaspectratio]{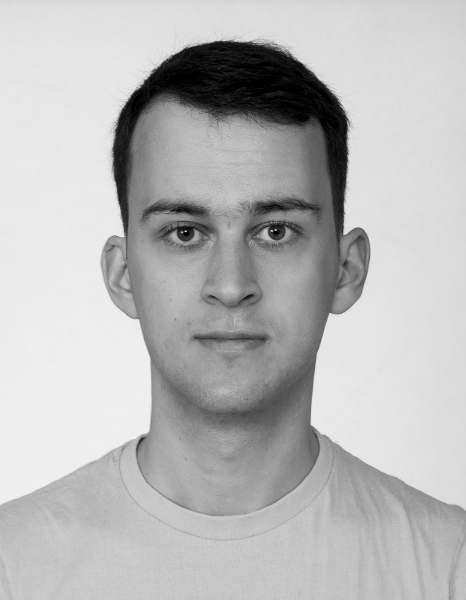}}]{Ondrej Klejch} (M'17) received a BSc (2013) an MSc (2015) in computer science from the Charles University in Prague, Czech Republic, and a PhD in automatic speech recognition (2020) from the University of Edinburgh. He is a post-doctoral researcher in automatic speech recognition at the University of Edinburgh. His research focuses on acoustic model adaptation using meta-learning approaches.
\end{IEEEbiography}
\vskip -2\baselineskip plus -1fil
\begin{IEEEbiography}[{\includegraphics[width=1.1in,height=2.25in,clip,keepaspectratio]{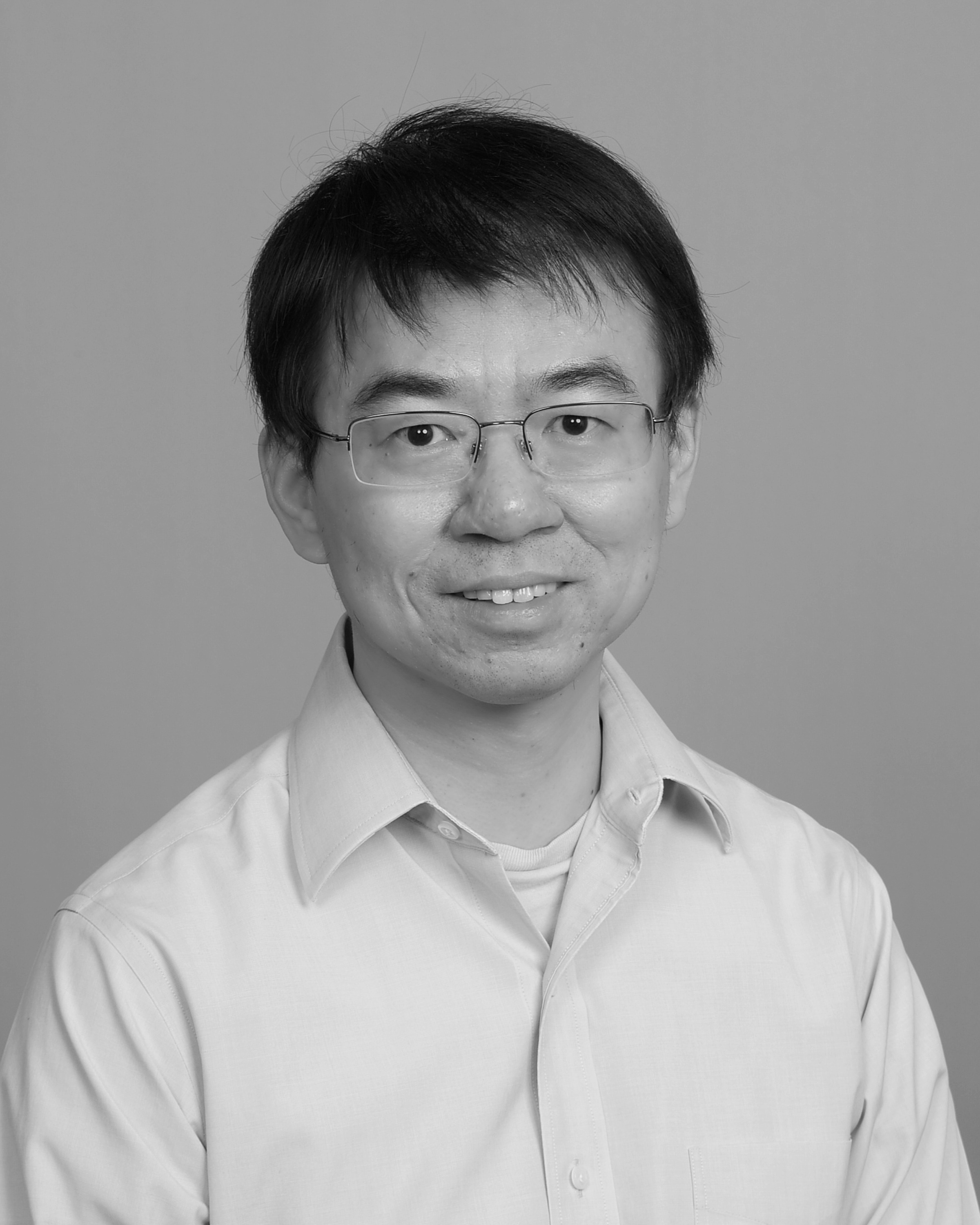}}]{Jinyu Li} (M’08) received the Ph.D. degree from the Georgia Institute of Technology in 2008. From 2000 to 2003, he was a Researcher with the Intel China Research Center and Research Manager in iFlytek, China. Currently, he is a Partner Applied Scientist in Microsoft Corporation, leading a team to design and improve speech modeling algorithms and technologies that ensure industry state-of-the-art speech recognition accuracy for Microsoft products.  Dr. Li is a member of the IEEE Speech and Language Processing Technical Committee. He also serves as Associate Editor of the IEEE/ACM Transactions on Audio, Speech, and Language Processing. 
\end{IEEEbiography}
\vskip -2\baselineskip plus -1fil
\begin{IEEEbiography}[{\includegraphics[width=1.1in,height=2.25in,clip,keepaspectratio]{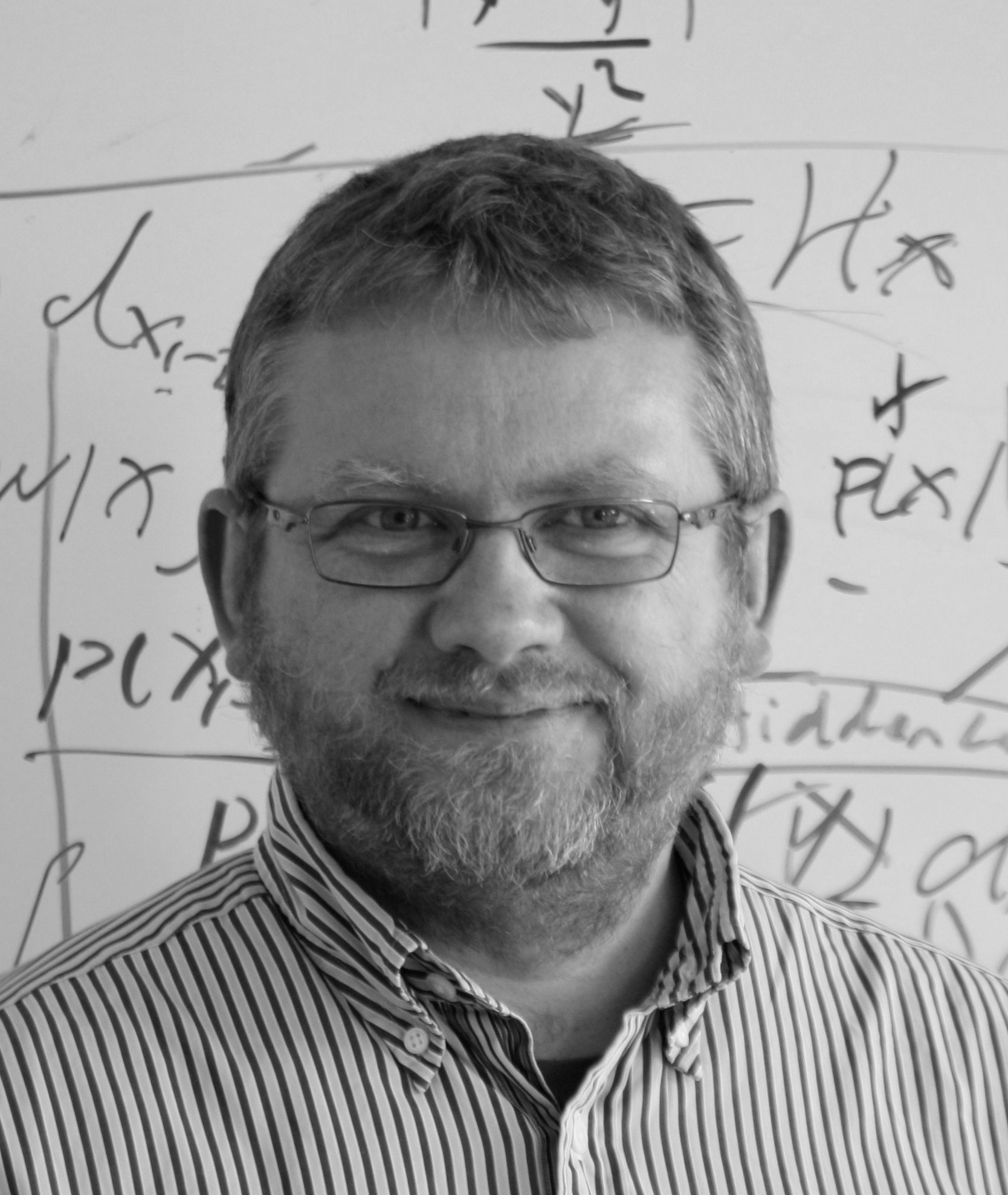}}]{Steve Renals} (M'91 -- SM'11 -- F'14) received a BSc in chemistry (1986) from the University of Sheffield, and an MSc in artificial intelligence (1987) and a PhD  in neural networks and speech recognition (1991) from the University of Edinburgh.  He is professor of speech technology at the School of Informatics, University of Edinburgh, having  previously held positions at ICSI Berkeley, the University of Cambridge, and the University of Sheffield. He has research interests in speech recognition, spoken language processing, neural networks, and machine learning.  Dr Renals is a fellow of ISCA (2016) and a senior area editor of the IEEE Open Journal of Signal Processing.
\end{IEEEbiography}
\vskip -2\baselineskip plus -1fil
\begin{IEEEbiography}[{\includegraphics[width=1.1in,height=2.25in,clip,keepaspectratio]{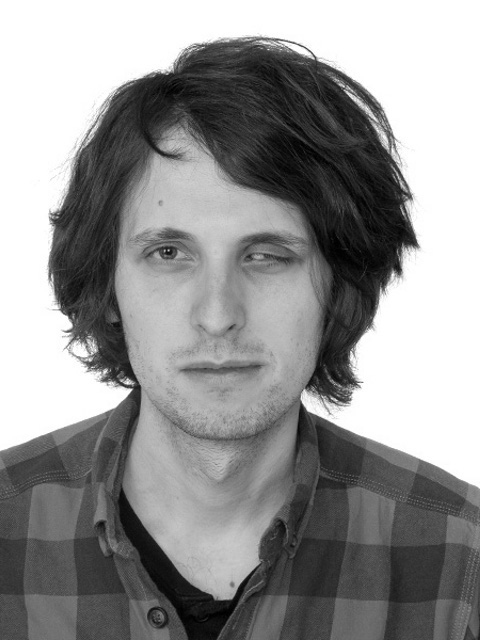}}]{Pawel Swietojanski} (M'12) received an MSc from AGH-UST in Cracow, Poland and a Ph.D. degree in computer science from the University of Edinburgh, U.K. He is currently a research scientist at Apple, and in the past held academic and research positions at UNSW Sydney and University of Edinburgh. His main research interests include machine learning and its applications in spoken language processing, with a particular focus on learning neural representations for acoustic modeling in speech recognition.
\end{IEEEbiography}


\end{document}